\documentclass[aps,pra,preprint,groupedaddress]{revtex4-2}

\usepackage{xcolor}					

\usepackage{placeins}
\usepackage{float}
\usepackage{graphicx}
\usepackage{gensymb}
\usepackage{amsmath,amssymb}
\usepackage{placeins}
\usepackage{multirow}
\usepackage{booktabs}
\usepackage[english]{babel}

\usepackage{subfigure}
\usepackage{overpic}
\usepackage{xfrac}

\begin{document}

\title{Unsteadiness of Shock-Boundary Layer Interactions \\ in a Mach 2.0 Supersonic Turbine Cascade}

\author{Hugo F. S. Lui}
\affiliation{University of Campinas, Campinas, S\~ao Paulo, Brazil}

\author{Tulio R. Ricciardi}
\affiliation{University of Campinas, Campinas, S\~ao Paulo, Brazil}

\author{William R. Wolf}
\email[]{wolf@fem.unicamp.br}
\affiliation{University of Campinas, Campinas, S\~ao Paulo, Brazil}

\author{James Braun}
\affiliation{Purdue University, West Lafayette, Indiana, USA}

\author{Iman Rahbari}
\affiliation{Purdue University, West Lafayette, Indiana, USA}

\author{Guillermo Paniagua}
\affiliation{Purdue University, West Lafayette, Indiana, USA}

\date{\today}

	\begin{abstract}
	
	The physics of shock-boundary layer interactions (SBLIs) in a supersonic turbine cascade at Mach 2.0 and Reynolds number 395,000, based on the axial chord, is investigated through a wall-resolved large eddy simulation. Special attention is given to the characterization of the low-frequency dynamics of the separation bubbles using flow visualization, spectral analysis, space-time cross correlations, and flow modal decomposition. The mean flowfield shows different shock structures formed on both sides of the airfoil. On the suction side, an oblique shock impinges on the turbulent boundary layer, whereas a Mach reflection interacts with the pressure side boundary layer. 
	The interactions taking place in the present turbine cascade show similarities and discrepancies with respect to more canonical cases. For example, the characteristic frequencies of the shock/bubble motions are comparable to those described in the literature of canonical cases. However, the suction side bubble leads to compression waves that do not coalesce into a separation shock, and a thin bubble forms on the pressure side despite the strong normal shock from the Mach reflection.
	Instantaneous flow visualizations illustrate elongated streamwise structures on the incoming boundary layers and their interactions with the shocks and separation bubbles. 
	The space-time cross-correlations reveal that the near-wall streaks drive the motion of the suction side separation bubble, which in turn promotes oscillations of the reattachment shock and shear layer flapping. 
	Organized motions in the SBLIs and their corresponding characteristic frequencies and spatial support are identified using proper orthogonal decomposition. 
	\end{abstract}

\maketitle

\section{Introduction}

Supersonic fluid machinery offers size and cost reduction in high-speed propulsion and power generation systems \citep{PANIAGUA201465, SOUSA2017247}, and for compact hydrocarbon cracking \citep{Rubini2021}. The detailed analysis of supersonic turbines is challenging predominantly due to the shock-boundary layer interactions (SBLIs) which arise when the detached oblique shock waves forming at the stator/rotor leading edges impinge on the boundary layers of the neighboring airfoils. The shocks impose intense adverse pressure gradients on the boundary layers that cause flow separation. This may lead to the formation of separation and reattachment shocks that interact with the turbulent boundary layers, resulting in strong pressure fluctuations and intense thermal loading which can compromise the turbine structural integrity. Moreover, the total pressure losses from the SBLIs also reduce the system overall efficiency 
\citep{DELERY1985,babinsky_harvey_2011,GAITONDE2015,Klinner2019,SPOTTSWOOD2019,Sandberg2022}.

When the boundary layer separates in a SBLI, a shear layer is developed to ensure a continuous variation of the flow at the recirculation region. Eventually, a Kelvin-Helmholtz (K-H) instability arises leading to high turbulence levels due to the formation of energetic coherent structures. A strong mixing also takes place where an energy transfer occurs from the external mean flow towards the separation region, leading to the flow reattachment \citep{babinsky_harvey_2011}. Typically, the separated flow has a ``low-frequency dynamics'' related to the oscillations of the separation bubble, motion of the separation and reattachment shocks, flapping of the shear layer, and the K–H instabilities. However, the incoming turbulent boundary layer excites a broad range of frequencies and combinations of all these features result in a broadband frequency spectrum superposed by low-frequency tonal peaks \citep{clemens2014,DUSSAUGE2006}.

Several authors have studied the low-frequency events taking place in SBLIs using experimental \citep{DUSSAUGE2006,dupont_haddad_debieve_2006, ganapathisubramani_clemens_dolling_2009, piponniau2009,Combs_2018,murphree_2021} and numerical \citep{pirozzoli2006, wu_martin_2008, Touber2009, priebe2012, morgan2013,Aubard2013,moreno_2014, agostini2015, adler_gaitonde_2018, vyas2019,adler_gaitonde_2020,hu_2021,Deshpande2021,bugeat_2022} techniques. Based on previous investigations, one can categorize the main flow phenomena in three different frequency bands: low frequency oscillations from the reflected shock motion and large-scale bubble breathing ($0.02 < St < 0.05 $), low to midfrequency motions of the separation bubble and flapping of the shear layer ($St \approx 0.1$), and the K-H instabilities ($0.3 < St < 0.5$).  Here, the Strouhal number $St = fL_{SB}/U_\infty$ is defined based on the length of the separation bubble $L_{SB}$ and the inlet velocity $U_\infty$. The sources of low-frequency dynamics have been the subject of debate because their driving mechanisms have not been fully characterized \citep{clemens2014}. 

Several studies have shown that the upstream boundary layer fluctuations are responsible for the low-frequency unsteadiness. In this context, 
\citet{beresh2002} experimentally investigated the relationship between upstream boundary layer properties and the separation shock foot motion in a Mach 5 compression ramp by using particle image velocimetry (PIV) and wall pressure measurements. They observed that the downstream shock motion is associated with positive velocity fluctuations in the incoming boundary layer while the upstream shock motion is related to negative velocity fluctuations. \citet{ganapathisubramani_clemens_dolling_2009} conducted PIV measurements to study the low-frequency dynamics of a Mach 2 SBLI in a compression ramp. Their measurements indicated the presence of low- and high-speed superstructures in the upstream turbulent boundary layer. These authors reported that the passage of high-speed (low-speed) structures through the separated flow leads to downstream (upstream) motion of the separation line. 

\citet{porter_2019} performed high-fidelity simulations of the flow over a Mach 2 compression ramp to investigate the influence of boundary layer disturbances on the low-frequency unsteadiness. Their statistical analysis revealed that the separation bubble responds selectively to specific large-scale, near-wall perturbations in the incoming boundary layer. They also suggested that the external forcing by the upstream turbulent boundary layer drives a weakened global mode of the separation region.
More recently, \citet{baidya_2020} studied a Mach 2 shock-boundary layer interaction using PIV. These authors reported that the passage of large-scale low (high) momentum structures through the interaction region causes the expansion (contraction) of the separation bubble which, in turn, leads to the reflected shock oscillations. 

While the previous authors indicated that upstream mechanisms may be the cause of the reflected shock and bubble motions, some studies have associated the low-frequency unsteadiness in SBLIs with downstream mechanisms. For example, \citet{pirozzoli2006} performed direct numerical simulations (DNS) to investigate a Mach 2.25 shock impinging on a turbulent boundary layer. They reported that the interaction of shear layer vortical structures with the incident shock generates acoustic disturbances that propagate upstream inside the subsonic region of the boundary layer. The authors concluded that such effect leads to motion of the separation bubble and the subsequent reflected shock oscillation. \citet{Touber2009} carried out a large eddy simulation (LES) of a Mach 2.3 oblique shock impinging on a flat plate. The authors performed a linear stability analysis and found an unstable global mode inside the bubble that can increase or decrease the separation region depending on the sign of the amplitude function. They also computed wall pressure space-time correlations and detected low-frequency waves in the initial portion of the bubble which propagate upstream. These authors suggest that this is an effect of the global mode on the separation bubble.

\citet{priebe2012} performed a DNS to study a Mach 2.9 flow over a compression ramp. Their statistical analysis indicated that the shock motion is related to breathing of the separation bubble and flapping of the shear layer. They also observed a low-frequency mode that may be related to the output response of the SBLI system, and this could be linked to a global instability in the separated flow. \citet{adler_gaitonde_2018} investigated the linear response to small perturbations in a Mach 2.3 shock-flat plate impingement. Their results showed that the SBLI promotes a global linear instability which is sustained through constructive feedback of perturbations inside the separation region. The self-sustaining perturbations have a significant influence on the shock motion and the bubble breathing. \citet{moreno_2014} and Adler \& Gaitonde \cite{adler_gaitonde_2020, adler2022} have also studied 3D effects on SBLIs which pose a new challenge for comprehension of the physical mechanisms taking place in complex high-speed flow configurations. This topic is relevant for the analysis of supersonic turbines where twist and tip effects are important, and it should receive further attention in future investigations.

In this work, a LES is performed to investigate the physics of SBLIs in a supersonic turbine cascade at Mach 2.0 and Reynolds number 395,000 based on the axial chord. In Secs. \ref{section:numerical_methodology} and \ref{section:configurations}, the numerical methods as well as the flow and mesh configurations are described. Results are then presented in Sec. \ref{section:results}.
First, the mean flow is analyzed in Sec. \ref{section:visualization1}, followed by a characterization of the spatial and temporal variations of the separation bubbles due to the passage of near-wall streaks from the upstream boundary layer, in Sec.  \ref{section:visualization2}. Next, spectral and statistical analyses are performed in Sec. \ref{section:spectral} to categorize the relevant frequencies involved in the flow, as well as to investigate the influence of upstream or downstream disturbances on the flow features appearing in the SBLIs. Finally, the filtered (spectral) proper orthogonal decomposition (POD) is used in Sec. \ref{section:POD} to identify the organized low-frequency motions in the flow. 

The present study provides insights on the SBLIs in a stator cascade, a configuration that is more complex than the canonical compression ramp and the frequently investigated oblique shock-boundary layer interaction on a flat plate.
Particular emphasis is placed on characterizing the low-frequency unsteadiness related to the separation bubbles, the shock waves and the flapping of the shear layers. In this context, we describe how the mechanisms responsible for the bubble and shock motions are related.
Understanding the physics of SBLI phenomena for the present configuration with wall curvature is essential to develop efficient supersonic fluid-machinery, including the development of effective cooling strategies and active flow control for multi-point operation.

\section{Numerical Methodology} 
\label{section:numerical_methodology}

\subsection{Governing equations}

A large eddy simulation is employed to solve the compressible Navier-Stokes equations in a curvilinear system with coordinates $\xi^i $, $i = 1,2,3$, written in terms of the contravariant velocity components $u^i$ as
\begin{equation}
\frac{\partial }{\partial t} \left( \sqrt{g} \rho \right) + \frac{\partial }{\partial \xi^i} \left(\sqrt{g} \rho  u^i \right) = 0 
\mbox{ ,}
\label{eq:continuity_curv3}
\end{equation}
\begin{eqnarray}
\frac{\partial }{\partial t} \left(\sqrt{g} \rho u^i\right) + \frac{\partial }{\partial \xi^j} \left[\sqrt{g} \left(\rho u^i u^j + g^{ij}p - \tau^{ij}  \right)\right] + \Gamma^{i}_{kj} \sqrt{g} \left(\rho u^k u^j + g^{kj}p - \tau^{kj}  \right)  = f_{b}^i
\label{eq:momentum_curv3}
\mbox{ ,}
\end{eqnarray}
and
\begin{eqnarray}
\frac{\partial}{\partial t} \left(\sqrt{g} E\right) +  \frac{\partial}{\partial \xi^i} \Bigg\{ \sqrt{g} \left[ \left(E + p \right) u^i -  \tau^{ij} g_{jk} u^k  + q_i \right] \Bigg\} = g_{ij} u^i f_{b}^j 
\mbox{ .}
\label{eq:energy_curv3}
\end{eqnarray}
The above set of equations represents the continuity, momentum and energy equations. The total energy $E$, the viscous stress tensor $\tau_{ij}$ and the heat flux $q_i$ for a fluid obeying Fourier's law are given, respectively, by
\begin{equation}
E = \frac{p}{\gamma - 1} + \frac{1}{2} \rho g_{ij} u^i u^j 
\mbox{ ,}
\end{equation}
\begin{equation}
\label{eq:stress_tensor}
\tau^{ij} = \frac{\mu}{Re_{a}} \left[g^{ik}  \left( \frac{\partial u^j}{\partial \xi^k}+ u^l  \Gamma^{j}_{lk} \right) + g^{jk}\left( \frac{\partial u^i}{\partial \xi^k} + u^m  \Gamma^{i}_{mk} \right) \right] + \left(\beta^{\ast} - \frac{2}{3} \frac{\mu}{Re_{a}} \right) \left[ g^{ij} \left(\frac{\partial u^k}{\partial \xi^k} + u^l \Gamma^{k}_{kl} \right)  \right] 
\mbox{ ,}
\end{equation}
and
\begin{equation}
\label{eq:heat_flux}
q_i = - \left(\frac{\mu}{Re_{a} Pr} + \kappa^{\ast}\right) g^{ij}  \frac{\partial T}{\partial \xi^j}
\mbox{ .}
\end{equation}
Assuming the gas to be calorically perfect, the set of equations is closed by the equation of state
\begin{equation}
p = \frac{\left(\gamma - 1 \right)}{\gamma} \rho T \mbox{ .}
\end{equation}

In the equations above, $t$ represents the time, $\rho$ is the density, $p$ is the pressure, $\mu$ is the dynamic viscosity coefficient, $T$ is the static temperature, and $\gamma$ is the ratio of specific heats. The equations are solved in nondimensional form where the length, velocity components, density, pressure, temperature and time are nondimensionalized by the axial airfoil chord $c_{x}$, inlet speed of sound $a_{\infty}$, inlet density $\rho_{\infty}$, $\rho_{\infty} a_{\infty}^2$, $\left(\gamma - 1 \right) T_{\infty}$ and $c_{x}/a_{\infty}$, respectively. 
Here, $Re_{a}$ is the Reynolds number based on the speed of sound, defined as $Re_{a} = Re/M_{\infty}$, where the Reynolds and Mach numbers are calculated as $Re = \rho_{\infty} U_{\infty} c_{x}/ \mu_{\infty}$ and $M_{\infty} = U_{\infty}/a_{\infty}$, respectively. The terms $U_{\infty}$, $T_{\infty}$ and $\mu_{\infty}$ represent the flow velocity, temperature and dynamic viscosity coefficient computed at the cascade inlet. 

The Prandtl number is given by $Pr = \mu_{\infty} c_{p} / \kappa_{\infty}$, where $c_p$ is the specific heat at constant pressure and $\kappa_{\infty}$ is the inlet thermal conductivity.  The viscosity is computed using the nondimensional Sutherland's law, written as
\begin{equation}
\mu = \left[(\gamma_{\infty} - 1) T\right]^\frac{3}{2}   \frac{1 + \frac{S_{\mu}}{T_{\infty}}}{T(\gamma_{\infty} - 1) + \frac{S_{\mu}}{T_{\infty}}} \mbox{ ,}
\end{equation}
where $S_{\mu}$ is Sutherland's constant. The parameters $\beta^{\ast}$ and $\kappa^{\ast}$ are the artificial bulk viscosity and thermal conductivity, computed using a shock capturing scheme. The term $f_{b}^i$ is the contravariant body force component employed for boundary layer tripping. In the governing equations, $g_{ij}$ and $g^{ij}$ are the covariant and contravariant metric tensors, respectively, $\sqrt{g}$ is the Jacobian of the covariant metric tensor, and $\Gamma^{i}_{jk}$ represents the Christoffel symbols of the second kind. Further details about the present formulation can be found in Refs. \citep{Aris:1989, Bhaskaran}. 

\subsection{Numerical schemes}

The spatial discretization of the governing equations is performed using a sixth-order accurate compact scheme \citep{Nagarajan2003} implemented on a staggered grid. A sixth-order compact interpolation method is also used to obtain fluid properties on the staggered nodes.
The sixth-order compact filter presented by \citet{Lele1992} is applied in flow regions far away from solid boundaries at each time step to control numerical instabilities which may arise from mesh nonuniformities and interpolations between overlapping grids.
An explicit subgrid scale model is not applied, however, the transfer function associated with compact filters has been shown to provide an approximation to subgrid scale models \citep{Mathew_etal_2003}. 

Two grids are employed in the present simulations: one is a body-fitted O-grid block which surrounds the vane and the other is a Cartesian block employed to enforce the pitchwise periodicity. In the O-grid, the time integration of the equations is carried out by the implicit second-order scheme of \citet{Beam1978} to reduce the stiffness problem typical of boundary layer grids. In the background Cartesian block, a third-order Runge-Kutta scheme is used for time advancement of the Navier-Stokes equations. A fourth-order Hermite interpolation scheme \citep{Delfis2001,Bhaskaran} is used to exchange information between grid blocks in the overlapping zones. The simulation time step is determined based on a variable time step algorithm \citep{Cook2007} considering the inviscid Courant–Friedrichs–Lewy (CFL) condition and the timescales associated with $\mu$, $\beta$ and $\kappa$. Further details about the numerical procedure can be found in Refs. \citep{Nagarajan2003, bhaskaran_thesis}. The code has been previously validated for simulations of unsteady compressible flows \cite{bhaskaran_thesis,Wolf2012}, including the flow through a turbine vane cascade \cite{Bhaskaran}. Further validation of the numerical tool for a supersonic flow involving a laminar SBLI is also provided in the Appendix.

Shock capturing schemes are required to introduce a minimal, yet sufficient, numerical dissipation in the vicinity of shock waves without damping the small scales of turbulence. In this work, the localized artificial diffusivity (LAD) \citep{Cook2007} is used to compute the artificial fluid properties, which are added to their physical transport counterparts in Eqs. \eqref{eq:stress_tensor} and \eqref{eq:heat_flux}. The specific implementation employed is the method LAD-D2-0 proposed by \citet{Kawai2010} with no artificial shear viscosity. 

To promote the transition to turbulence on the airfoil boundary layers, an artificial body force has been included in the momentum and energy equations [see the right-hand side of Eqs. (\ref{eq:momentum_curv3}) and (\ref{eq:energy_curv3})] as described by \citet{Sansica}, where a time-periodic unsteady actuation and random spanwise treatment are assumed. The body-force tripping is applied at $0.22 < x < 0.27 $ on the suction side, and at $0.10 < x < 0.15 $ on the pressure side. The tripping is applied along the wall-normal region up to a distance of $0.001c_{x}$. The magnitude of the forcing is chosen experimentally to guarantee a bypass transition.

\subsection{Proper orthogonal decomposition}

Identification and analysis of the organized motions related to the separation bubbles, shock waves and shear layers are of utmost importance to understand their overall role in the SBLIs. Modal decomposition techniques have emerged as powerful tools for analyzing the unsteady flows \citep{Taira_AIAAJ2017} facilitating the assessment of organized structures which may play an important role in the flow dynamics. In recent years, these techniques have been applied for the investigation of low-frequency unsteadiness in SBLIs \citep{humble_2009,grilli_2012,priebe_2016,pasquariello_2017,Nichols_2017,Vanstone_2019,Hu_2019,hu_2021,sasaki_2021,Hoffman_2022}. 

Coherent flow structures can be defined in terms of fluctuation quantities, such that a Reynolds decomposition is employed to split the instantaneous flow field $\mathbf{Q}(\mathbf{x},t)$ into an averaged time-independent component $\mathbf{\bar{Q}}(\mathbf{x})$ and its fluctuation field $\mathbf{Q}^{'}(\mathbf{x},t)$. Among the various modal decomposition techniques, the proper orthogonal decomposition is a data-driven method used to extract organized motion \citep{Sirovich1990} based on data correlation by seeking the best $L_2$-norm representation of the fluctuation flowfield. It requires the calculation of the covariance matrix $\mathbf{C}$, defined as
	\begin{equation}
	\mathbf{C} = \mathbf{Q'\,}^{T} \mathbf{W} \ \mathbf{Q'} \mbox{ ,}
	\end{equation}
	where the matrix $\mathbf{W}$ contains the integration weights based on the spatial discretization.
	The singular value decomposition (SVD) is used to decompose the matrix $\mathbf{C}$, which results in orthonormal empirical functions $\boldsymbol{\phi}_{i}(\mathbf{x})$ and $\mathbf{a}_{i}(t)$, respectively the spatial and temporal POD modes, with an amplitude scaling factor given by the singular values $\boldsymbol{\sigma}_i$. 
	In this decomposition, the maximum number of modes $M$ is equal to the rank of $\mathbf{Q}'$ which, for the present high-fidelity simulation, is given by the number of time samples (snapshots).
	A linear superposition of the modes can be performed without loss of information according to
	\begin{equation}
	\mathbf{Q}^{'}(\mathbf{x},t) = \sum_{i=1}^{M} \mathbf{a}_{i}(t) \boldsymbol{\sigma}_i \boldsymbol{\phi}_{i}(\mathbf{x}) \mbox{.}
	\end{equation} 

The POD modes are ranked solely based on their energetic content and a single mode may contain a broad range of frequencies. Since the dominant flow features in SBLIs have distinct tonal peaks \citep{clemens2014,adler_gaitonde_2018}, the filtered/spectral POD technique \citep{sieber2016,sieber2017nature} is employed to remove broadband content while still retaining the spectral information around a central frequency. This method allows for a continuous shifting from the energetically optimal POD to the standard Fourier decomposition. This is achieved by a convolution of the covariance matrix $C$ with a filter function $g$ as
\begin{equation}
\tilde{\mathbf{C}}(t) = \mathbf{C}(t) * g(\tau) \mbox{ ,}
\end{equation}
and the SVD is applied to the filtered matrix $\tilde{\mathbf{C}}$. 

This approach is beneficial since the energy-ranking property of the POD is still retained. Also, it benefits from a more organized temporal dynamics since the filtering operation ultimately works as a band-pass filter of the temporal modes. This is important since turbulent flows are composed of multiple frequencies and chaotic dynamics, which may impact the coherent structures and lead to intermittent events or flow modulation. 
With the filtered POD, it is possible to analyze specific frequency bands rather than a single frequency content, as is the case of the Fourier transform. One important remark of this methodology is that the temporal modes depend on the filter transfer function, including its width and shape such that the filter parameters are case-dependent \citep{Jean_2017}.

\section{Flow and Mesh Configurations}
\label{section:configurations}

\begin{figure}
	
	\begin{overpic}[trim = 1mm 17mm 1mm 15mm, clip,width=0.99\textwidth]{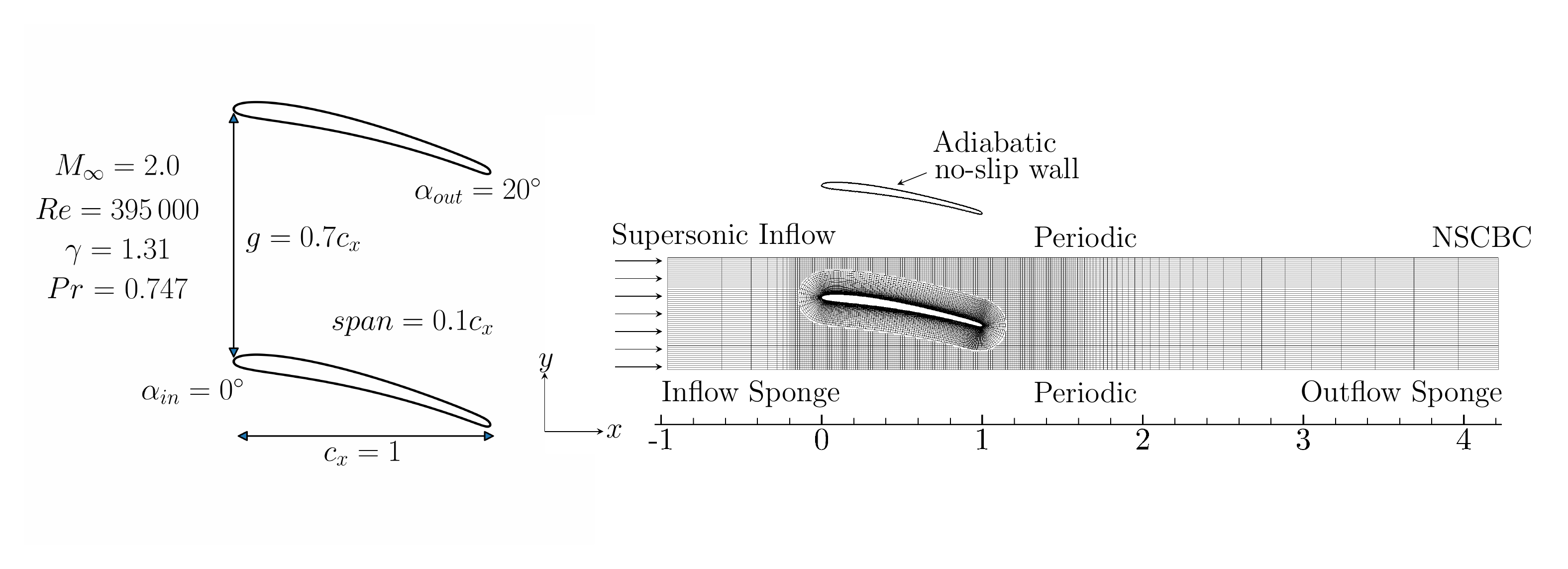}
		\put(0,23){(a)}
		\put(35,23){(b)}		
	\end{overpic} 
	\caption{Schematics of (a) flow configuration and geometrical parameters, and (b) computational domain skipping every $5$ grid points.}
	\label{fig:schematic}
\end{figure}

This section presents details of the flow configuration investigated and describes the computational grids employed in the LES calculations. Figure \ref{fig:schematic} (a) shows the flow conditions and geometrical parameters. The inlet Mach number is $M_{\infty}$ = 2.0 and the Reynolds number based on the inlet velocity and axial airfoil chord is $Re$ = 395,000. The fluid is assumed to be a calorically perfect gas, where the ratio of specific heats is $\gamma = 1.31$, the Prandlt number is $Pr = 0.747$ and the ratio of the Sutherland constant over inlet temperature is 0.07182. 
A realistic turbine cascade would be subjected to incoming turbulence. However, in the present analysis, the inlet consists of a ``clean'' inflow. This setup allows a more direct comparison against other SBLI studies available in the literature for canonical flow configurations.

The turbine geometry is designed to ensure the starting process of the passage at the specified inlet Mach Number and flow turning \citep{PANIAGUA201465}. More details about the stator geometry and the inlet flow conditions can be found in Ref. \citep{LIU2019}. This flow configuration was also studied in Ref. \citep{Lui2021}, where a comparison between the present LES and Reynolds-averaged Navier-Stokes (RANS) solutions presented good agreement in terms of the mean flow quantities, especially on the suction side. Further comparisons against RANS solutions are presented in the Appendix in terms of mean flow quantities. These comparisons allow an assessment of the overall flow topology including the shock waves and their interactions with the boundary layers.

Figure \ref{fig:schematic} (b) shows a schematic of the overset grid employed in the LES along with the employed boundary conditions. The body-fitted O-grid block has $1280 \times 300 \times 144$ points in the streamwise, wall-normal and spanwise directions, respectively. The O-grid is embedded in a background Cartesian grid block of size $960 \times 280 \times 72$. Overall, the computational grid is composed of approximately $75\times10^6$ points. No-slip adiabatic boundary conditions are applied along the airfoil surface. A supersonic inflow boundary condition is employed, while the Navier-Stokes characteristic boundary conditions (NSCBC) \citep{Poinsot1992} are applied at the outflow. A damping sponge is also applied near the inflow and outflow boundaries to minimize reflection of numerical disturbances \citep{ISRAELI1981,Nagarajan2003}. Periodic boundary conditions are used in the $y$-direction of the background grid to simulate a linear cascade of vanes. Periodic boundary conditions are also applied in the spanwise direction to avoid the modeling complexities near the tip and end-wall in the present analysis.

The standard recommendations in terms of grid resolution \citep{Georgiadis2010} are followed in the present wall-resolved LES. The near-wall resolution is kept in the range given by $10<\Delta s^+<40$, $0.2<\Delta n^+<0.4$, and $5<\Delta z^+<12$, where $s$, $n$, and $z$ represent the tangential, wall-normal and spanwise flow coordinates. These numbers are computed for regions where the boundary layers are fully developed and in equilibrium, away from the tripping and separation regions. Therefore, the present grid resolution is similar or higher than those employed by other high-fidelity simulations of SBLIs \cite{loginov_adams_zheltovodov_2006,Touber2009, pasquariello_2017,hu_2021}.

The spanwise extent of the computational domain is equal to 10$\%$ of the axial chord, which is equivalent to 14 incoming boundary layer thickness $\delta_0$ on the suction side. This value is similar or higher than those employed by previous numerical studies \cite{loginov_adams_zheltovodov_2006,Touber2009, pasquariello_2017,hu_2021}. In addition, two-point correlations were computed along the span at various chord locations and wall-normal distances for pressure and velocity fluctuations. Results (not presented here for brevity) showed that these quantities have a rapid correlation decay along the span, thus ensuring that the computational domain is sufficiently wide to not affect the turbulence dynamics.

The simulation employs a variable time step $\Delta t \approx 2.6 \times 10^{-5}$, nondimensionalized with respect to the inlet velocity and computed based on an inviscid CFL number of 1.2.
After the initial transient period, the simulation is run for approximately 26 nondimensional time units and it covers around $12$ cycles of the low-frequency motion ($St = 0.045$). The 3D flow data and the spanwise-averaged flowfields are recorded every $0.006$ and $0.0012$ time units, respectively, for the calculation of statistics.

\section{Results}
\label{section:results}

In this section, the physical mechanisms taking place on the suction and pressure sides of the airfoil are investigated using flow visualization and statistical analysis which allow for a comparison between the different features of the SBLIs. 

\subsection{Visualization of instantaneous and mean flows}
\label{section:visualization1}

\begin{figure}
	\centering	
	\begin{overpic}[trim = 1mm 1mm 1mm 1mm, clip,width=0.48\textwidth]{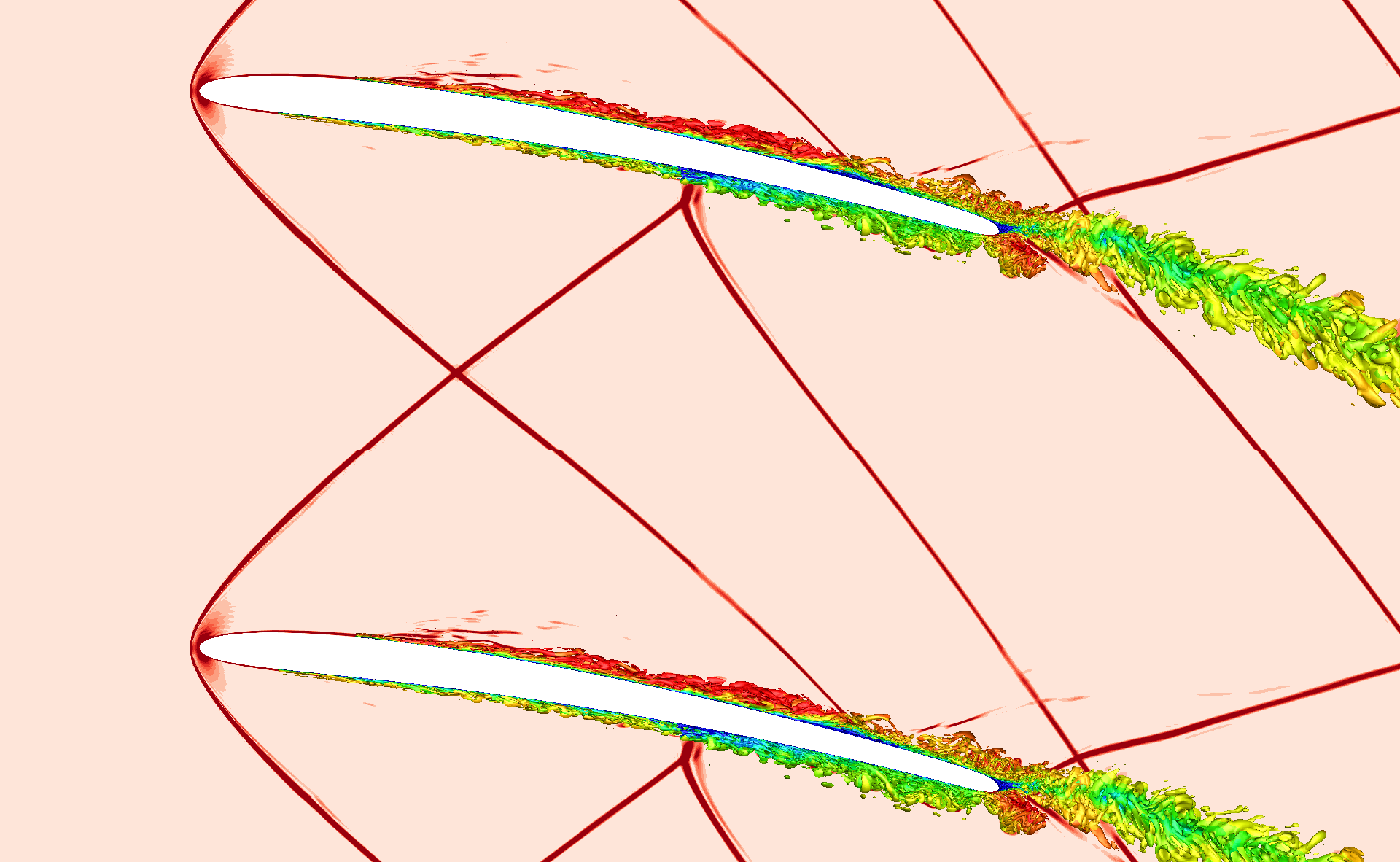}
		\put(3,55){(a)}
	\end{overpic} 
	\begin{overpic}[trim = 1mm 1mm 1mm 1mm, clip,width=0.48\textwidth]{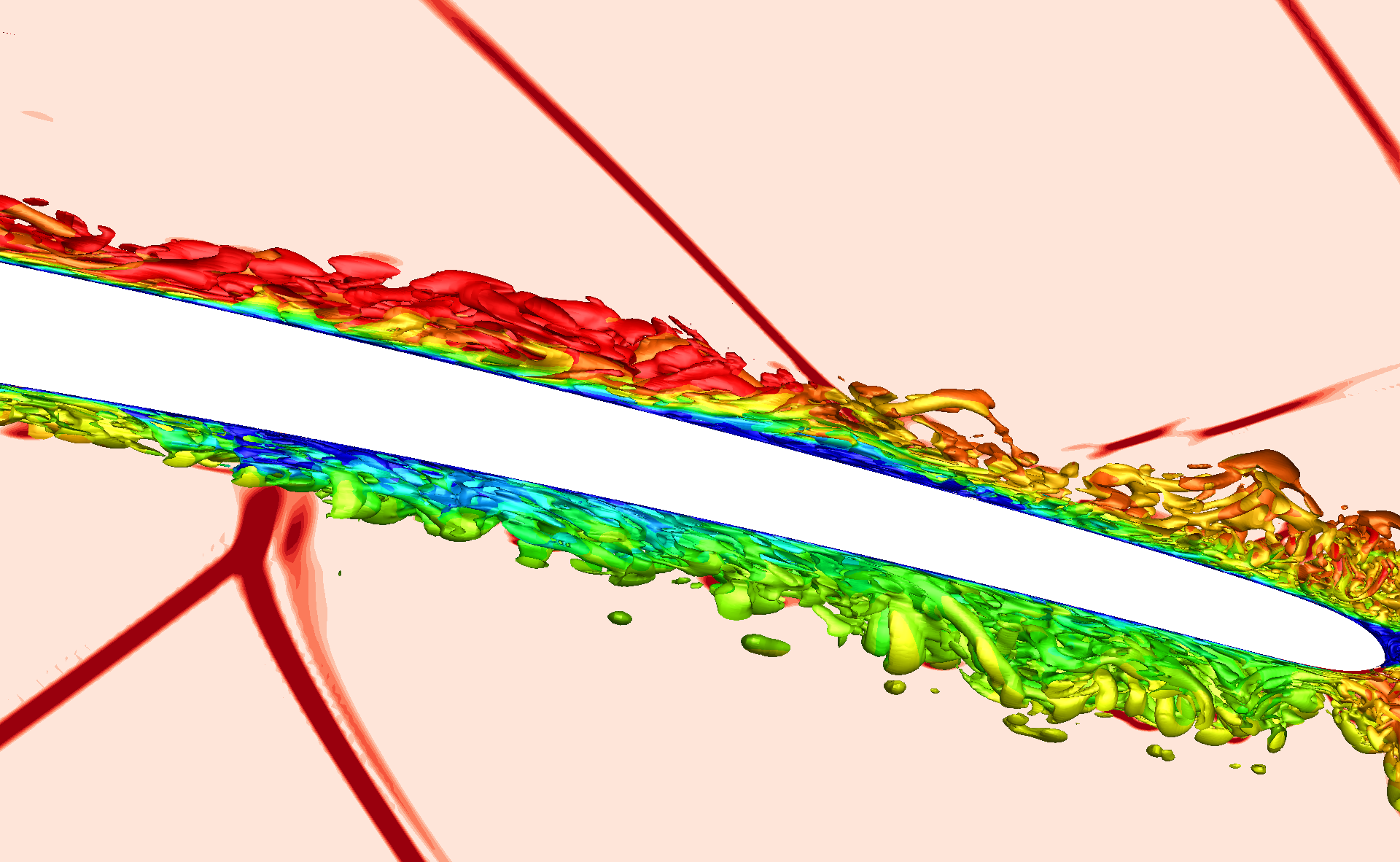}
		\put(3,55){(b)}
	\end{overpic} \\
	\vspace{2mm}
	\begin{overpic}[trim = 1mm 1mm 1mm 1mm, clip,width=0.48\textwidth]{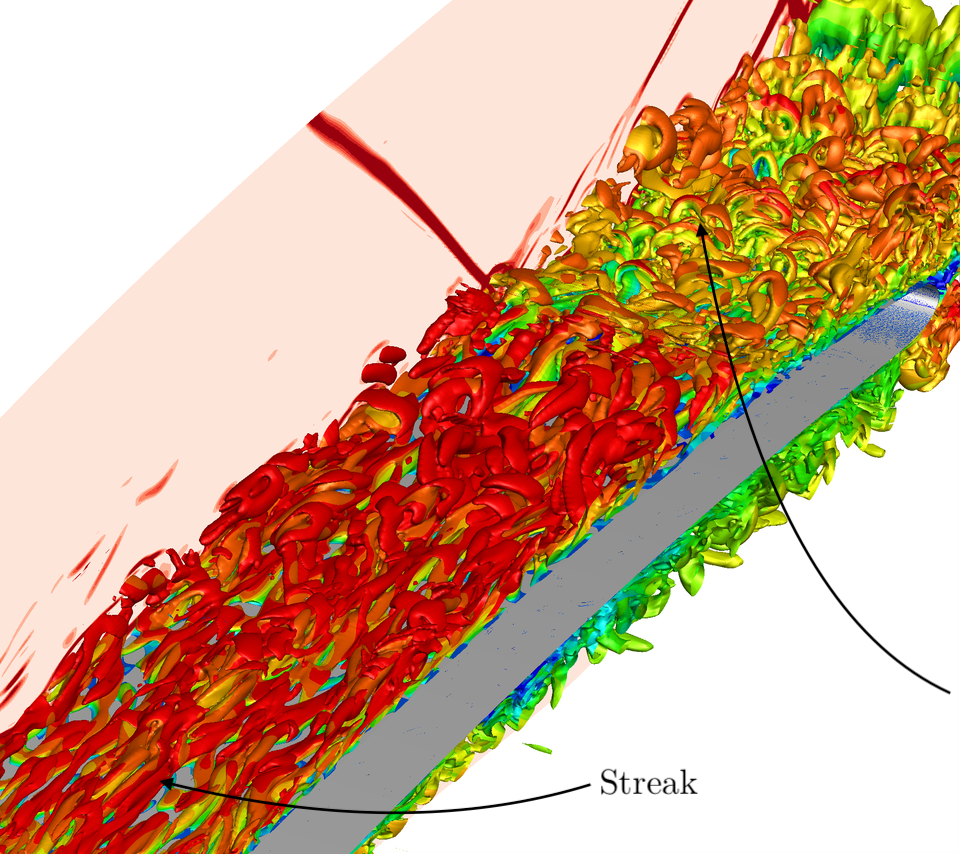}
		\put(3,83){(c)}
	\end{overpic} 
	\begin{overpic}[trim = 1mm 1mm 1mm 1mm, clip,width=0.48\textwidth]{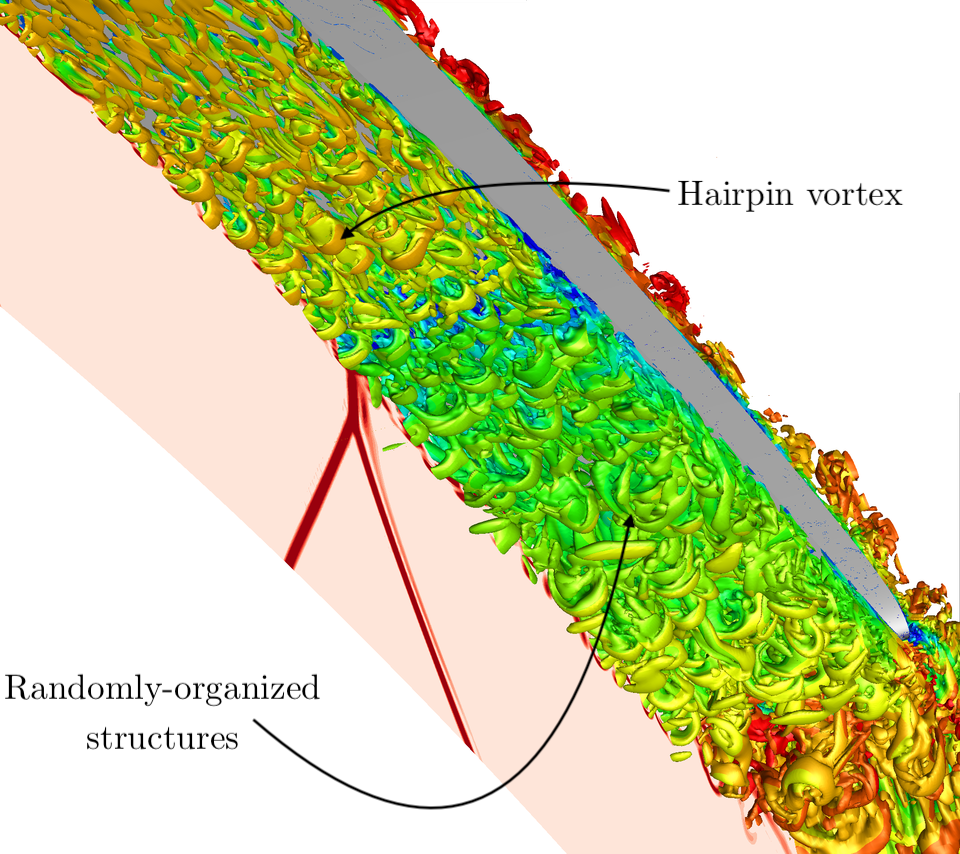}
		\put(90,83){(d)}
	\end{overpic}
	\caption{Iso-surfaces of $Q$-criterion colored by $u$-velocity component with red contours, in the background, revealing the shock wave structure by displaying the magnitude of density gradient $|\nabla \rho|$: (a) view of linear cascade, (b) detail view of the SBLIs, (c) view of suction side boundary layer, (d) view of pressure side boundary layer.}
	\label{fig:q_criterion}
\end{figure}

A snapshot of the flow is shown in Figs. \ref{fig:q_criterion}(a)$-$\ref{fig:q_criterion}(d) displaying the isosurfaces of $Q$-criterion colored by the $u$-velocity component together with a background view of magnitude of density gradient $|\nabla \rho|$.  In Fig. \ref{fig:q_criterion}(a), we can observe the oblique shock waves that are generated at the airfoils' leading edges, and their interactions with the turbulent boundary layers of the neighboring vanes. Oblique shocks are also displayed at the trailing edges.
A detailed view of the flowfield is presented in Fig. \ref{fig:q_criterion}(b), where one can observe the separation bubbles (blue color contours) induced by the SBLIs, as well as different shock structures formed on both sides of the airfoil. 

On the suction side, an oblique shock impinges on the turbulent boundary layer and leads to the formation of compression waves near the reattachment location, which coalesce into an intermittent reattachment shock.
However, on the pressure side a more pronounced reflected shock structure is generated due to a Mach reflection. This latter condition occurs when a regular reflected shock is unable to turn the flow to be tangent to the wall, i.e., the maximum flow turning angle at the given Mach number downstream of the incident shock is smaller than the wall angle. In this case, a normal shock is generated to ensure that the flow is tangent to the wall surface \citep{anderson2003modern,matheis_hickel_2015}. In Figs. \ref{fig:q_criterion}(c) and \ref{fig:q_criterion}(d), one can see the vortical structures on the suction and pressure side boundary layers, respectively. Figure \ref{fig:q_criterion}(c) reveals the presence of streaks in the region further upstream of the SBLI, while downstream of the shock, the formation of randomly organized structures is observed similar to Ref. \citep{fang2020}. On the pressure side, Fig. \ref{fig:q_criterion}(d) shows that hairpin vortices appear along the boundary layer upstream of the SBLI, whereas downstream of the shock, the vortical structures are similar to those presented on the suction side. A movie displaying the time evolution of the flowfield, including the flow features described above, is provided as supplemental material (movie 1).

\begin{figure}
	\begin{center}
		{\includegraphics[trim = 1mm 3mm 1mm 10mm, clip, width=0.90\textwidth]{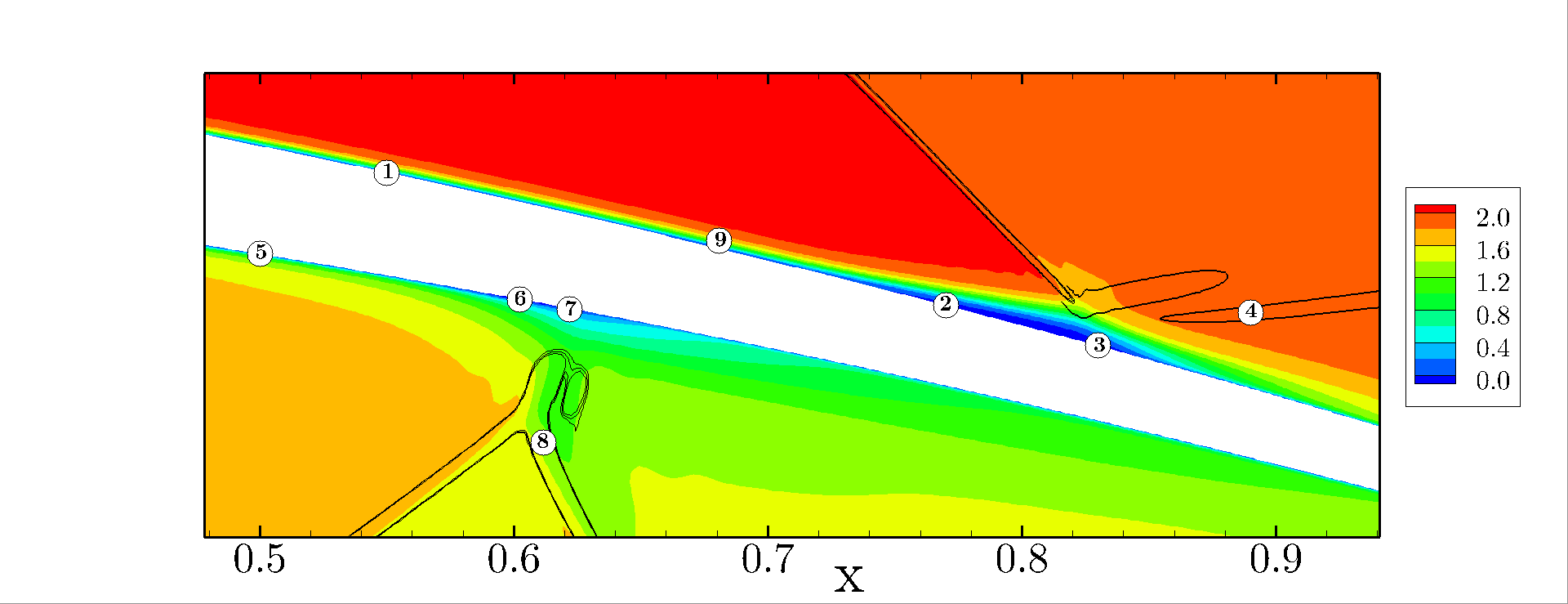}}
	\end{center}
	\caption{Spanwise averaged contour of time-averaged $u$-velocity. The black lines display the shock waves visualized by pressure gradient magnitude and white circles indicate the probe locations used for calculation of the flow statistics.}
	\label{fig:mean_flow_probes}
\end{figure}

The spanwise and time averaged $u$-velocity contour is presented in Fig. \ref{fig:mean_flow_probes}. The shock waves are displayed in terms of pressure gradient and appear as black lines in the figure. Important regions of the flow are identified by the nine points marked in the plot. The suction side separation bubble is observed between markers 2 and 3 while the pressure side bubble is smaller, falling between markers 6 and 7. These locations are also used for extraction of time signals which are postprocessed for spectral analysis.
\begin{figure}
	\begin{overpic}[trim = 1mm 1mm 1mm 1mm, clip,width=.49\linewidth]{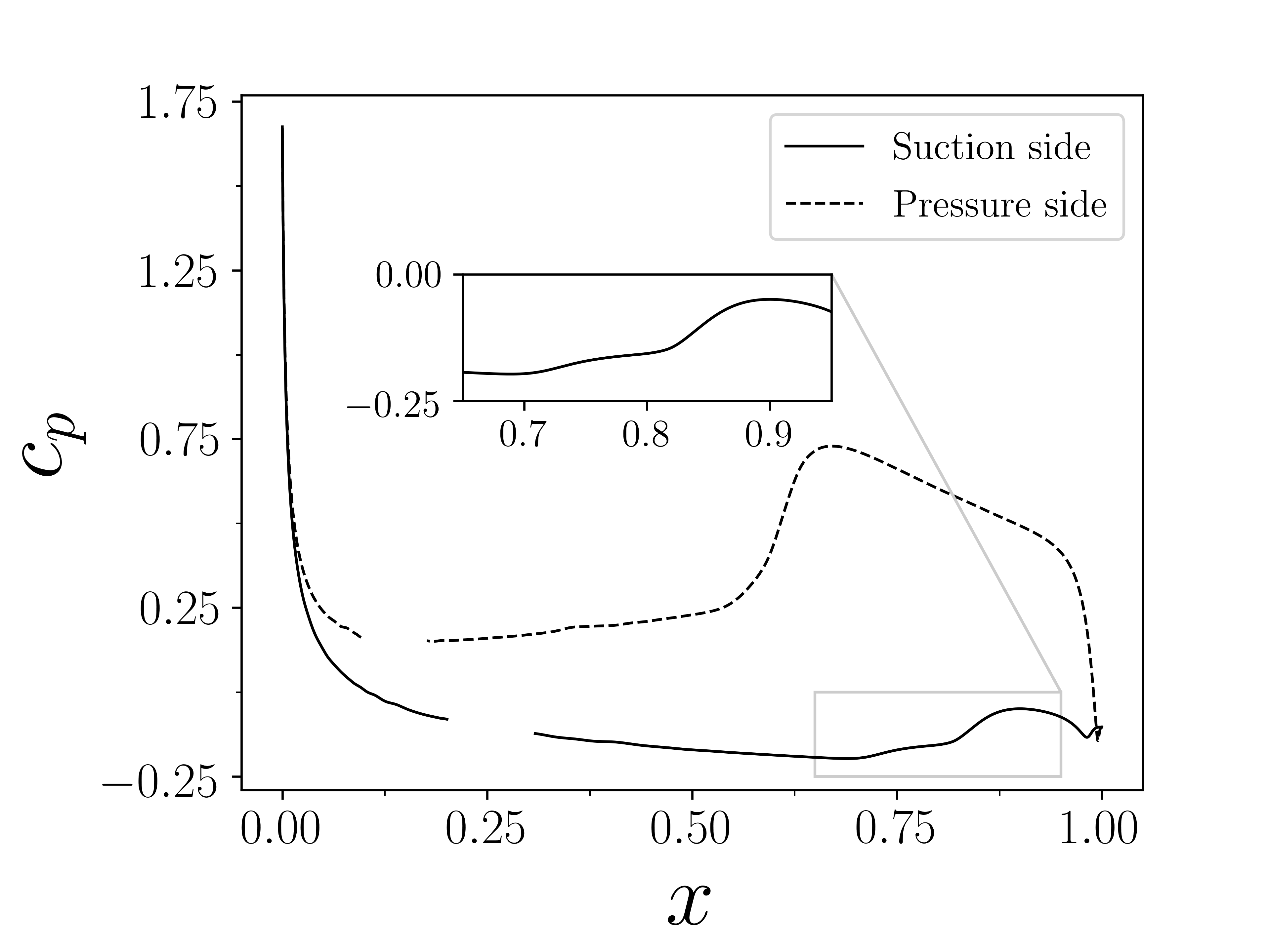}
		\put(1,67){(a)}
	\end{overpic}
	\begin{overpic}[trim = 1mm 1mm 1mm 1mm, clip,width=.49\linewidth]{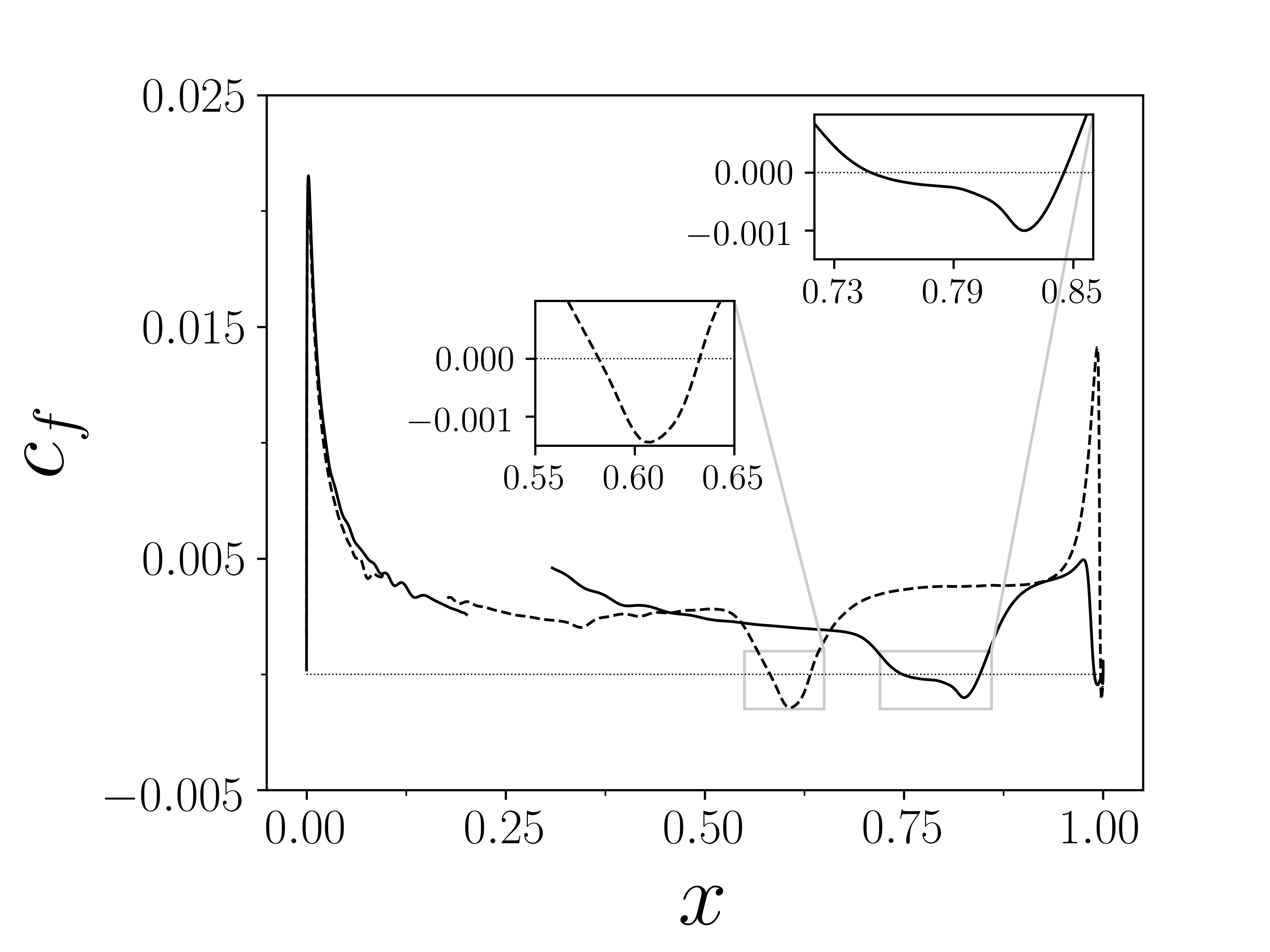}
		\put(1,67){(b)}
	\end{overpic}
	\caption{Spanwise and time averaged (a) pressure coefficient and (b) skin-friction coefficient distributions. The insets show (a) the pressure variation along the suction side bubble and (b) the regions with separated boundary layers.}
	\label{fig:coefficients}
\end{figure}

The distribution of the spanwise and time averaged pressure coefficient $c_p = \frac{p - p_\infty}{0.5 \rho_\infty U_{\infty}^2}$ along the airfoil chord is shown in Fig. \ref{fig:coefficients}(a). The regions where the boundary layer tripping is applied are removed from the plot. One can notice a smoother rise in $c_p$ on the suction side while a steeper rise, due to the normal shock, is observed on the pressure side. Figure \ref{fig:coefficients}(b) presents the spanwise and time averaged skin-friction coefficient distribution $c_f = \frac{\tau_w}{0.5 \rho_\infty U_{\infty}^2}$, where $\tau_w$ represents the wall shear stress. Again, the regions where the boundary layer tripping is applied are removed from the figure and a horizontal dashed line delimits the $c_f=0$ value. The friction coefficient distribution highlights the presence of separation bubbles on both sides of the airfoil, characterized by locations where $c_f < 0$. On the suction side, the mean separation bubble is formed between $0.75 < x < 0.85$ and, hence, its length is $\langle L_{SB} \rangle = 0.10c_{x}$. For the pressure side, the mean bubble size is $\langle L_{SB} \rangle = 0.05c_{x}$, and it forms between $0.58 < x < 0.63$. 

\subsection{Spatiotemporal analysis of the separation bubbles}
\label{section:visualization2}

\begin{figure}
	\centering
	\begin{overpic}[trim = 70mm 1mm 1mm 1mm,clip,width=0.48\textwidth]{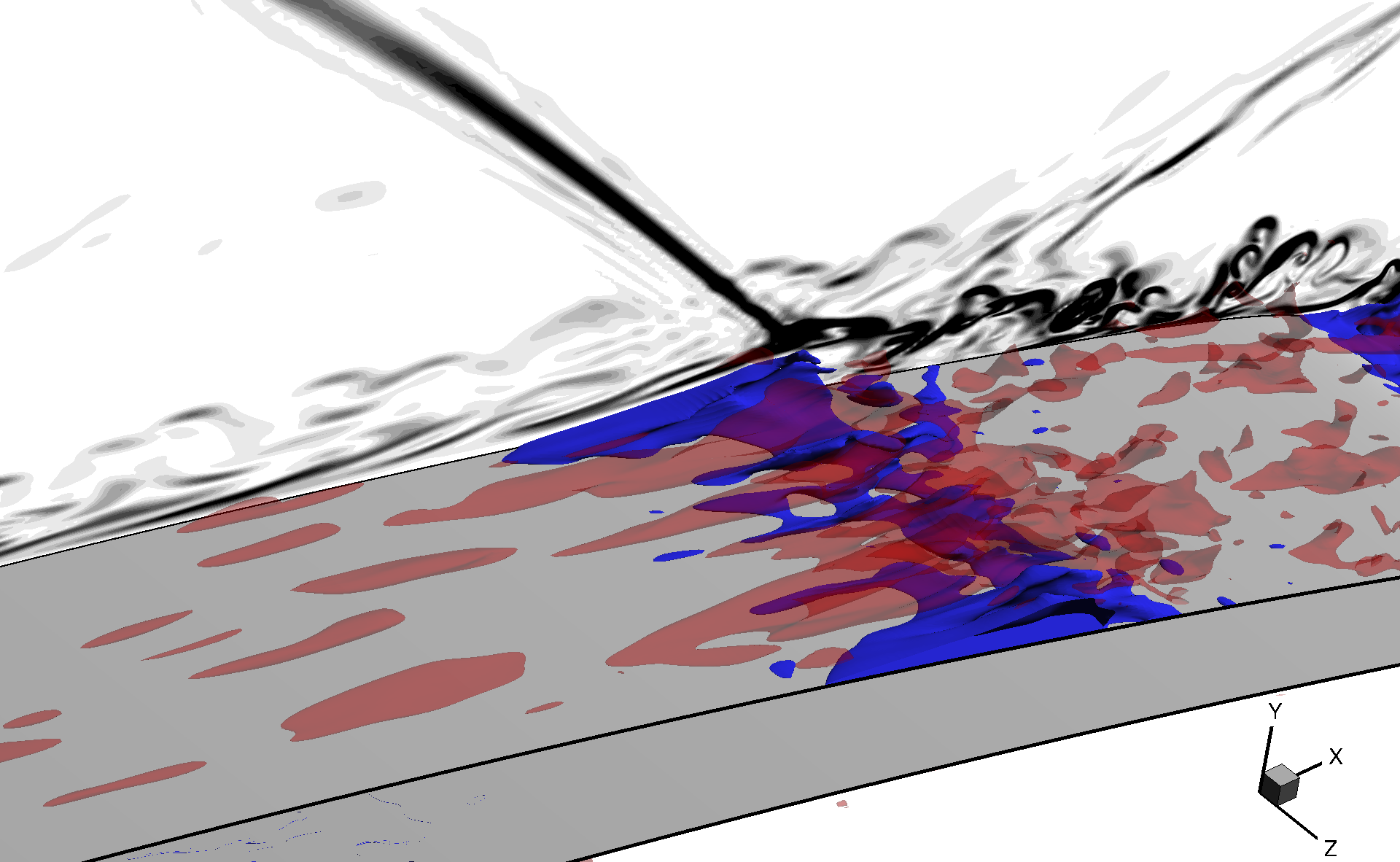}
		\put(0,55){(a)}
	\end{overpic} 
	\begin{overpic}[trim = 70mm 1mm 1mm 1mm,clip,width=0.48\textwidth]{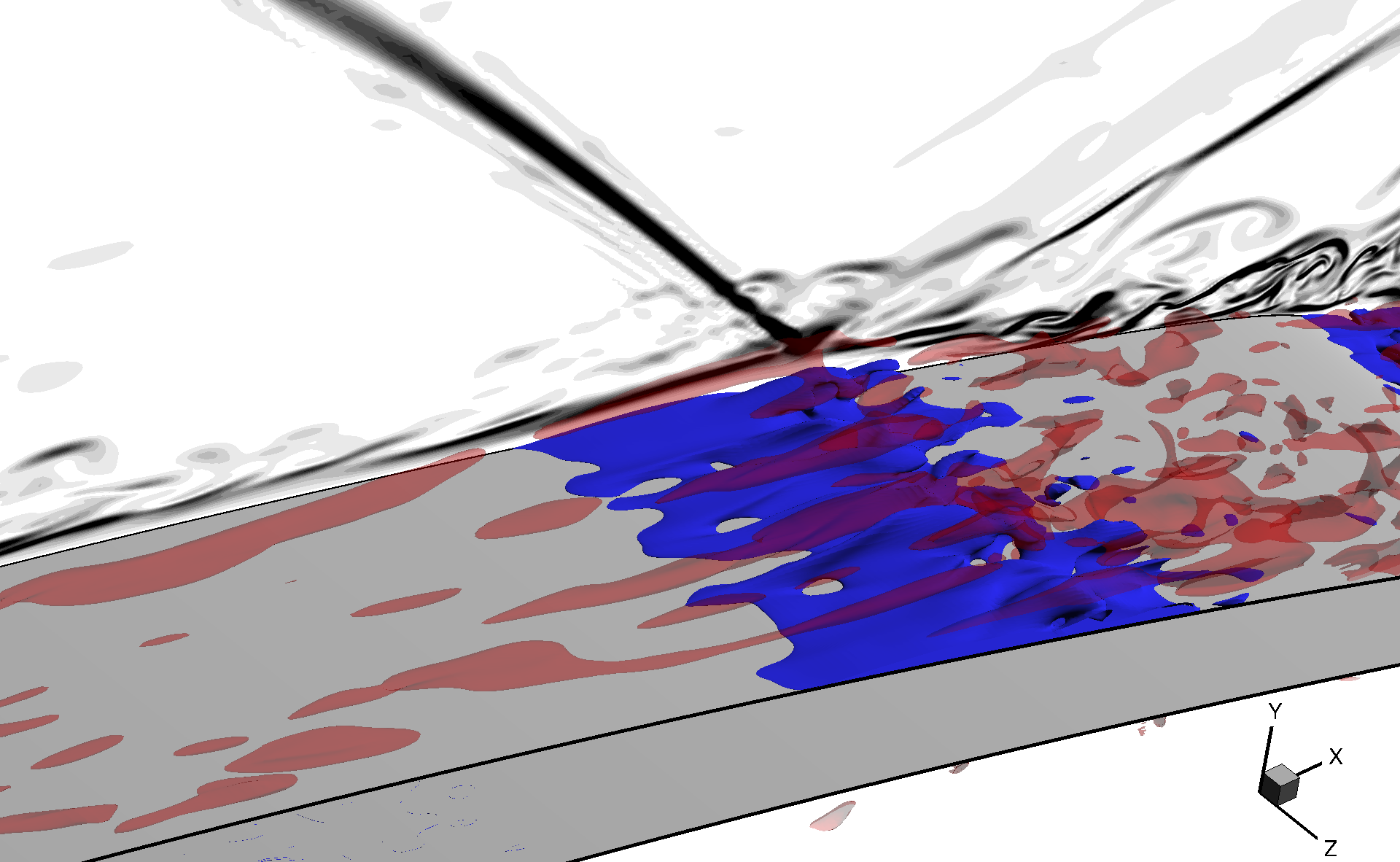}
		\put(0,55){(b)}
	\end{overpic} 
	
	\vskip 0.4cm
	
	\begin{overpic}[trim = 8mm 10mm 10mm 2mm,clip,width=0.49\textwidth]{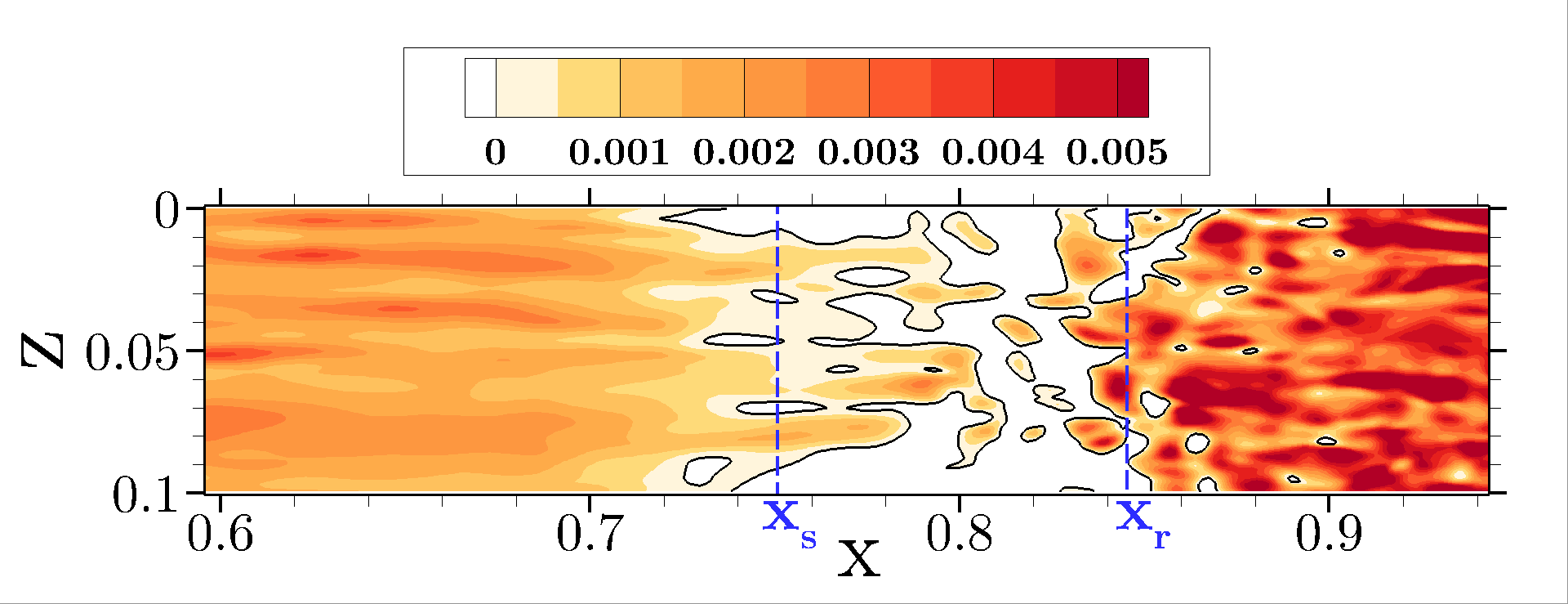}
		\put(0,30){(c)}
	\end{overpic} 
	\begin{overpic}[trim = 8mm 10mm 10mm 2mm,clip,width=0.49\textwidth]{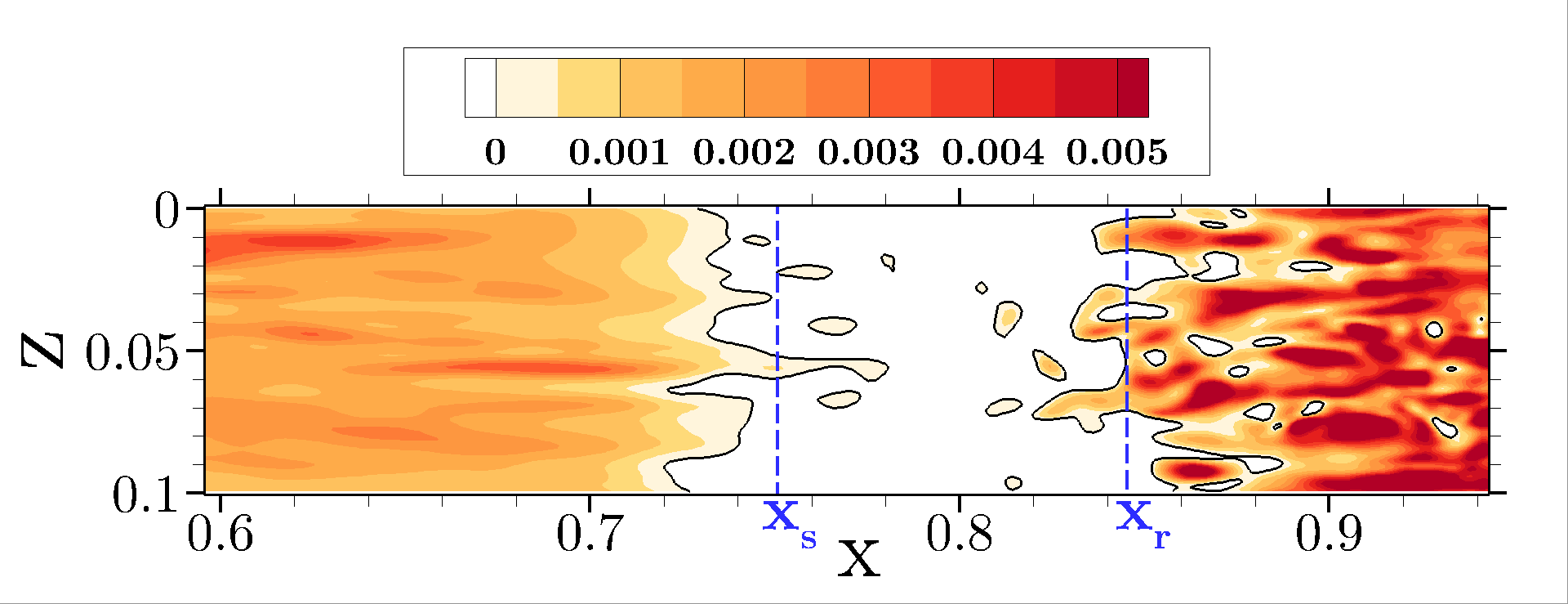}
		\put(0,30){(d)}
	\end{overpic} 
	
	\vskip 0.05cm
	
	\begin{overpic}[trim = 8mm 10mm 10mm 2mm,clip,width=0.49\textwidth]{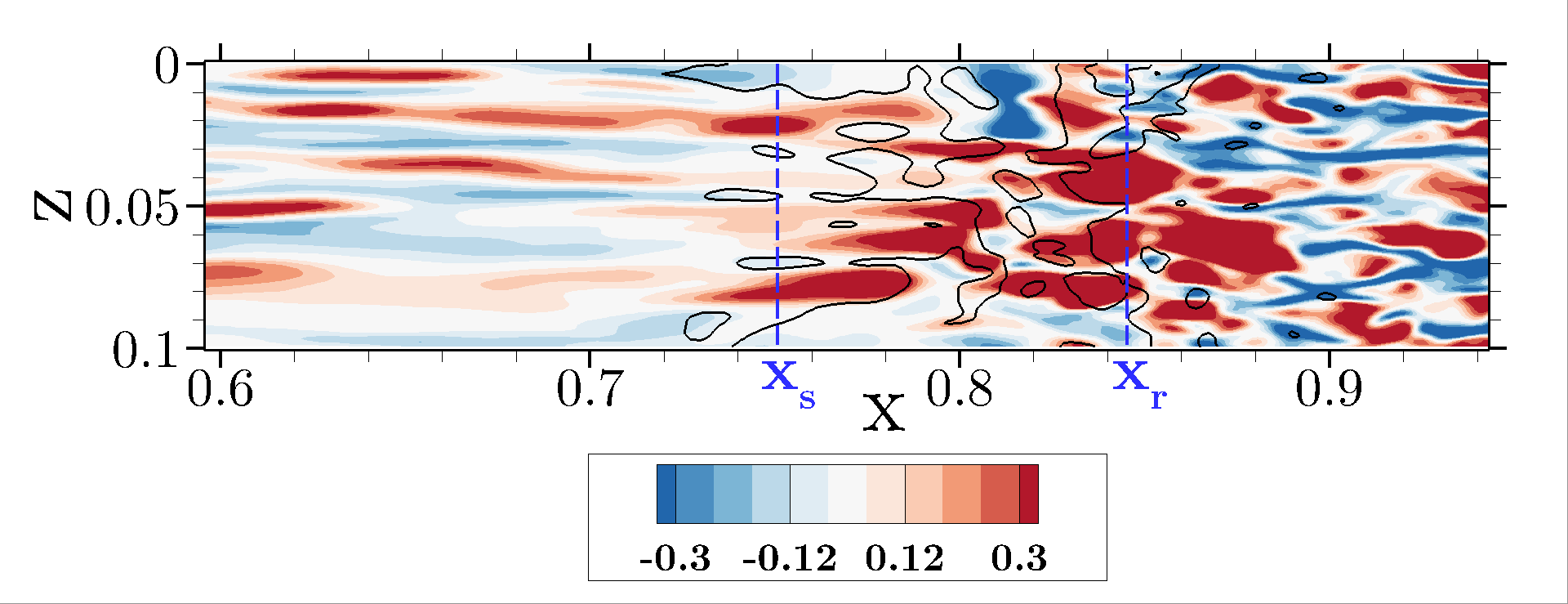}
		\put(0,32){(e)}
	\end{overpic} 
	\begin{overpic}[trim = 8mm 10mm 10mm 2mm,clip,width=0.49\textwidth]{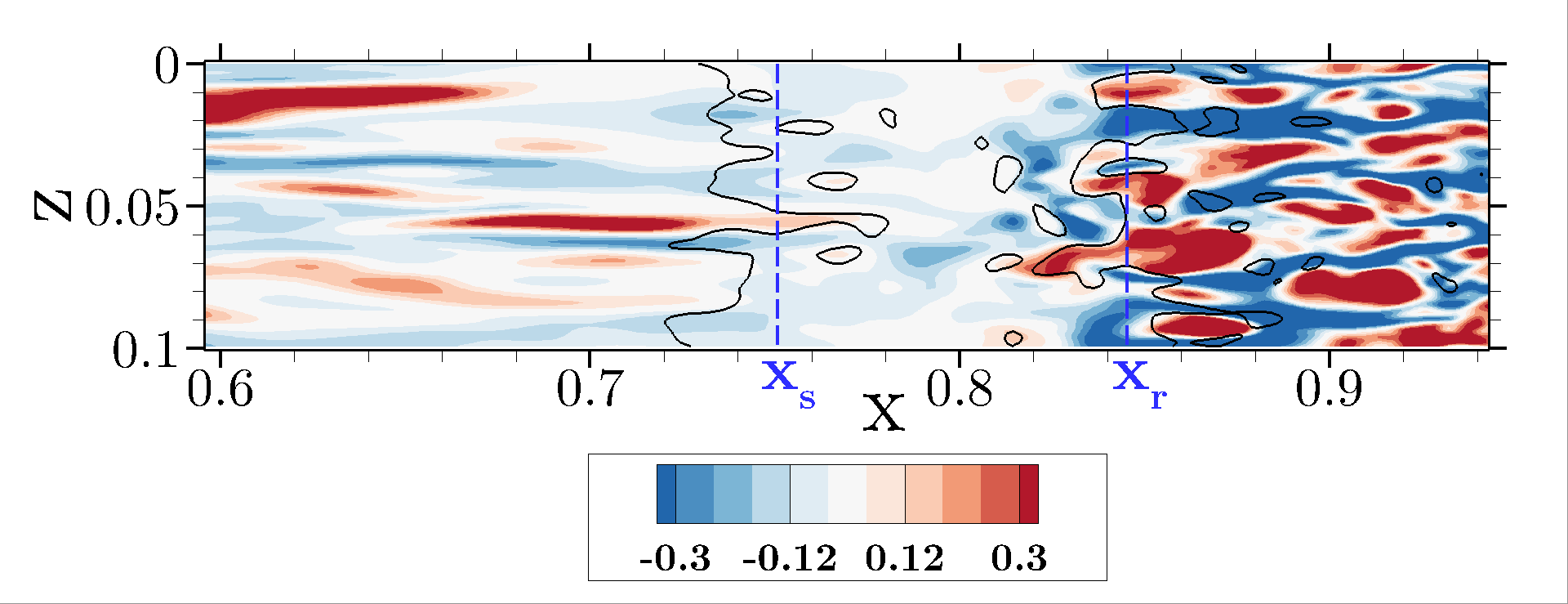}
		\put(0,32){(f)}
	\end{overpic} 
	\caption{Suction side separation bubble topology at different time instants with (a,b) isosurfaces of instantaneous $u=0$ velocity (blue color) and velocity fluctuations $u^{\prime}=0.3$ (red color). The shocks are displayed in the background by grayscale contours of $|\nabla \rho|$. Planes of (c,d) skin-friction coefficient and (e,f) tangential velocity fluctuations at $y^{+} \approx $ 6. The black lines display the $c_f = 0$ contour level while the blue lines represent the mean separation and reattachment positions.}
	\label{fig:separation_bubble_3D_SS}
\end{figure}

Figure \ref{fig:separation_bubble_3D_SS} shows the spatial variations of the suction side separation bubble for two time instants. The left-hand side plots represent a moment when the overall bubble size is small, while the right-hand side plots correspond to an instant when the separation bubble is large. Figures \ref{fig:separation_bubble_3D_SS}(a) and \ref{fig:separation_bubble_3D_SS}(b) display the isosurfaces of instantaneous $u=0$ velocity in blue color, which delimit the regions of flow recirculation. Isosurfaces of positive $u$ velocity fluctuations are also shown in red color to highlight the organized flow structures that interact with the bubble. We can observe the existence of elongated streamwise structures in the upstream boundary layer, some of which penetrate the bubble causing the flow to locally reattach at its leading edge, as highlighted in Fig. \ref{fig:separation_bubble_3D_SS}(a). Some structures, however, are advected over the bubble, bursting downstream and causing reattachment at its trailing edge. A movie showing the unsteady three-dimensional behavior of the suction side separation bubble is provided as supplemental material (movie 2). 

To better visualize the spatial variations of the separation bubble, the instantaneous skin-friction coefficient is presented along the $x$-$z$ plane in Figs. \ref{fig:separation_bubble_3D_SS}(c) and \ref{fig:separation_bubble_3D_SS}(d). In these figures, the white zones display the separated flow region. It can be observed that the bubble has undulations along the spanwise direction, with spots of positive $c_f$ where the flow locally reattaches. The figures also indicate that some boundary layer streaks imprint a signature on the wall, upstream of the bubble. Downstream of the reattachment, the wall friction is larger due to streak bursting which occurs in the SBLI. 
Figures \ref{fig:separation_bubble_3D_SS}(e) and \ref{fig:separation_bubble_3D_SS}(f) exhibit the tangential velocity fluctuations computed in the $x$-$z$ plane near the wall, at $y^+ \approx 6$. This location in terms of wall units is computed based on mean flow properties from the attached boundary layer, upstream of the bubble. One can observe that the passage of high-speed (low-speed) streaks near the recirculation region leads to downstream (upstream) motion of the separation point. These findings are consistent with previous studies from Refs. \citep{beresh2002,ganapathisubramani_clemens_dolling_2009,porter_2019, baidya_2020}. 
A movie displaying the temporal variation of the suction side separation bubble in terms of instantaneous friction coefficient, as well as the response of the bubble to the passage of low and high-speed streaks near the wall is provided as supplemental material (movie 3).

\begin{figure}
	
	\centering
	\begin{overpic}[trim = 1mm 1mm 1mm 1mm,width=0.48\textwidth]{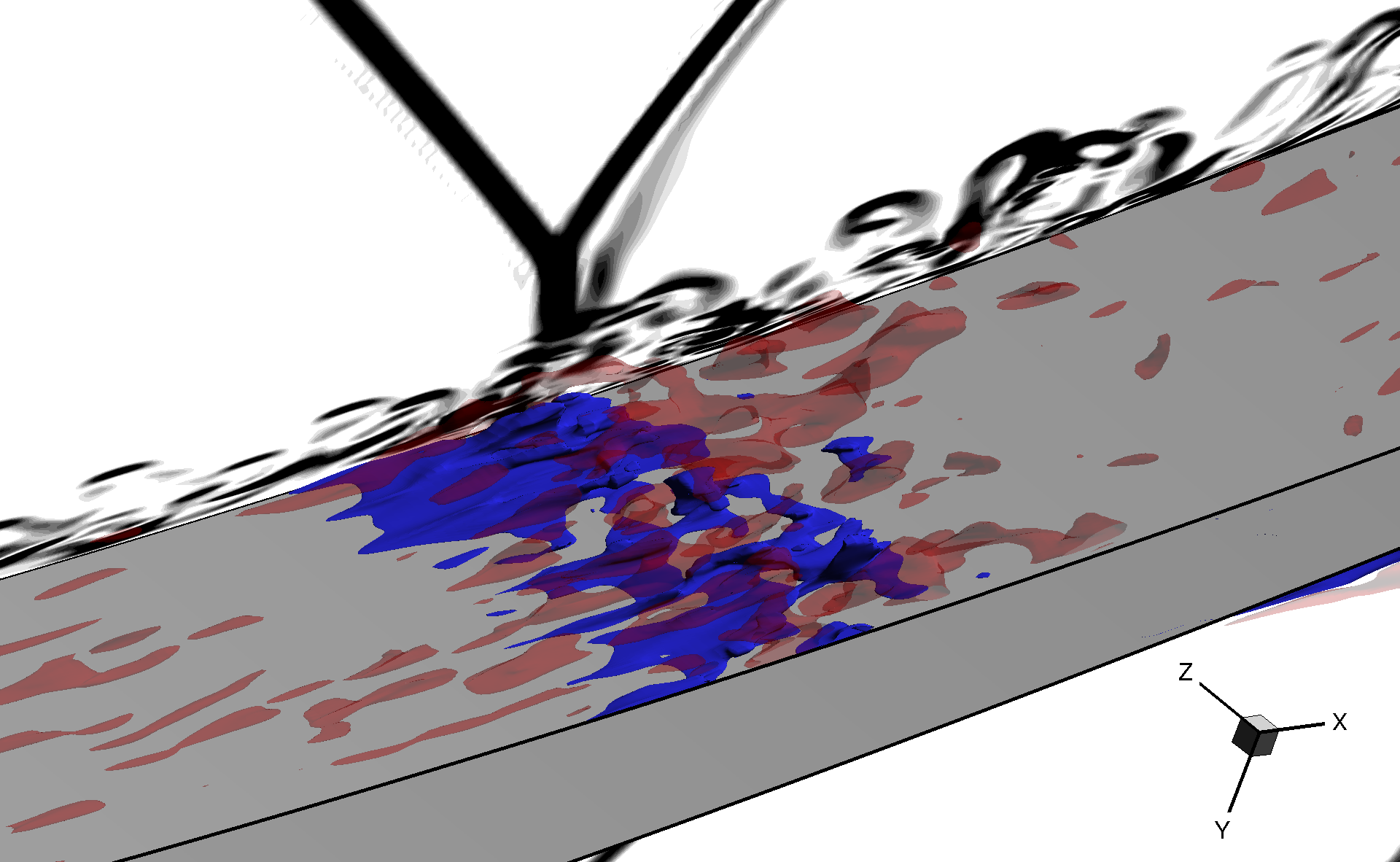}
		\put(0,55){(a)}
	\end{overpic} 
	\begin{overpic}[trim = 1mm 1mm 1mm 1mm,width=0.48\textwidth]{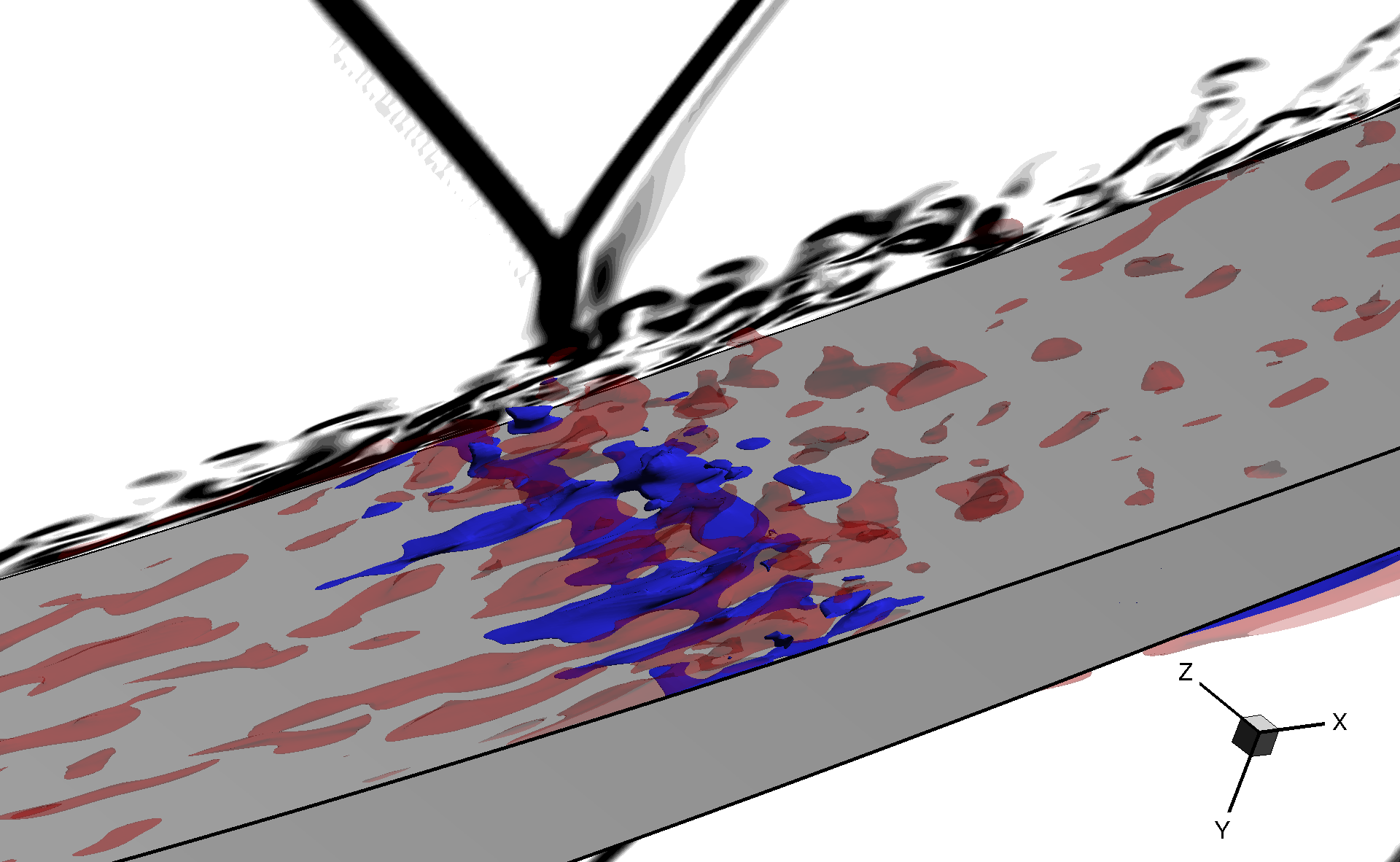}
		\put(0,55){(b)}
	\end{overpic} 
	
	\vskip 0.4cm
	
	\begin{overpic}[trim = 8mm 5mm 10mm 2mm,clip,width=0.49\textwidth]{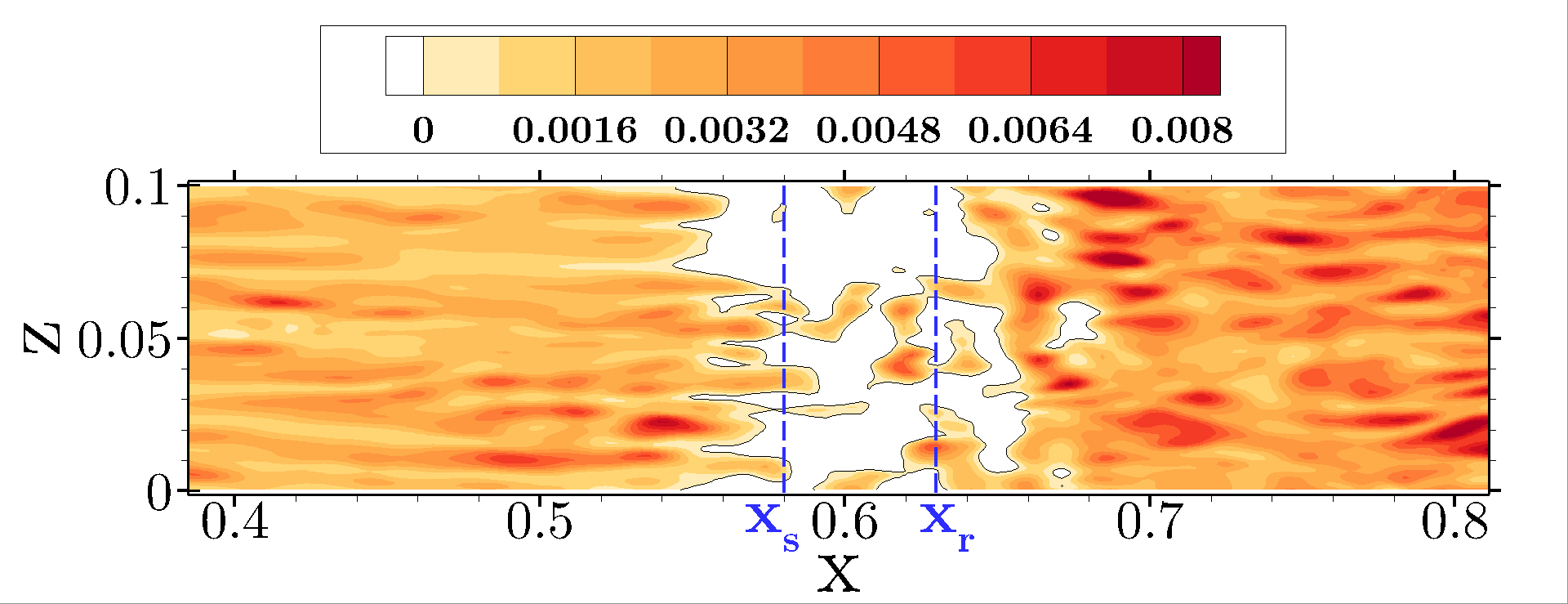}
		\put(0,30){(c)}
	\end{overpic} 
	\begin{overpic}[trim = 8mm 5mm 10mm 2mm,clip,width=0.49\textwidth]{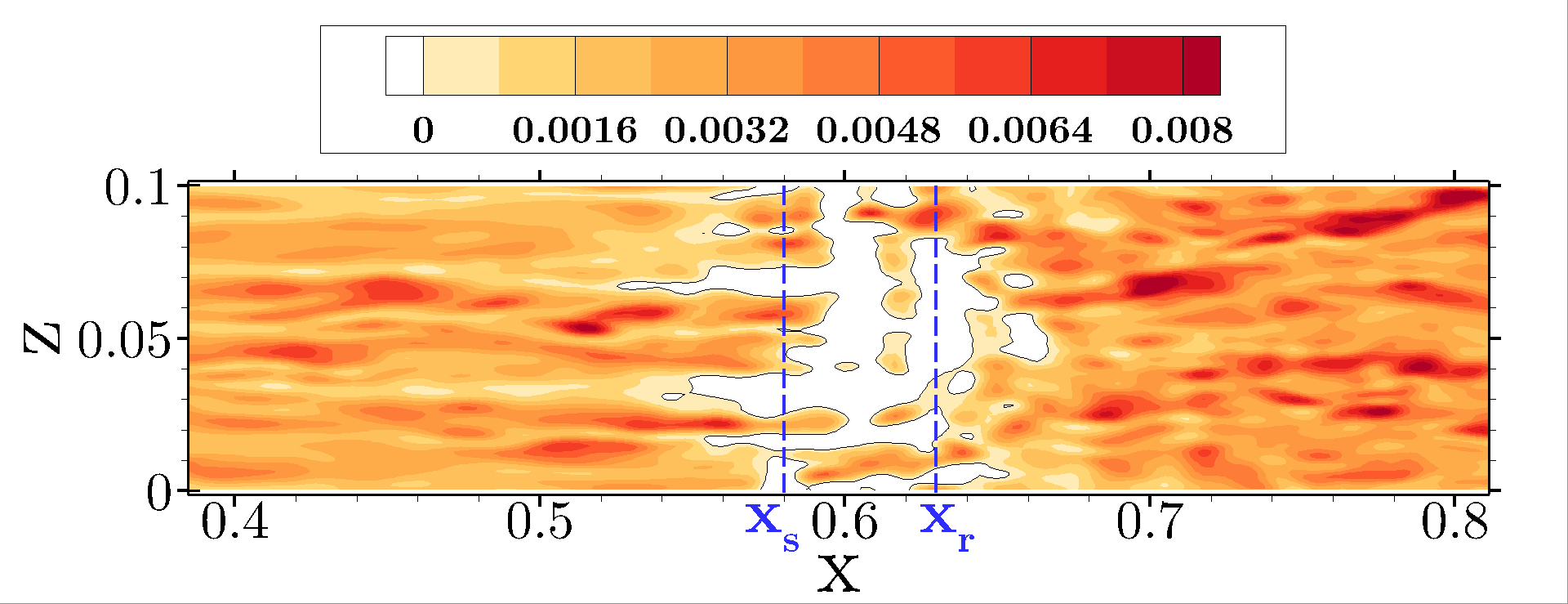}
		\put(0,30){(d)}
	\end{overpic}
	
	\vskip 0.1cm
	
	\begin{overpic}[trim = 8mm 8mm 10mm 2mm,clip,width=0.49\textwidth]{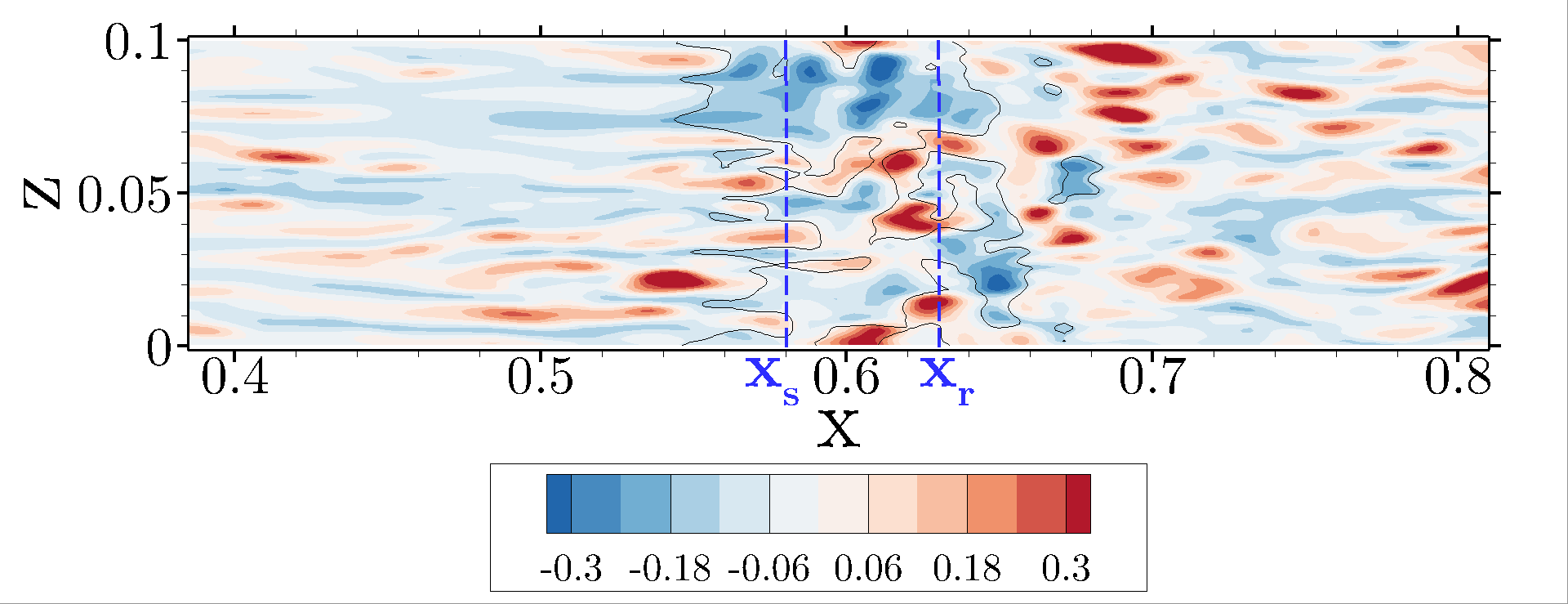}
		\put(-1.5,34){(e)}
	\end{overpic} 
	\begin{overpic}[trim = 8mm 8mm 10mm 2mm,clip,width=0.49\textwidth]{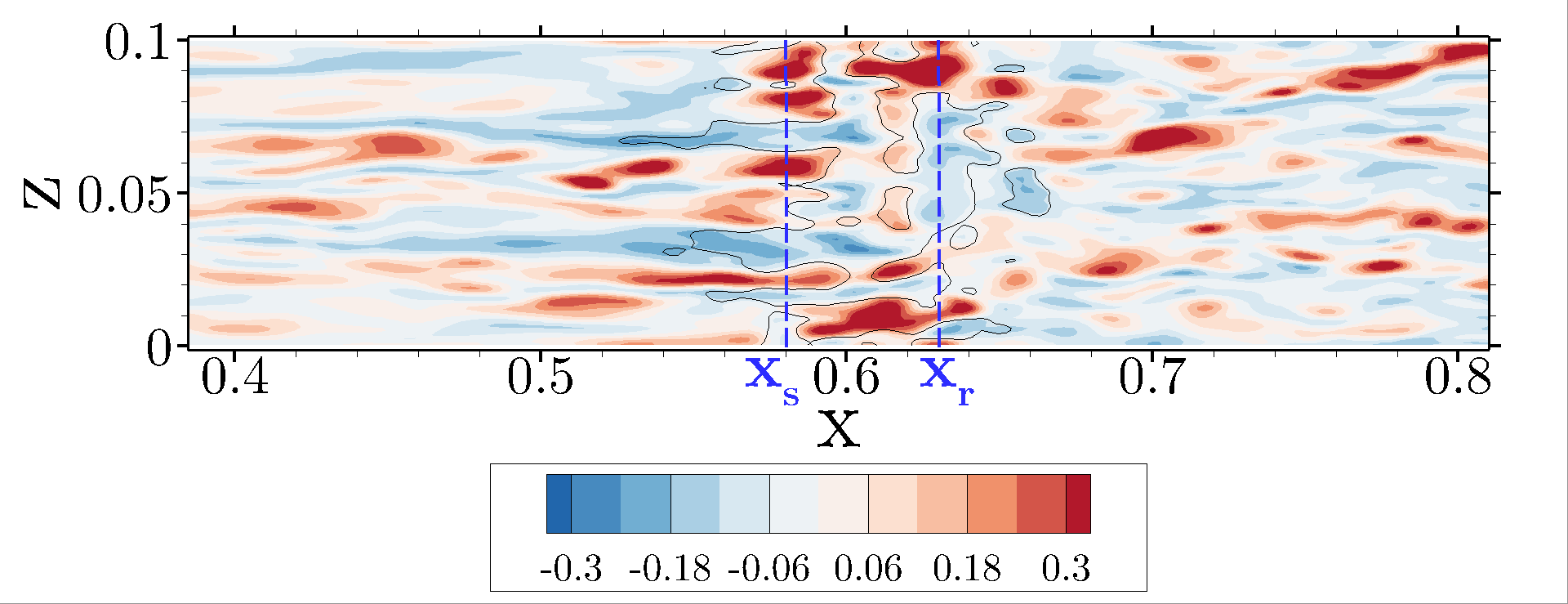}
		\put(-1.5,34){(f)}
	\end{overpic} 
	
	\caption{Pressure side separation bubble topology at different time instants with (a,b) isosurfaces of instantaneous $u=0$ velocity (blue color) and velocity fluctuations $u^{\prime}=0.3$ (red color). The shocks are displayed in the background by grayscale contours of $|\nabla \rho|$. Planes of (c,d) skin-friction coefficient and (e,f) tangential velocity fluctuations at $y^{+} \approx $ 5. The black lines display the $c_f = 0$ contour level while the blue lines represent the mean separation and reattachment positions.}
	\label{fig:separation_bubble_3D_PS}
\end{figure}

Figure \ref{fig:separation_bubble_3D_PS} presents the spatial distribution of the pressure side separation bubble at two time instants. The left-hand side figures correspond to the instant when the bubble is large, while the right-hand side plots are associated with an instance with a small separation bubble. Similar to the observations made for the suction side, the streamwise structures in the incoming boundary layer may penetrate the bubble, locally reattaching the flow, as seen in Figs. \ref{fig:separation_bubble_3D_PS}(a) and \ref{fig:separation_bubble_3D_PS}(b). A movie showing the unsteady three-dimensional behavior of the pressure side bubble is provided as supplemental material (movie 4). 
The skin-friction coefficient computed on the $x$-$z$ plane is displayed in Figs. \ref{fig:separation_bubble_3D_PS}(c) and \ref{fig:separation_bubble_3D_PS}(d) and, once again, these plots show the modulated spanwise variation of the separation bubble with internal spots of reattached flow. On the pressure side, the streaky signatures of the positive $c_f$ values have similar levels upstream and downstream of the bubble, contrary to the observations made for the suction side. In this case, the bubble is thinner and the streaks transported over its surface do not interact with the normal shock. Figures \ref{fig:separation_bubble_3D_PS}(e) and \ref{fig:separation_bubble_3D_PS}(f) display the tangential velocity fluctuations near the wall at $y^+ \approx 6$. It can be seen that the separation region conforms to the presence of negative velocity fluctuations, while reattached regions have positive velocity fluctuations. A movie showing the temporal variation of the pressure side separation bubble in terms of instantaneous friction coefficient and near-wall velocity fluctuations is provided as supplemental material (movie 5).

\begin{figure}
	
	\begin{overpic}[trim = 1mm 1mm 1mm 1mm, clip,width=.49\linewidth]{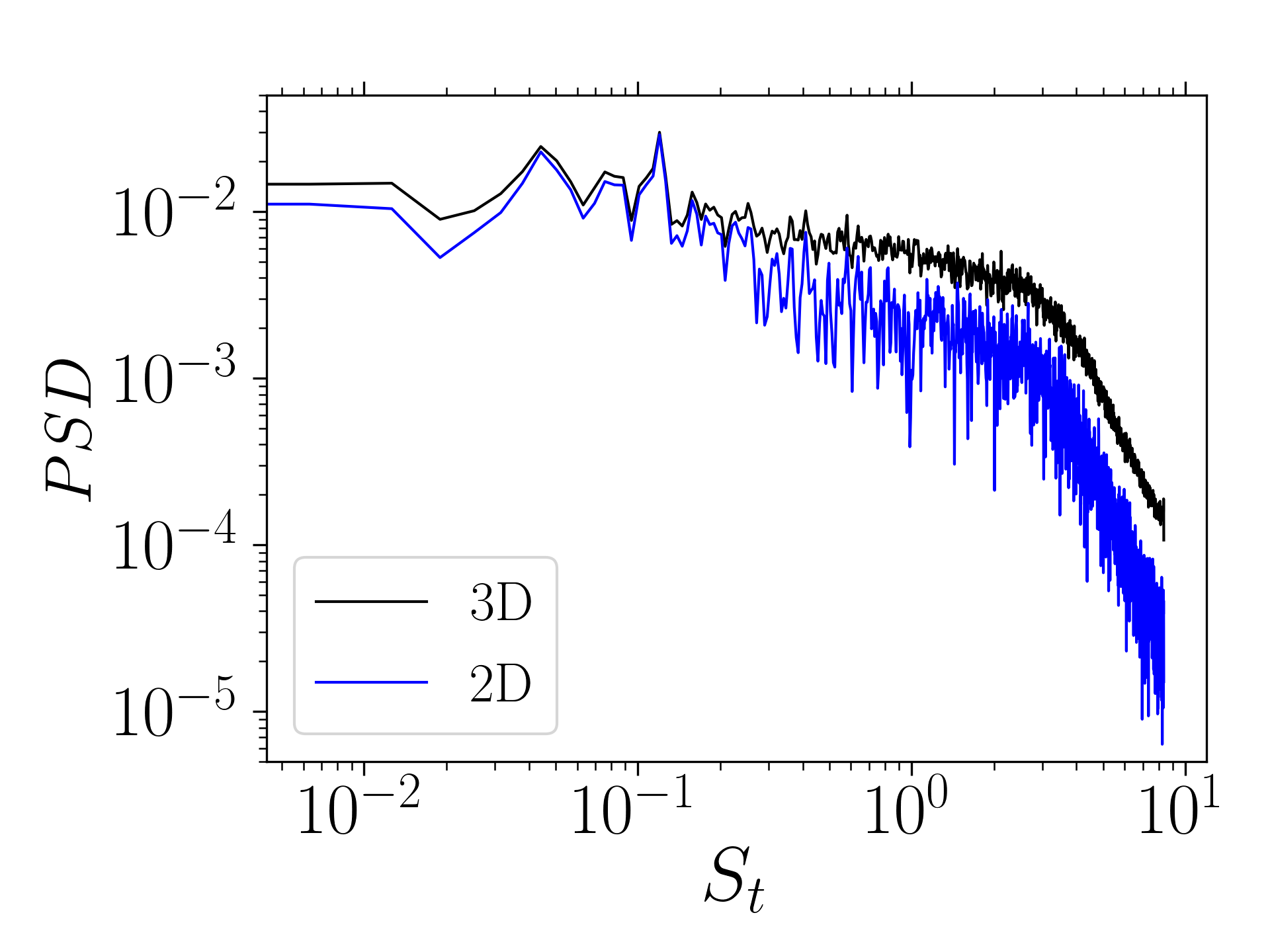}
		\put(1,64){(a)}
	\end{overpic}
	\begin{overpic}[trim = 1mm 1mm 1mm 1mm, clip,width=.49\linewidth]{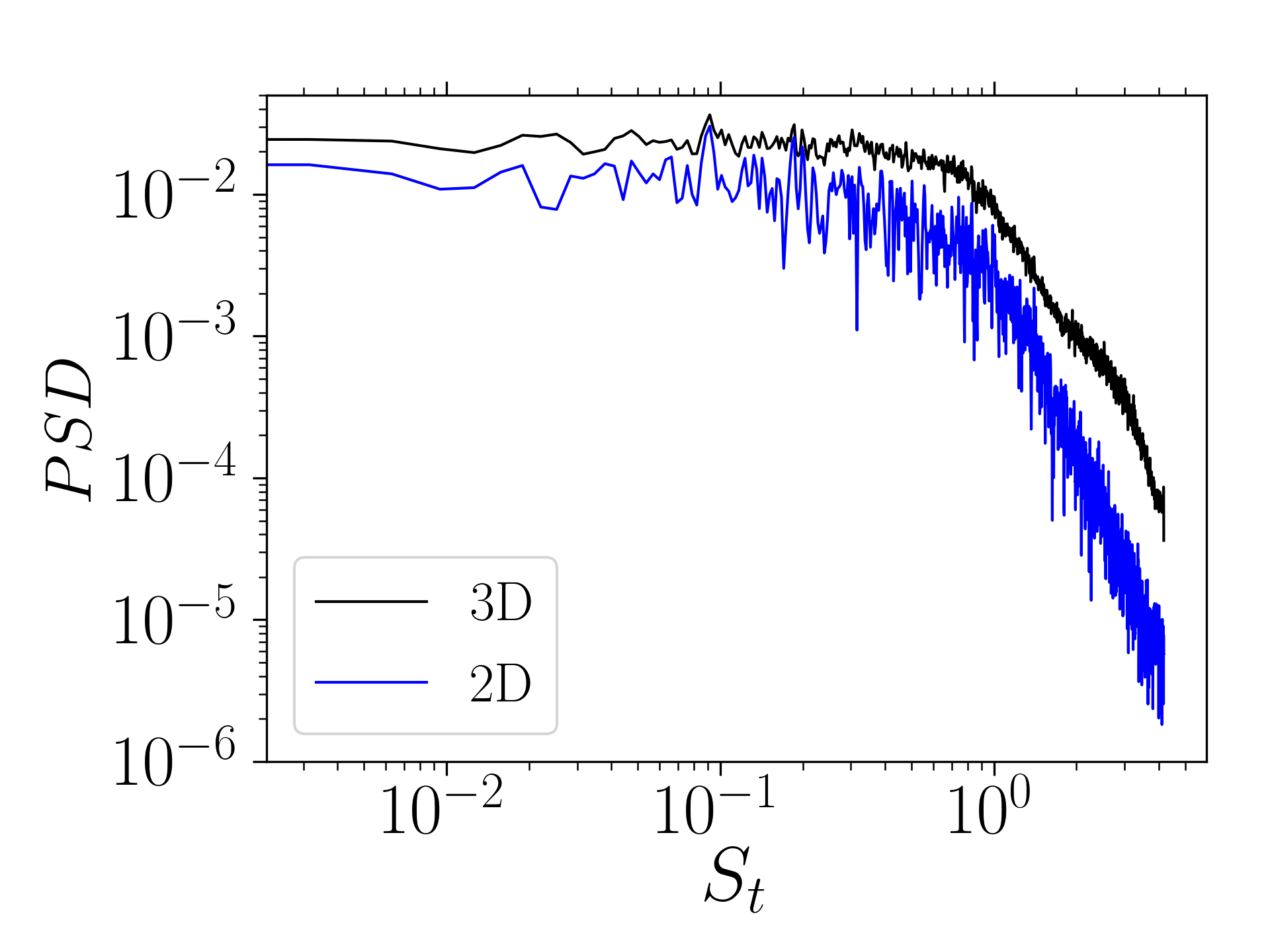}
		\put(1,64){(b)}
	\end{overpic}
	\vskip -0.7cm
	\caption{Comparison between the PSD of spanwise averaged wall pressure (2D) and the spanwise averaged PSD of wall pressure (3D) at (a) suction side (probe 3) and (b) pressure side (probe 6).}
	\label{fig:comp_2D_3D_PSD}
\end{figure}

Although the present results indicate a three-dimensional behavior of the separation bubbles, other studies  \citep{priebe2012,sasaki_2021} show that their low-frequency dynamics are essentially two-dimensional, justifying the investigation of spanwise averaged results. In this context, Fig. \ref{fig:comp_2D_3D_PSD} presents a comparison between the wall pressure power spectral density (PSD) obtained using two different approaches: in the first, the PSD is computed using spanwise averaged pressure fluctuations, as if the simulations were effectively 2D. In the second, the PSD is computed in a 3D fashion separately for all the points in the spanwise direction and then averaging the results over all these points. In both cases, the PSDs are calculated using Welch's method, where the signal is divided into 3 bins with 66$\%$ overlap using a Hanning window function. In Figs. \ref{fig:comp_2D_3D_PSD}(a) and \ref{fig:comp_2D_3D_PSD}(b), the results are presented for probe 3 and 6 from Fig. \ref{fig:mean_flow_probes}, which are located at the suction and pressure side separation bubbles, respectively. As it can be seen from the plots, the peaks associated with the low-frequency unsteadiness, ranging from $St = 0.04$ to $0.12$, are captured by both 2D and 3D approaches. This result confirms that the low-frequency dynamics is essentially two-dimensional in the present flow configuration. Therefore, in the following analyses, the spanwise averaged data will be used to investigate the low-frequency events taking place in the SBLIs.

\begin{figure}
		
	\centering
	\begin{overpic}[trim = 1mm 5mm 1mm 1mm, clip ,width=0.99\textwidth]{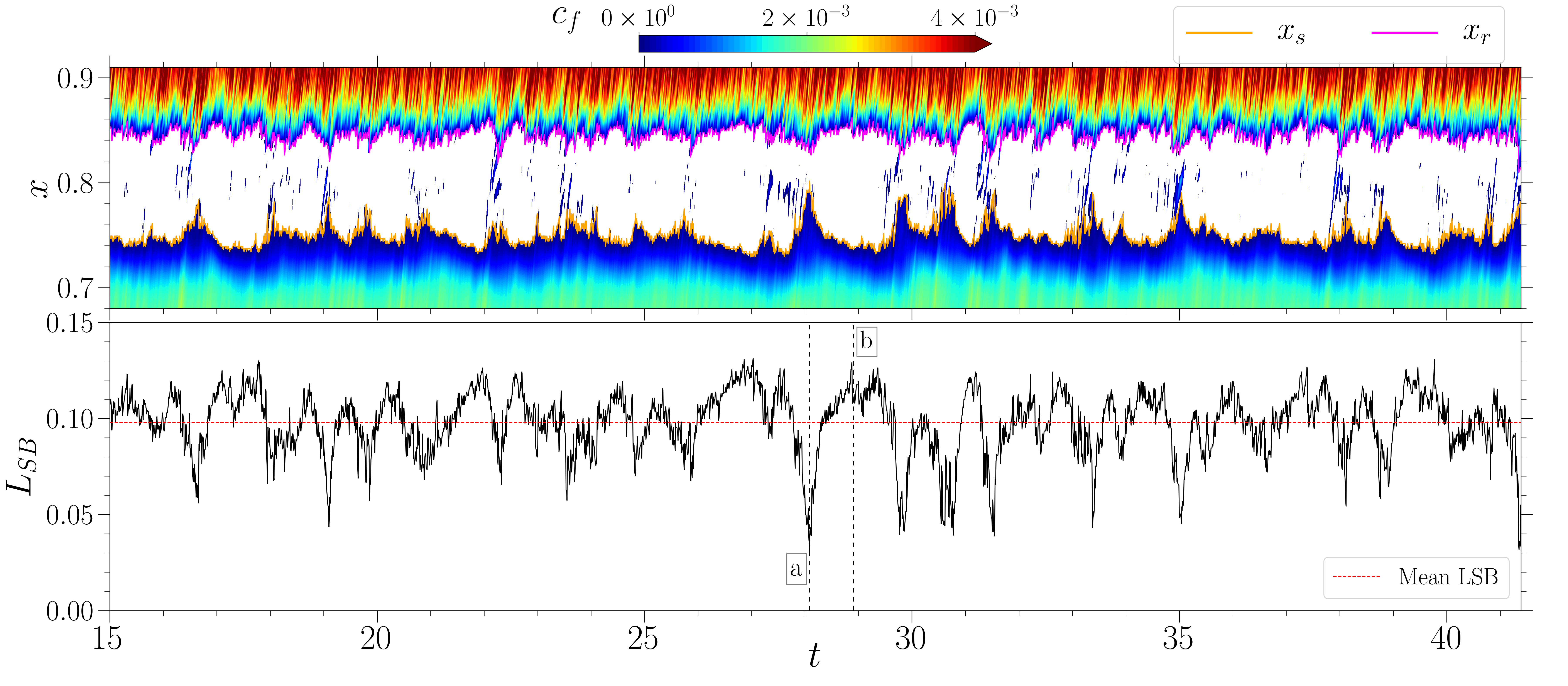}
		\put(-2,37){(a)}
		\put(-2,21){(b)}
	\end{overpic} 		
	\caption{Skin friction analysis of the suction side separation bubble with (a) space-time $c_f$ variation, (b) temporal variation of separation bubble length $L_{SB}$.}
	\label{fig:cf_contour_SS}		
			
\end{figure}

Further analysis of the suction side separation bubble is presented in Figure \ref{fig:cf_contour_SS} by inspecting the skin friction distribution. Its space-time variation is presented in Fig. \ref{fig:cf_contour_SS}(a) in terms of $c_f$, where the white area displays the separation region. The orange and magenta lines depict the locations of the separation and reattachment points, $x_{s}(t)$ and $x_{r}(t)$, respectively. One can observe that these points are most of the time out of phase, i.e., when $x_s$ moves upstream (downstream), $x_r$ moves downstream (upstream). One can also note that, for some instants, the separation point may move further downstream, while the instantaneous reattachment point oscillates closer to its mean value.

\begin{figure}
	\begin{overpic}[trim = 1mm 1mm 1mm 1mm, clip,width=.49\linewidth]{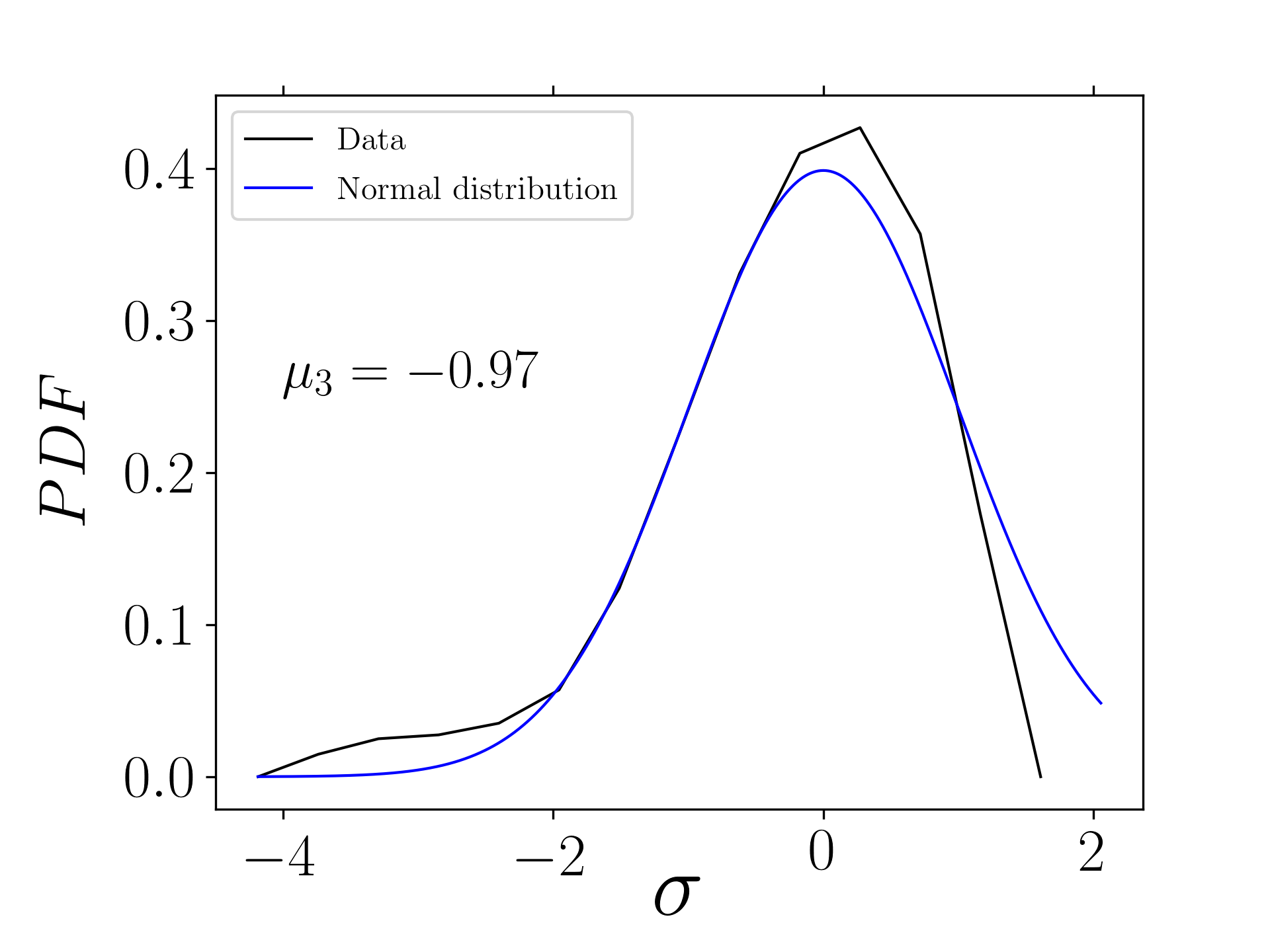}
		\put(1,67){(a)}
	\end{overpic}
	\begin{overpic}[trim = 1mm 1mm 1mm 1mm, clip,width=.49\linewidth]{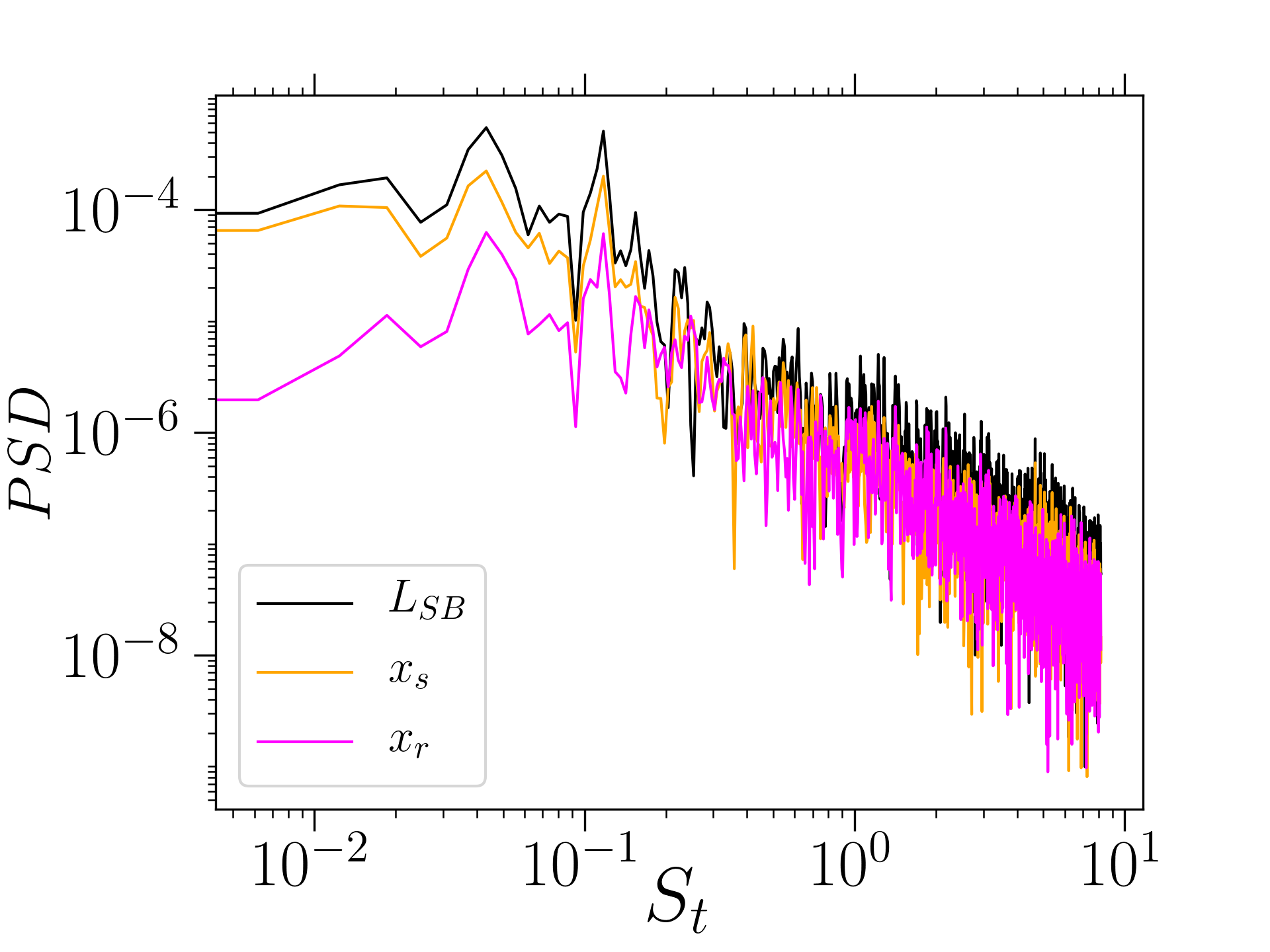}
		\put(1,67){(b)}
	\end{overpic}
	\caption{Statistical and spectral analyses of the suction side separation bubble with (a) PDF of $L_{SB}$ as a function of its standard deviation $\sigma$, and (b) PSD of $L_{SB}$, $x_s$ and $x_r$.}
	\label{fig:psd_pdf_LSB_SS}
\end{figure}

\begin{figure}
		
	\centering
	\begin{overpic}[trim = 1mm 5mm 1mm 1mm, clip ,width=0.99\textwidth]{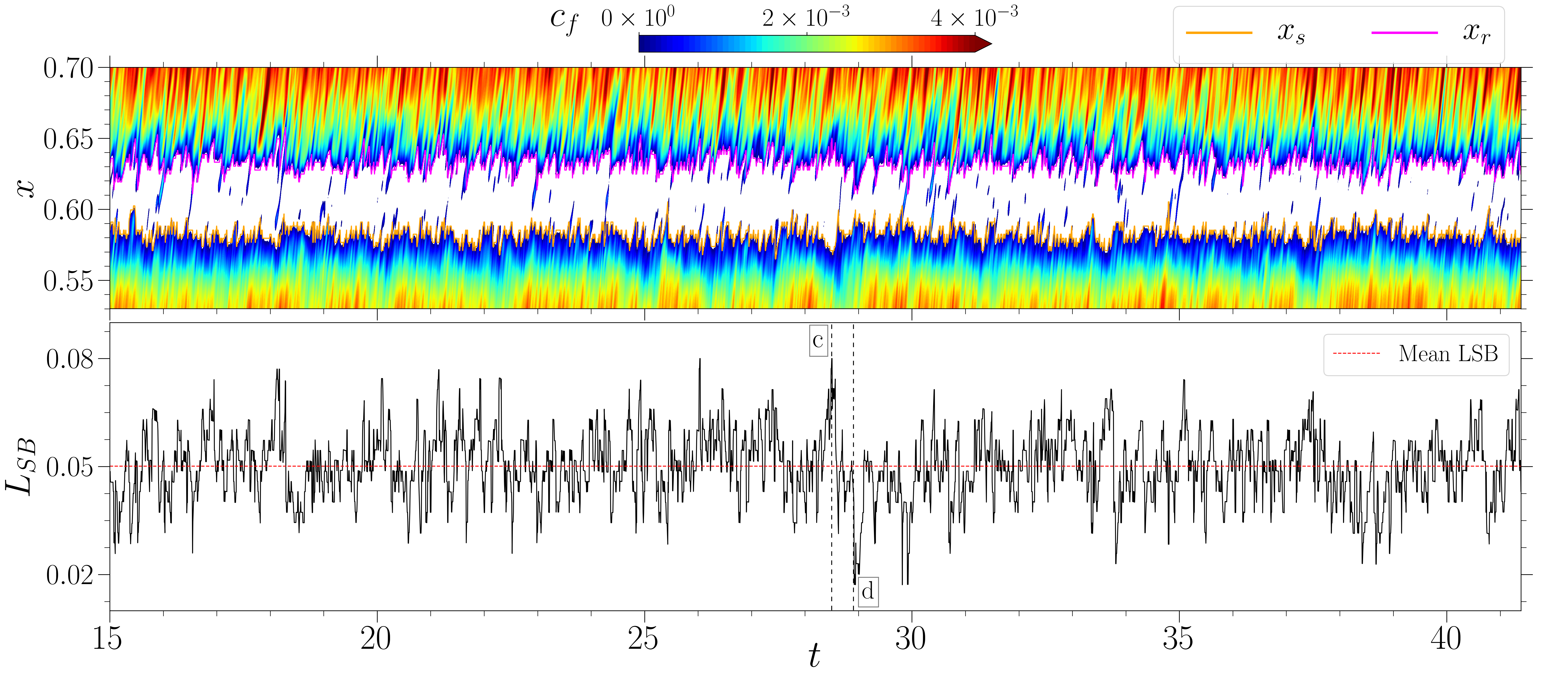}
		\put(-2,37){(a)}
		\put(-2,21){(b)}
	\end{overpic} 		
	\caption{Skin friction analysis of the pressure side separation bubble with (a) space-time $c_f$ variation, (b) temporal variation of separation bubble length $L_{SB}$.}
	\label{fig:cf_contour_PS}		
			
\end{figure}

\begin{figure}
	\begin{overpic}[trim = 1mm 1mm 1mm 1mm, clip,width=.49\linewidth]{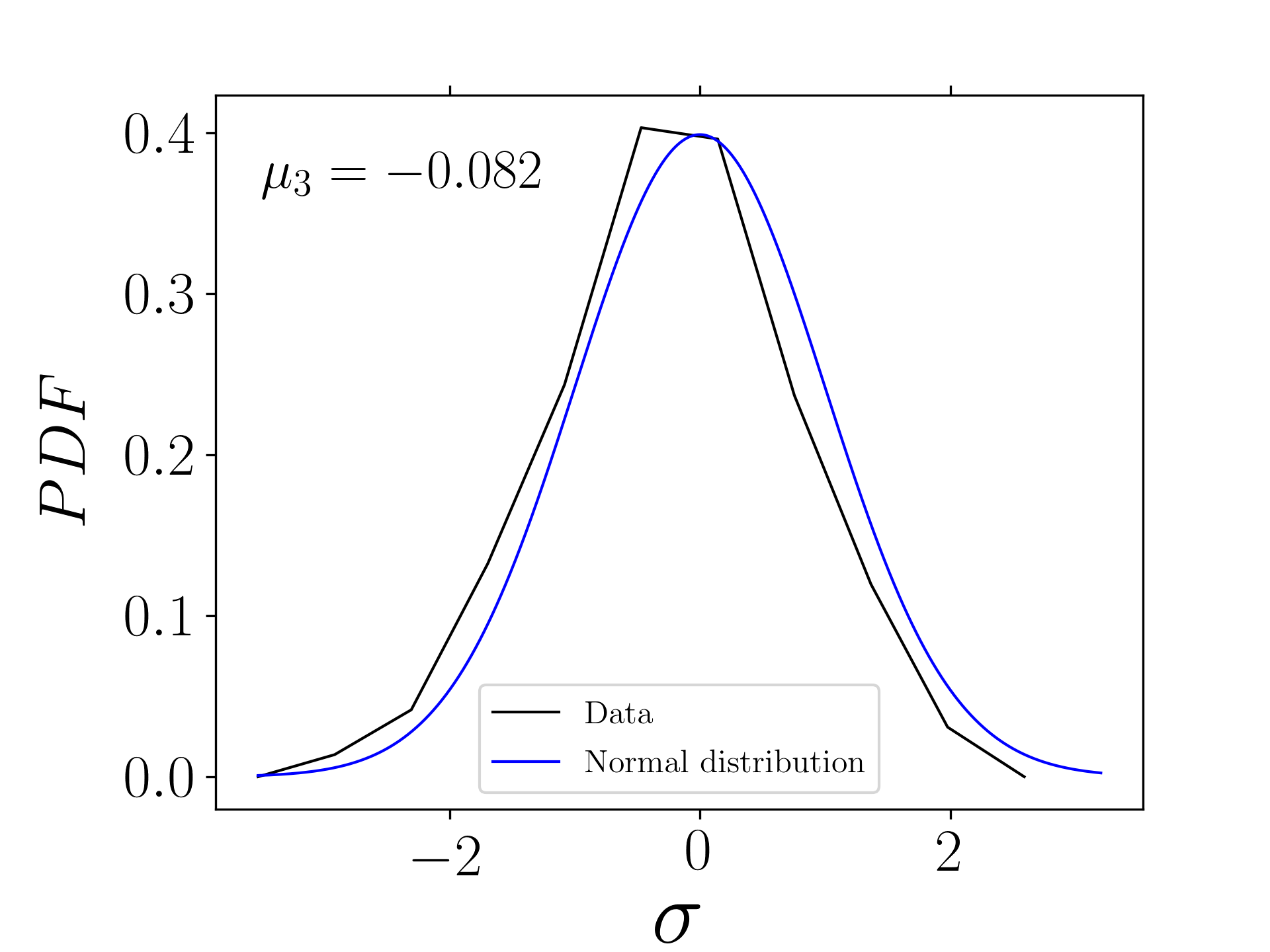}
		\put(1,67){(a)}
	\end{overpic}
	\begin{overpic}[trim = 1mm 1mm 1mm 1mm, clip,width=.49\linewidth]{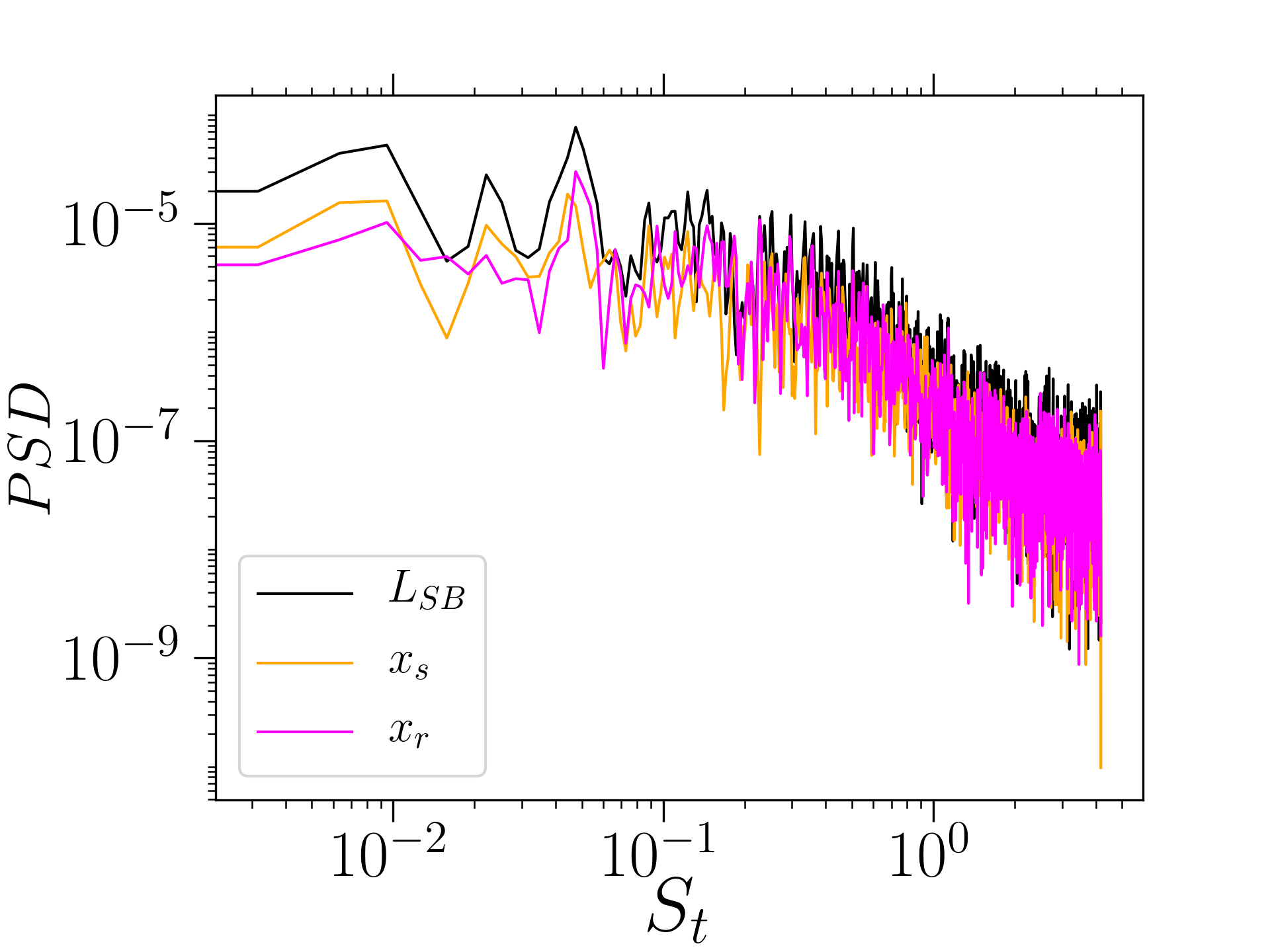}
		\put(1,67){(b)}
	\end{overpic}
	\caption{Statistical and spectral analyses of the pressure side separation bubble with (a) PDF of $L_{SB}$ as a function of its standard deviation $\sigma$, and (b) PSD of $L_{SB}$, $x_s$ and $x_r$.}
	\label{fig:psd_pdf_LSB_PS}
\end{figure}

To better characterize the separation bubble dynamics, the temporal evolution of the bubble length $L_{SB}$ and its probability density function (PDF) are shown in Figs. \ref{fig:cf_contour_SS}(b) and \ref{fig:psd_pdf_LSB_SS}(a), respectively. The instantaneous length of the separation bubble is defined as $L_{SB}(t) = x_{r}(t) - x_{s}(t)$. The low-frequency unsteadiness is apparent in the $L_{SB}$ signal, meaning that the separation bubble undergoes a 
low-frequency contraction/expansion motion. One can also observe that, in some instants, the separation bubble has very small sizes which coincide with the moments when the separation point moves further downstream. This effect leads to a negative skewness $\mu_{3}$ in the PDF shown in Fig. \ref{fig:psd_pdf_LSB_SS}(a).
The PSDs of $L_{SB}$, $x_s$ and $x_r$ are presented in Fig. \ref{fig:psd_pdf_LSB_SS}(b). For all spectra, two energetic low-frequency peaks are captured at $St \approx 0.045$ and $St \approx 0.12$. According to previous studies \citep{dupont_haddad_debieve_2006,Touber2009,adler_gaitonde_2018}, the first peak may be associated with the bubble breathing and shock oscillations, while the second peak should corresponds to flapping of the shear layer.

Figure \ref{fig:cf_contour_PS} presents a similar analysis of the skin friction coefficient for the pressure side separation bubble. In Fig. \ref{fig:cf_contour_PS}(a), one can see that the separation and reattachment points tend to oscillate around their mean values, without large excursions. This trend is confirmed by Figs. \ref{fig:cf_contour_PS}(b) and \ref{fig:psd_pdf_LSB_PS}(a), which show the separation bubble length $L_{SB}$ and the PDF of this signal. Differently from the suction side, the PDF computed for the pressure side is symmetric about the mean, resembling a Gaussian distribution. The spectral analyses of the separation and reattachment points, as well as of the bubble length, are shown in \ref{fig:psd_pdf_LSB_PS}(b) as the PSD of the $x_s$, $x_r$ and $L_{SB}$ signals. All spectra show a dominant peak at $St \approx 0.045$, which is most likely associated with the breathing motion of the separation bubble and the reflected shock oscillation, as previously discussed. It is important to remind that the Strouhal numbers for the pressure and suction sides are computed by the respective mean bubble lengths $\langle L_{SB} \rangle$.
\begin{figure}
	
	\centering
	\begin{overpic}[trim = 5mm 5mm 5mm 5mm,clip,width=0.48\textwidth]{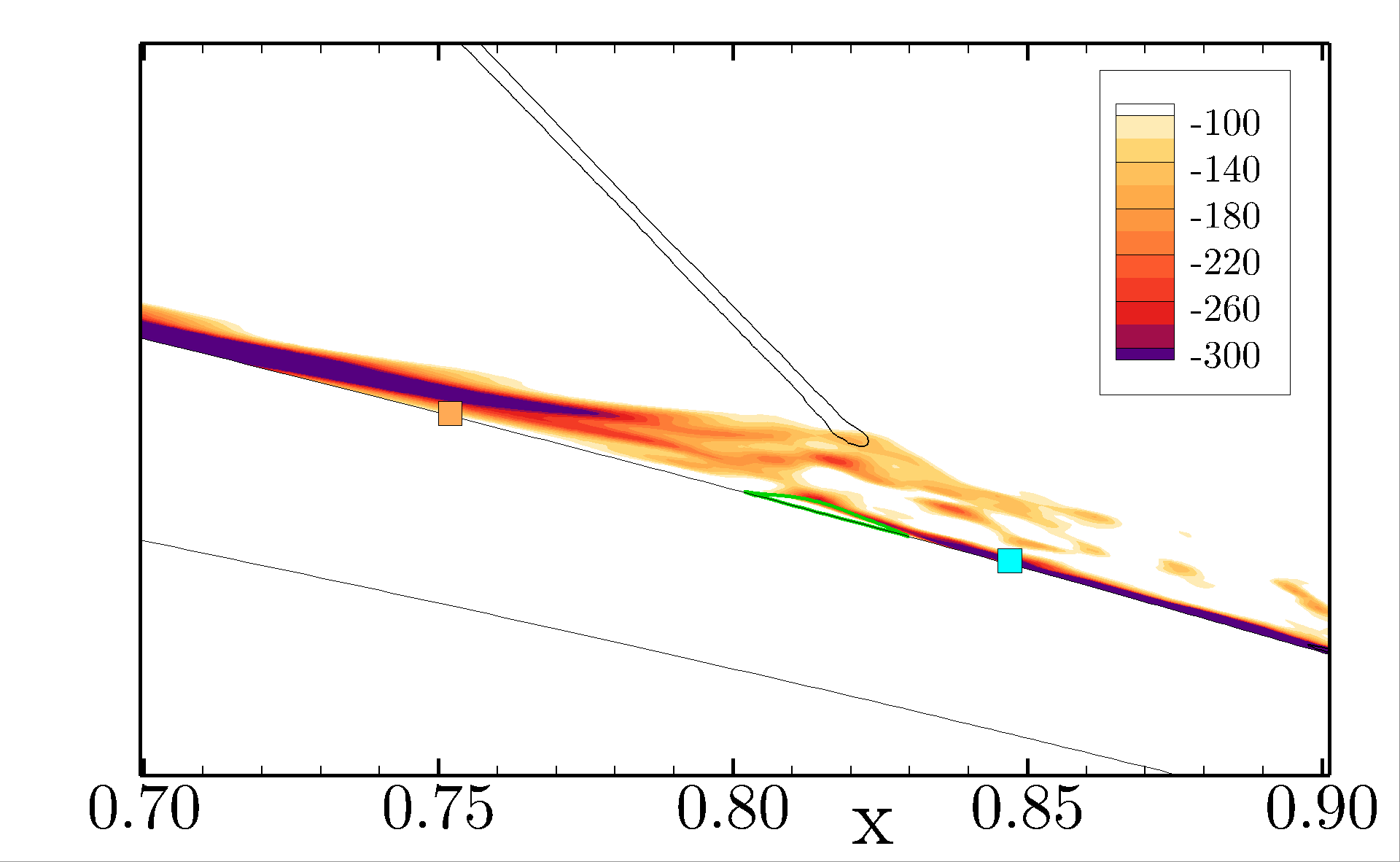}
		\put(12,12){(a)}
	\end{overpic} 
	\begin{overpic}[trim = 5mm 5mm 5mmm 5mm,clip,width=0.48\textwidth]{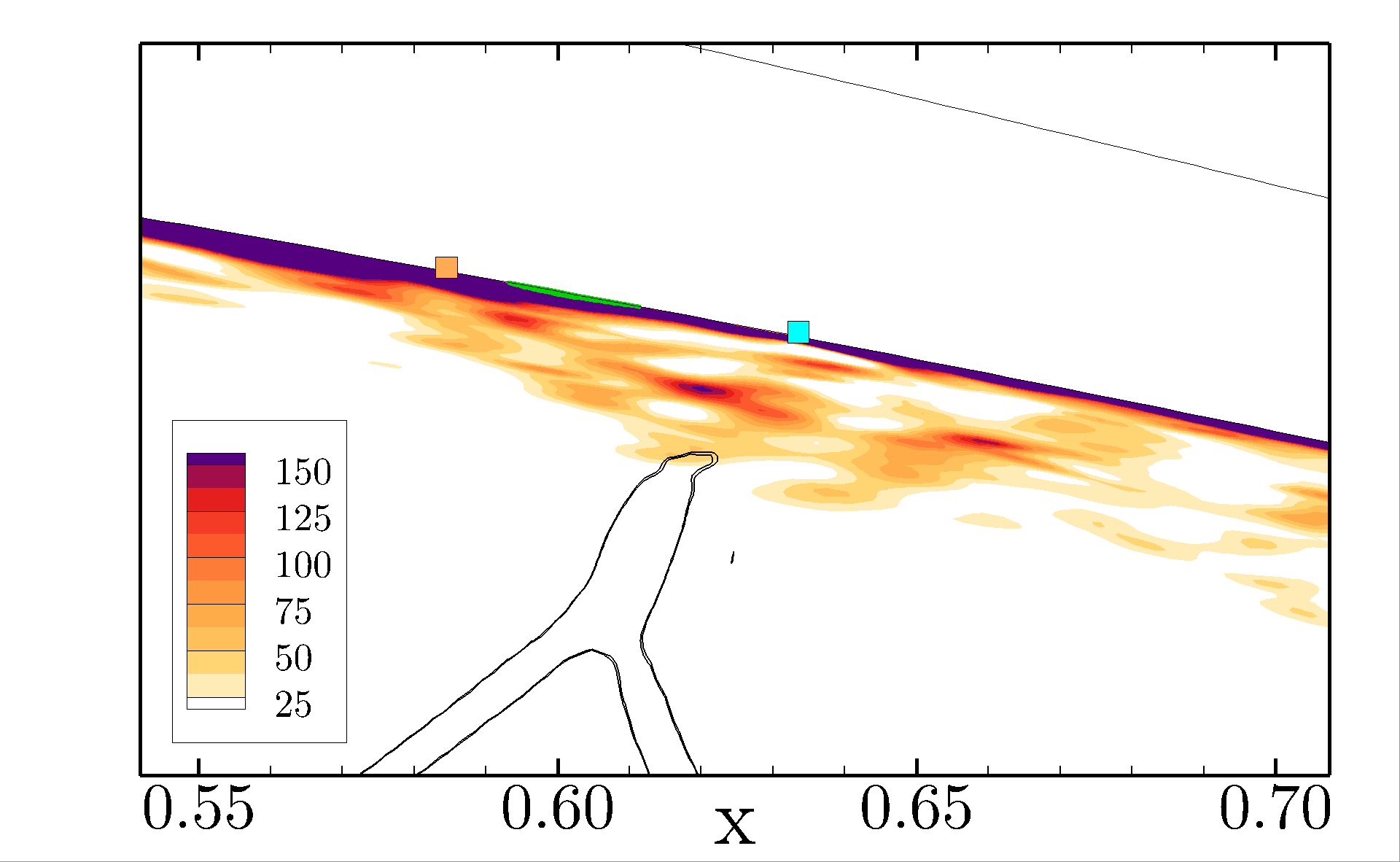}
		\put(85,42){(c)}
	\end{overpic} 
	\begin{overpic}[trim = 5mm 5mm 5mm 5mm,clip,width=0.48\textwidth]{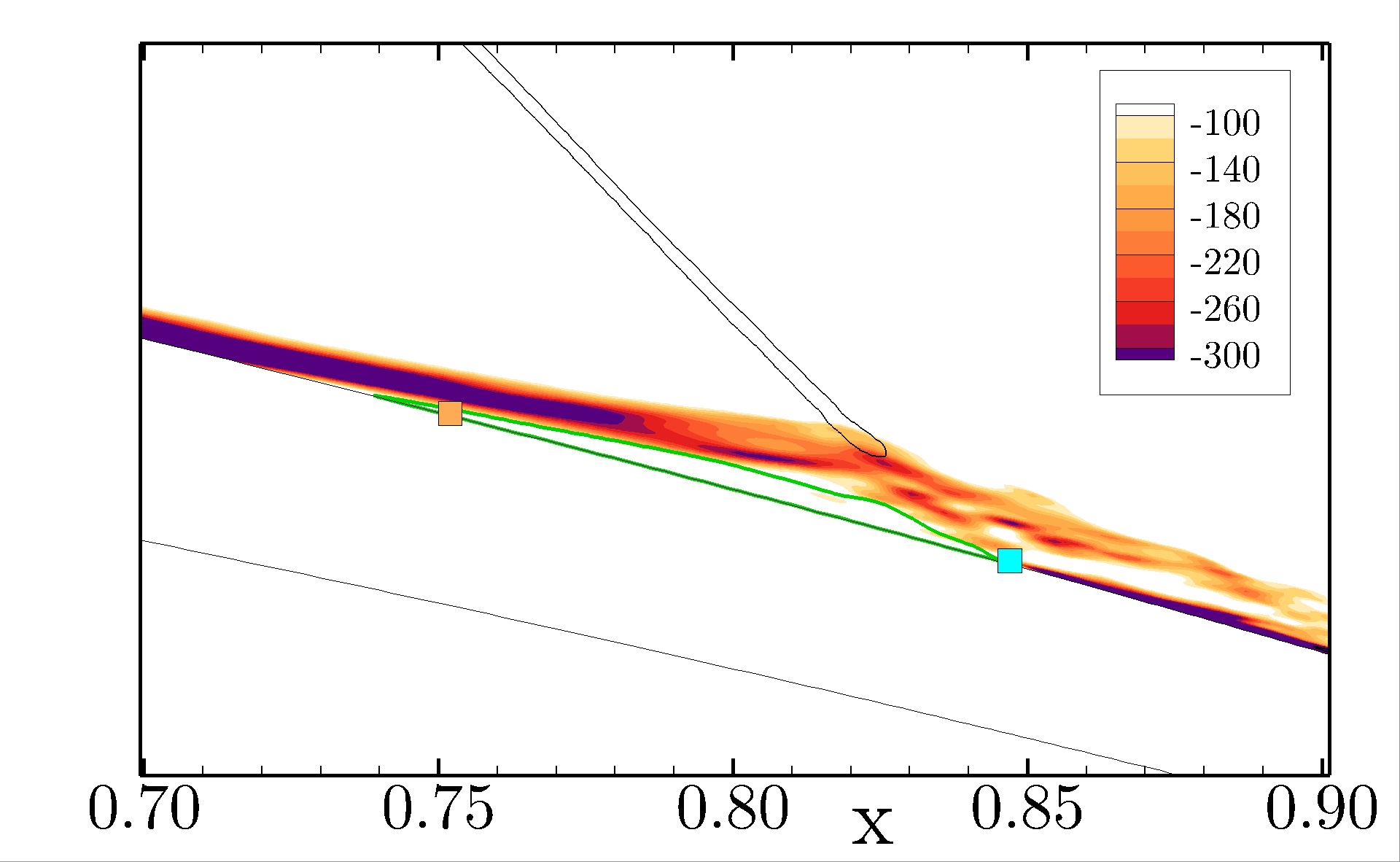}
		\put(12,12){(b)}
	\end{overpic} 
	\begin{overpic}[trim = 5mm 5mm 5mm 5mm,clip,width=0.48\textwidth]{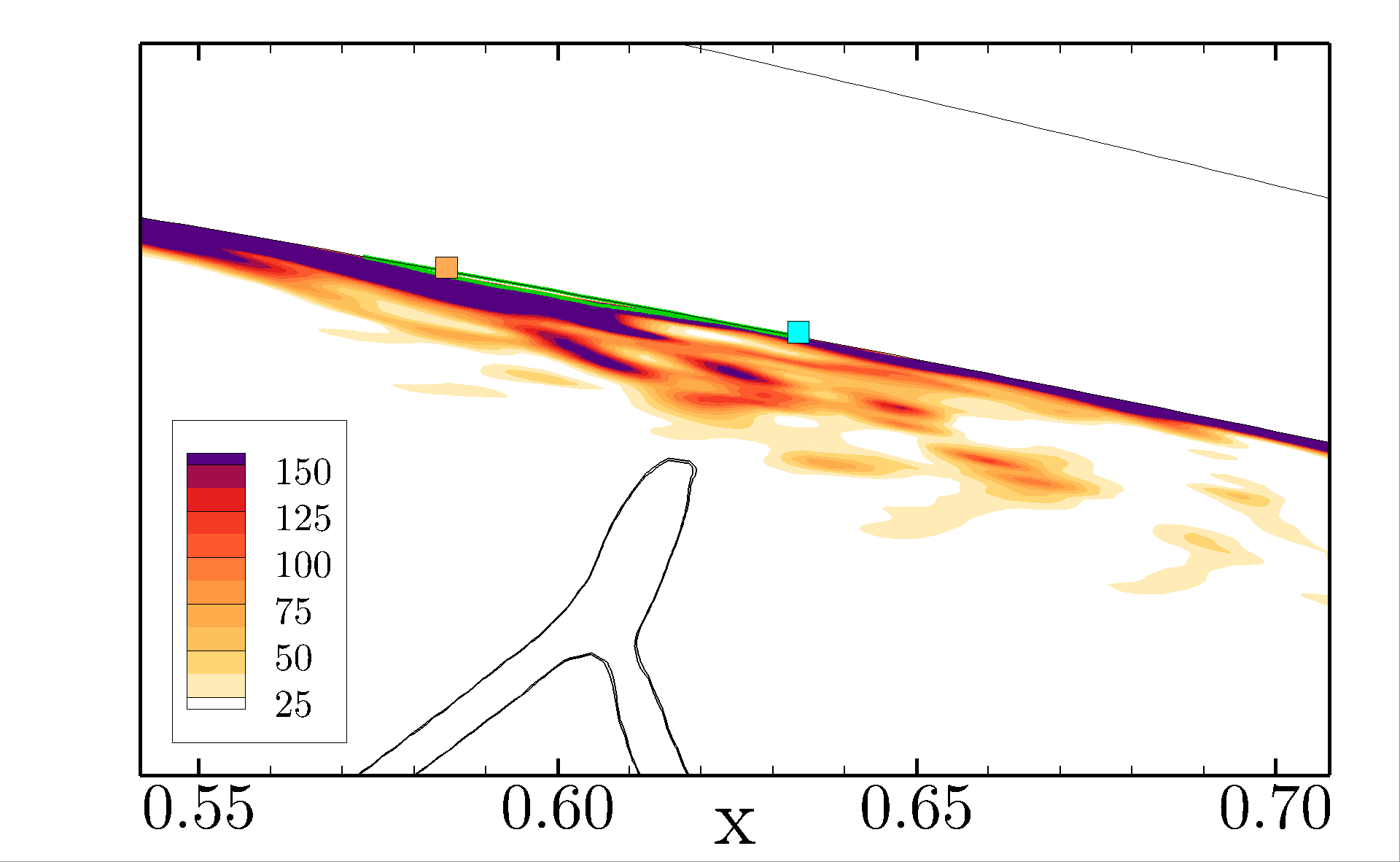}
		\put(86,42){(d)}
	\end{overpic} 
	\caption{Spanwise and time averaged $z$-vorticity contours at different time instants on the (a,b) suction side and (c,d) pressure side. The green lines depict the separation bubble and the black lines show the incident shock visualized by the pressure gradient magnitude. The colored squares represent the mean separation and reattachment positions.}
	\label{fig:z_vorticity}
\end{figure}

To illustrate the 2D structure of the separation bubble and shear layer at different instants, snapshots of $z$-vorticity are shown on the suction and pressure sides in Fig. \ref{fig:z_vorticity}. These snapshots correspond to the instants (a) and (b) indicated by vertical dashed lines in Fig. \ref{fig:cf_contour_SS}(b), and instants (c) and (d) marked in Fig. \ref{fig:cf_contour_PS}(b). In Fig. \ref{fig:z_vorticity}, the region enclosed by the green line shows the recirculation bubble and the black lines display the incident shocks. In addition, the mean separation and reattachment locations are denoted by the orange and cyan squares, respectively. In Fig. \ref{fig:z_vorticity}(a), the bubble suffers a contraction and the instantaneous separation point moves considerably downstream with respect to its mean value, while the reattachment point moves slightly upstream from the mean. Moreover, it can also be observed that the shear layer around and downstream of the bubble becomes more diffused.

The snapshot displayed in Fig. \ref{fig:z_vorticity}(b) shows an instant when the recirculation bubble expands. In this case, one can observe the upstream movement of the instantaneous separation point, whereas the reattachment point is near its mean position. At this time instant, the shear layer around the bubble is elongated and has higher vorticity levels compared to the smaller bubble. A movie displaying the time evolution of the 2D structure of the separation bubble and shear layer on the suction side is provided as supplemental material (movie 6). 

Figures \ref{fig:z_vorticity}(c) and \ref{fig:z_vorticity}(d) present the snapshots of the pressure side separation bubble at time instants when the bubble is small and large, respectively. For the former case, one can observe a small downstream displacement of the instantaneous separation point with respect to the mean. However, the reattachment point moves upstream from its reference location. When the separation bubble is larger, the instantaneous separation point has moved slightly upstream with respect to the mean, while the reattachment position is near its mean location. These trends are similar to those observed for the suction side bubble, although the incident shock structures are different. For the pressure side, the shear layers are more diffused near the incident shock. A movie showing the time evolution of the 2D structure of the separation bubble and shear layer on the pressure side is provided as supplemental material (movie 7). 

\subsection{Dynamics of shock-boundary layer interactions}
\label{section:spectral}

\begin{figure}
	
	\centering
	\begin{overpic}[trim = 1mm 1mm 1mm 1mm,clip,width=0.99\textwidth]{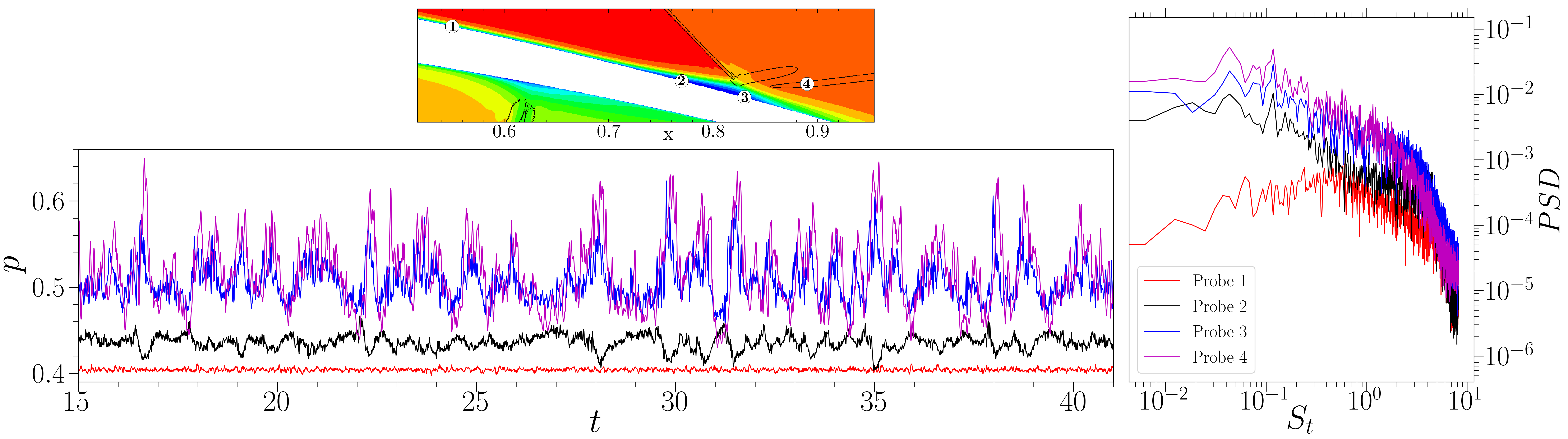}
		\put(2.5,25.0){(a)}
		\put(90,23.5){(b)}
	\end{overpic} 
	\caption{Spectral analysis of the spanwise averaged wall pressure at different locations on the suction side with (a) pressure fluctuations and (b) the corresponding PSDs.}
	\label{fig:PSD_p_wall_SS}
	
\end{figure}

The unsteadiness of the separation bubble and reattachment shock is investigated by analyzing the pressure signals and their corresponding PSDs. First, the temporal evolution of pressure signals at probes 1-4 on the suction side are shown in Fig. \ref{fig:PSD_p_wall_SS}(a), where the probe locations are indicated in Fig. \ref{fig:mean_flow_probes}. The incoming boundary layer (probe 1) shows smaller pressure fluctuations compared to the other locations. For probes 2 and 3, positioned inside the separation bubble, one can observe an increase in the mean pressure as well as an amplification of the pressure fluctuations. These effects are more pronounced for probe 3 due to the incident shock. Moreover, the signals from probes 2 and 3 appear to be anti-correlated while those from probes 3 and 4 are highly correlated. In this previous case, the strong pressure variations extracted at probe 4 exhibit a phase shift, being preceded by those from probe 3. One can also notice that the high pressure fluctuations coincide with those instants when the separation bubble is contracted, as can be seen in Fig. \ref{fig:cf_contour_SS}(b). 

To further characterize the SBLI dynamics, the PSDs of the pressure signals at probes 1-4 are presented in Fig. \ref{fig:PSD_p_wall_SS}(b). The spectral analysis yields results similar to those from Fig. \ref{fig:cf_contour_SS}(d). The same low-frequency peaks previously observed in the motion of the separation bubble are captured by probes 2 to 4. Two dominant peaks appear at $St \approx 0.045$ and $St \approx 0.12$. However, the pressure spectrum computed for the upstream turbulent boundary layer (probe 1) does not present the tones and is more broadband. Furthermore, the energetic content of this spectrum is contained at higher Strouhal numbers, while the opposite is true for the probes at the recirculation and shock regions.
\begin{figure}
	
	\centering
	\begin{overpic}[trim = 1mm 1mm 1mm 1mm,clip,width=0.99\textwidth]{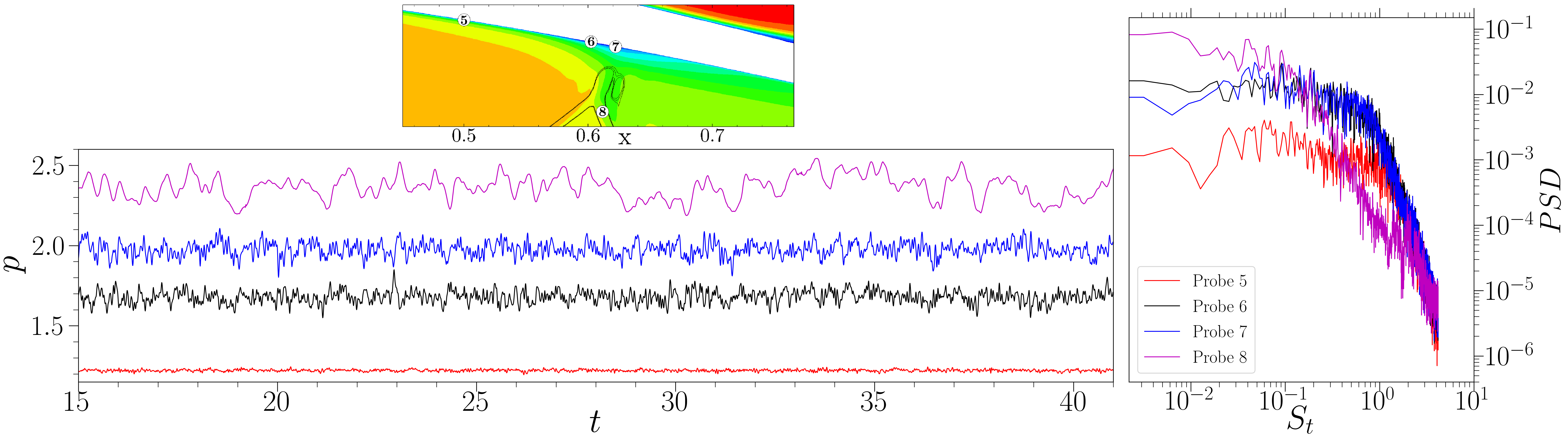}
		\put(2.5,25.0){(a)}
		\put(90,23.5){(b)}
	\end{overpic} 
	\caption{Spectral analysis of the spanwise averaged wall pressure at different locations on the pressure side with (a) pressure fluctuations and (b) the corresponding PSDs.}
	\label{fig:PSD_p_wall_PS}
	
\end{figure}

The pressure signals and their PSDs at probes 5-8 (pressure side) are shown in Figs. \ref{fig:PSD_p_wall_PS}(a) and \ref{fig:PSD_p_wall_PS}(b), respectively, where the probe locations are presented in Fig. \ref{fig:mean_flow_probes}.
Similarly to the suction side, the upstream boundary layer signal from probe 5 exhibits smaller pressure fluctuations compared to the other signals. 
For this probe, the signal energy is spread across a broad range of frequencies ($St \approx 0.01-1.0$). 
Differently from the suction side, the pressure fluctuations have similar amplitudes along the bubble at probes 6 and 7. From a visual inspection, the signals from these two probes do not seem correlated. 
Probe 8 exhibits lower frequency pressure fluctuations. Since this probe is located on the reflected shock, which is further away from the turbulent boundary layer, its spectrum shows a fast decay at the higher Strouhal numbers. The PSDs computed for the probes inside the bubble and at the shock display a tonal peak at $St \approx 0.045$ and its harmonic $St \approx 0.09$.

\begin{figure}
	\centering
	\begin{overpic}[trim = 1mm 1mm 1mm 1mm,clip,width=0.48\textwidth]{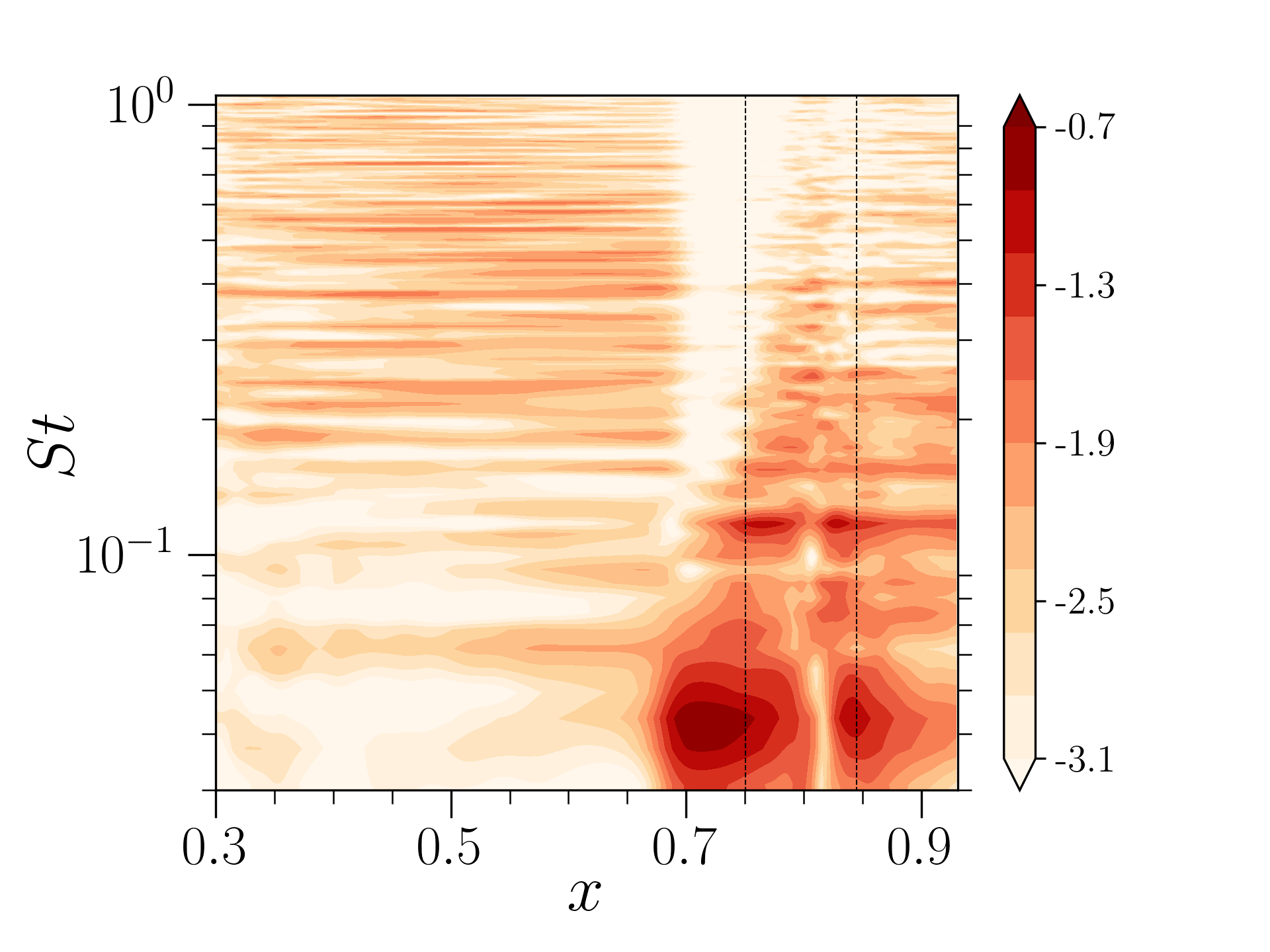}
		\put(0,65){(a)}
	\end{overpic} 
	\begin{overpic}[trim = 1mm 1mm 1mm 1mm,clip,width=0.48\textwidth]{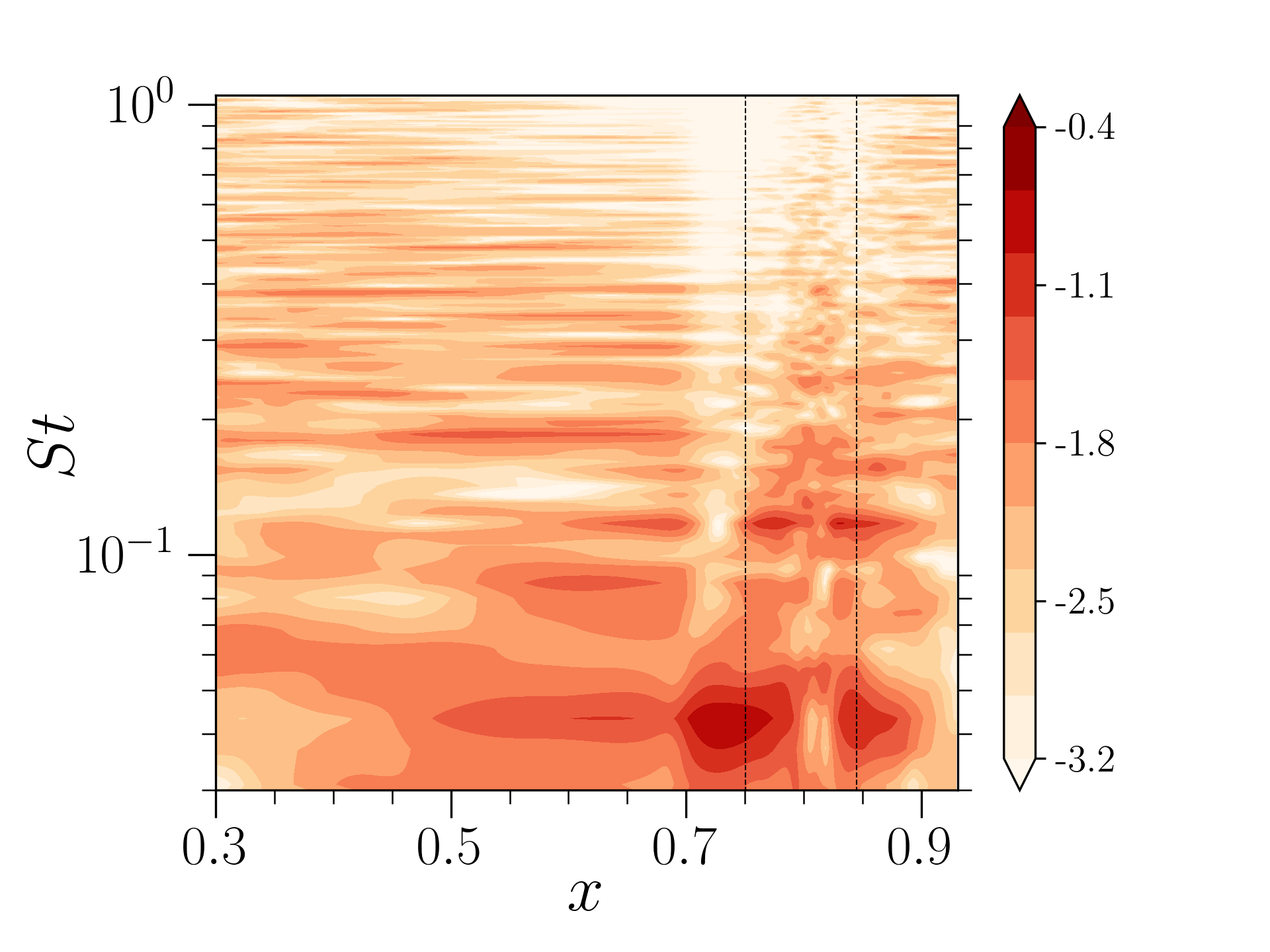}
		\put(-1,65){(b)}
	\end{overpic} 
	\begin{overpic}[trim = 1mm 1mm 1mm 1mm,clip,width=0.48\textwidth]{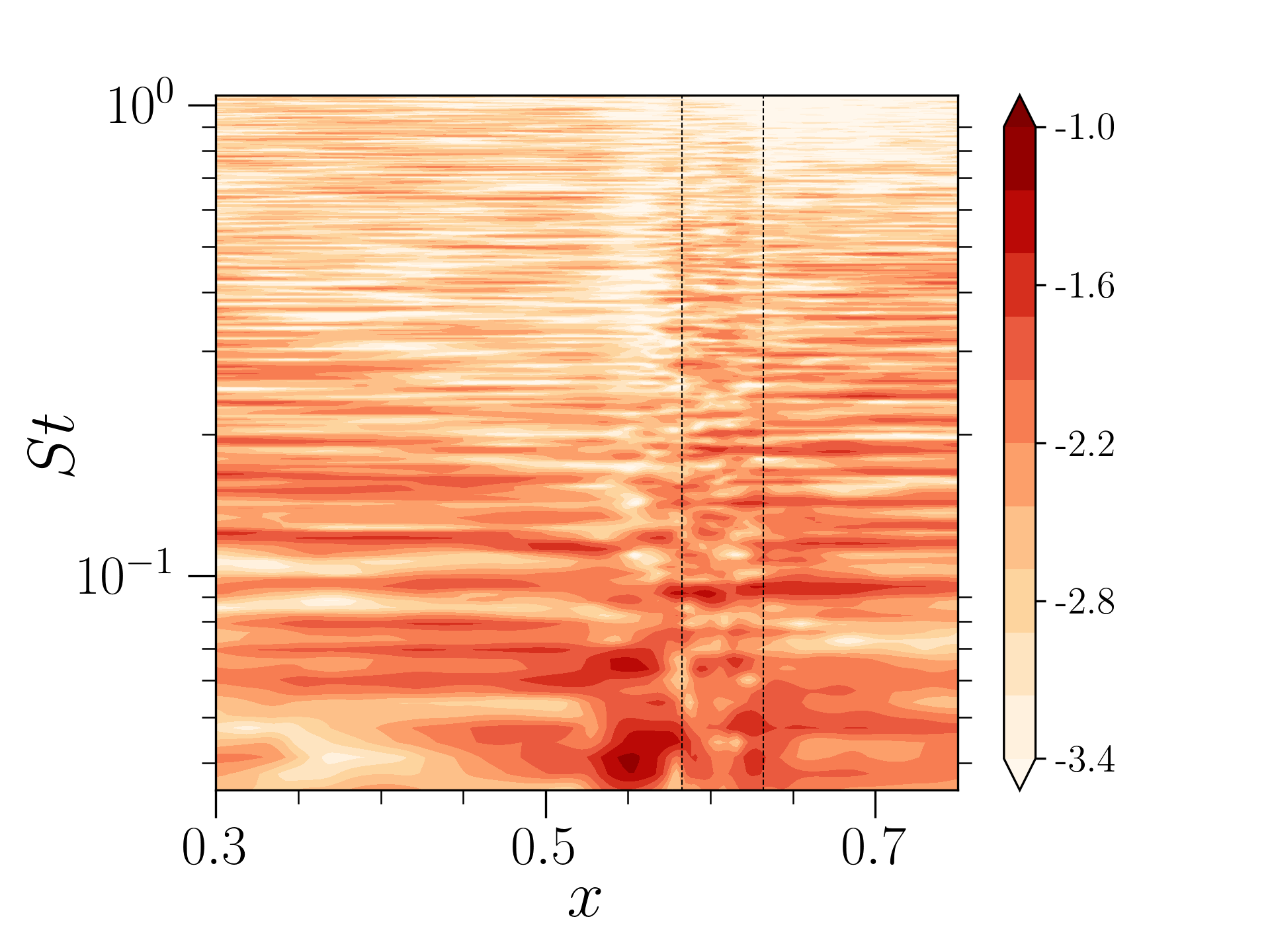}
		\put(0,65){(c)}
	\end{overpic} 
	\begin{overpic}[trim = 1mm 1mm 1mm 1mm,clip,width=0.48\textwidth]{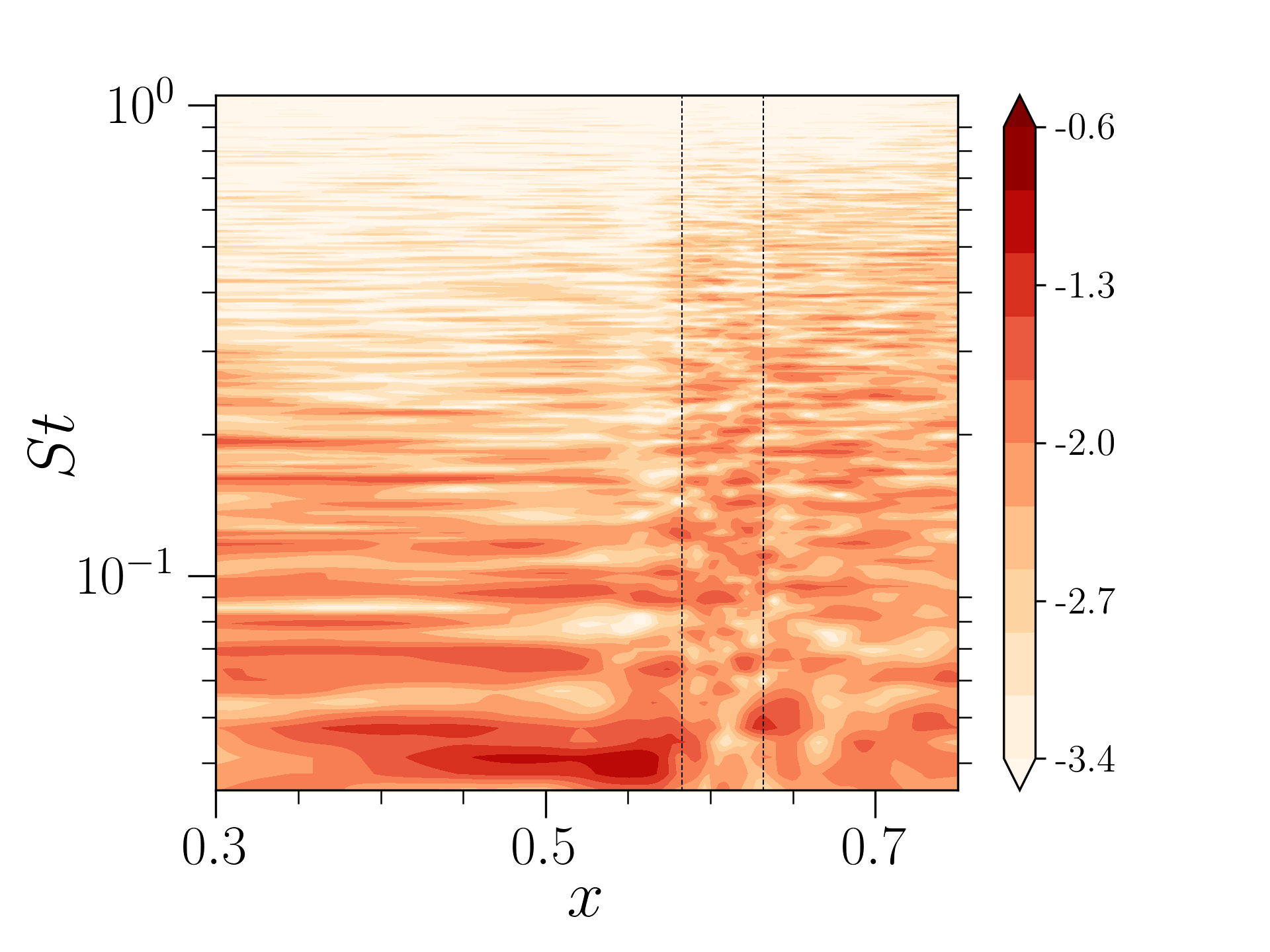}
		\put(-1,65){(d)}
	\end{overpic} 
	
	\caption{Power spectral density maps on the suction (top row) and pressure (bottom row) sides. Results are presented for (a,c) wall pressure and (b,d) tangential velocity at $y^{+} \approx 6$. Black dashed lines indicate the mean separation and reattachment locations.}
	\label{fig:PSD_maps}
\end{figure}

Figure \ref{fig:PSD_maps} shows the contours of PSD in terms of wall pressure (left column) and tangential velocity at $y^{+} \approx 6$ (right column) as a function of the frequency ($St$) and the airfoil chord location ($x$). The top and bottom rows present results for the suction and pressure sides, respectively. The PSD at each streamwise position is normalized by its integrated value for all frequencies and color levels are plotted in log-scale. The black vertical dashed lines indicate the mean separation and reattachment positions. In Fig. \ref{fig:PSD_maps}(a), one can observe two pronounced peaks at $St \approx 0.045$ and $St \approx 0.12$. For the lower frequency, the peaks appear just upstream the mean separation point and at the mean reattachment position. However, the higher frequency peaks are excited inside the bubble, slightly downstream the separation point and upstream of the reattachment location. 

The lower frequency content ($St \approx 0.045$) found near the separation region in Fig. \ref{fig:PSD_p_wall_SS} agrees well with previous studies \cite{dupont_haddad_debieve_2006,Touber2009,priebe2012,adler_gaitonde_2018}. However, the low-frequency peak present near the reattachment location is not so commonly observed in canonical cases of oblique SBLIs. Similarly to the current wall pressure PSD, \citet{adler_gaitonde_2018} reported the presence of low-frequency content ($St \approx 0.03$) near the separation and reattachment locations, where the low-frequency peak at reattachment is less intense compared to that near the separation point. These authors suggested that normalizing the PSD independently at each streamwise position (as also done in the present work) highlights the low-frequency fluctuations at reattachment. In the present SBLI, we believe that the cause of the reattachment peak is due to the streaks transported over the bubble, which burst impinging the reattachment shock. They lead to the low-frequency motion of the bubble in this region and this feature can be seen in Figs. \ref{fig:separation_bubble_3D_SS}(a) and \ref{fig:separation_bubble_3D_SS}(b), besides the movie submitted as supplemental material.

The PSD map of the tangential velocity near the wall is presented in Fig. \ref{fig:PSD_maps}(b) and allows one to infer about the upstream boundary layer influence on the low-frequency unsteadiness. Results are obtained for the same $y$-position as in Figs. \ref{fig:separation_bubble_3D_SS}(e) and \ref{fig:separation_bubble_3D_SS}(f), which show streaks impinging on the separation bubble. The PSD maps of the streamwise velocity show that the incoming boundary layer excites several frequencies of the spectrum. 
However, the separation region responds selectively to an excitation at $St \approx 0.045$. This result suggests that near-wall streaks in the upstream boundary layer drive the separation bubble motion, which is in close agreement with the observations in Fig. \ref{fig:separation_bubble_3D_SS} and conforms with the findings from Refs. \citep{ganapathisubramani_clemens_dolling_2009, porter_2019}.

Results for the pressure side are plotted in Figs. \ref{fig:PSD_maps}(c) and \ref{fig:PSD_maps}(d). The wall pressure PSD displays two regions with high spectral energy at $St \approx 0.045$, as can be seen in Fig. \ref{fig:PSD_maps}(c). The first one appears upstream of the separation region and the second spot is observed near the reattachment location. A higher frequency at $St \approx 0.09$ is also excited along the recirculation bubble and downstream of the SBLI. The PSD map of the tangential velocity near the wall is shown in Fig. \ref{fig:PSD_maps}(d) and results are obtained for the same wall-normal position as in Figs. \ref{fig:separation_bubble_3D_PS}(e) and \ref{fig:separation_bubble_3D_PS}(f). This map illustrates that the upstream boundary layer carries some content in the lower band of the spectrum, where a region of high spectral energy is observed at $St \approx 0.04$. Differently from the suction side, the recirculation bubble is not excited by velocity fluctuations at low-frequencies. That occurs because most streaks are transported above the pressure side bubble, which is thinner than that formed on the suction side.
\begin{figure}
	\centering
	\begin{overpic}[trim = 1mm 16mm 1mm 5mm,clip,width=0.48\textwidth]{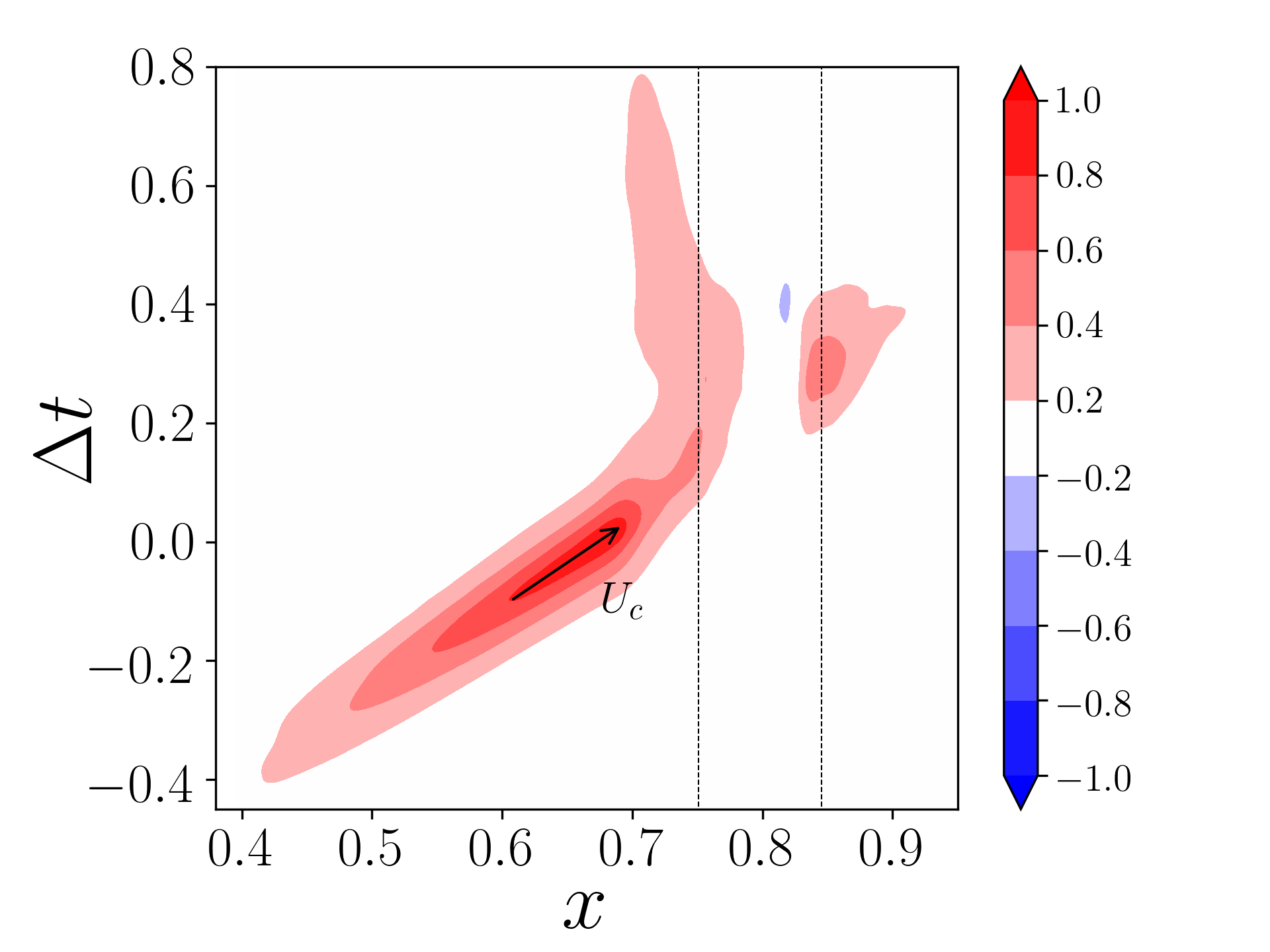}
		\put(0,57){(a)}
	\end{overpic} 
	\begin{overpic}[trim = 1mm 16mm 1mm 5mm,clip,width=0.48\textwidth]{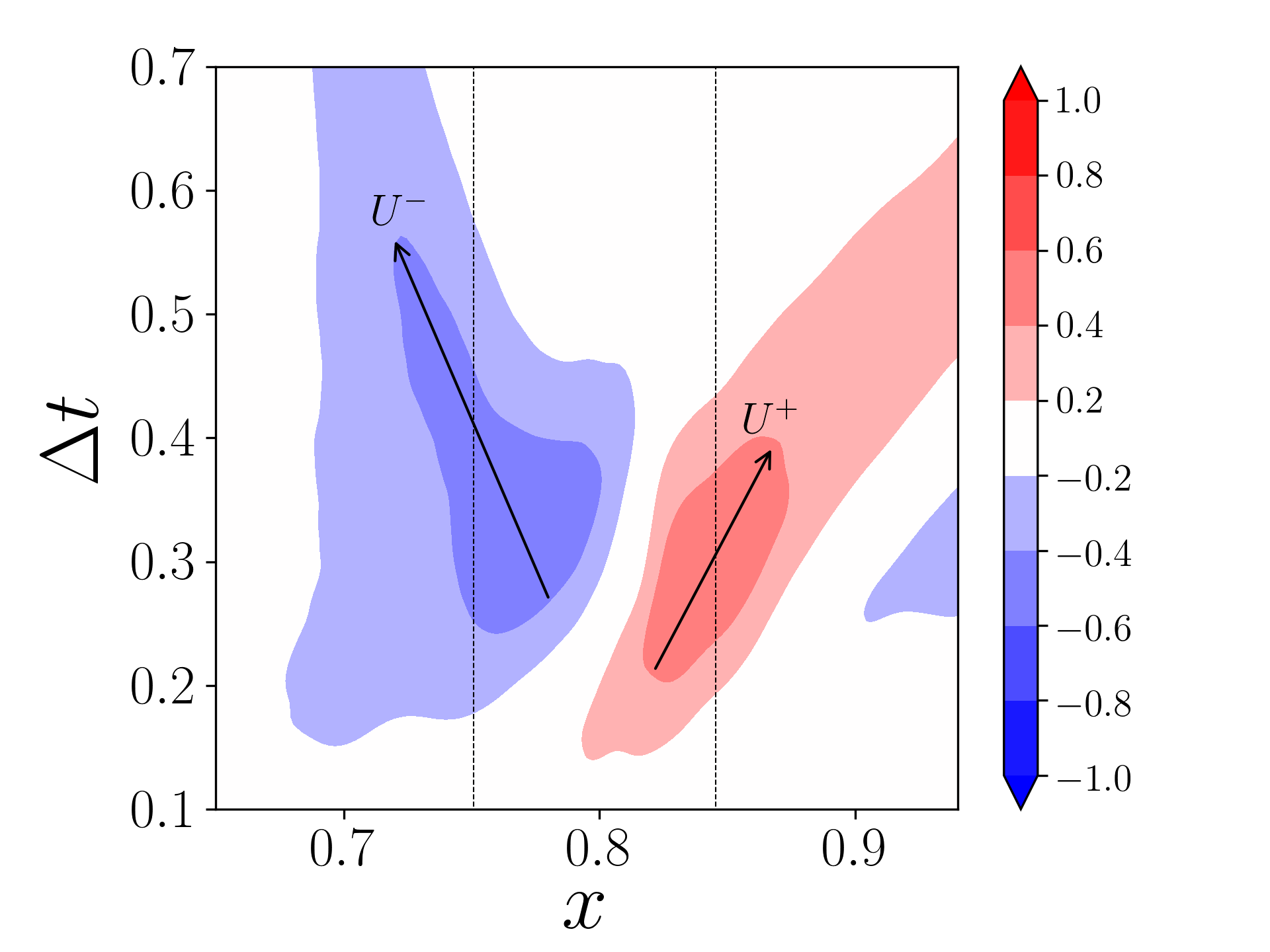}
		\put(0,57){(b)}
	\end{overpic} 
	\begin{overpic}[trim = 1mm 1mm 1mm 5mm,clip,width=0.48\textwidth]{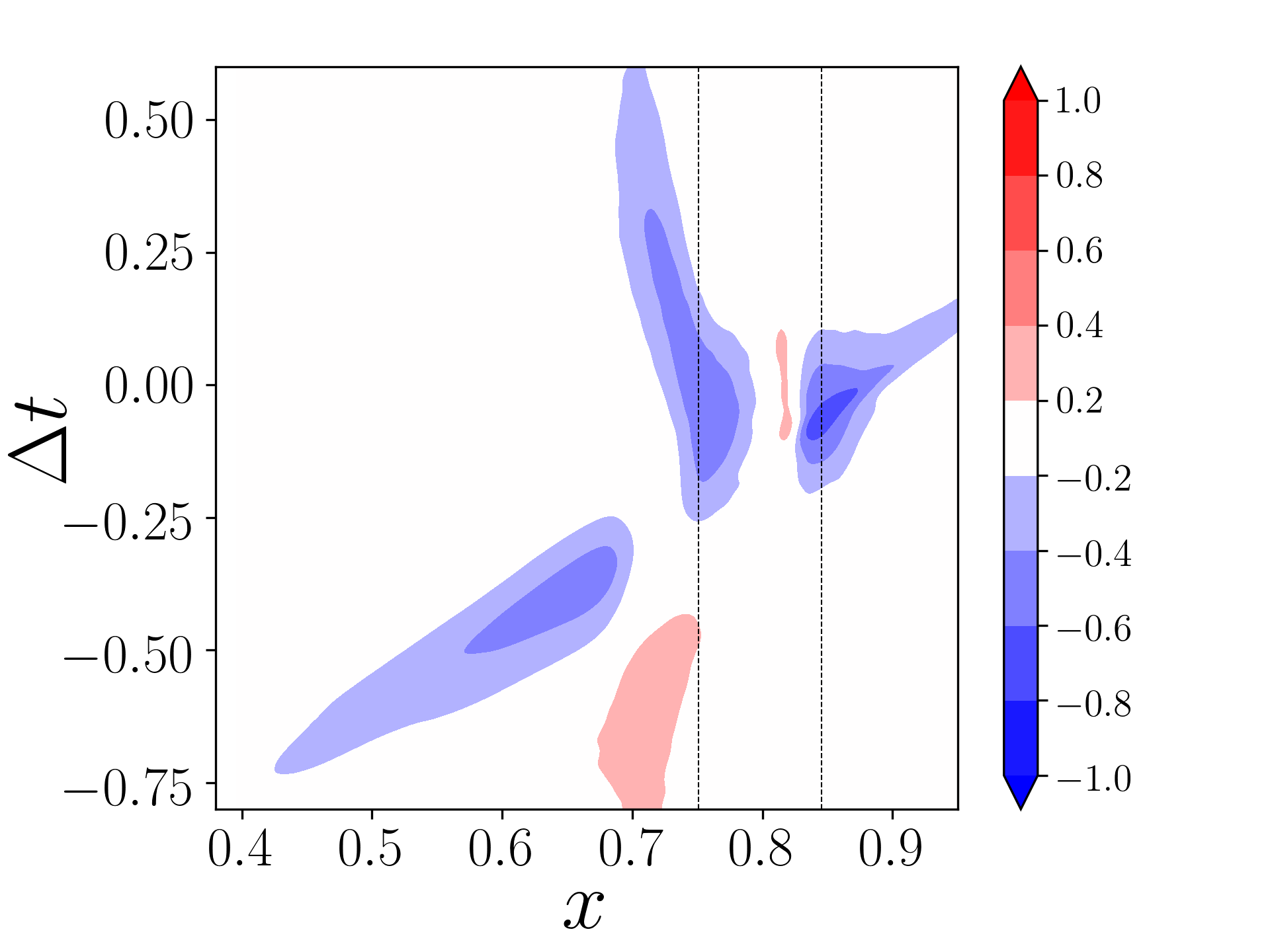}
		\put(0,65){(c)}
	\end{overpic} 
	\begin{overpic}[trim = 1mm 1mm 1mm 5mm,clip,width=0.48\textwidth]{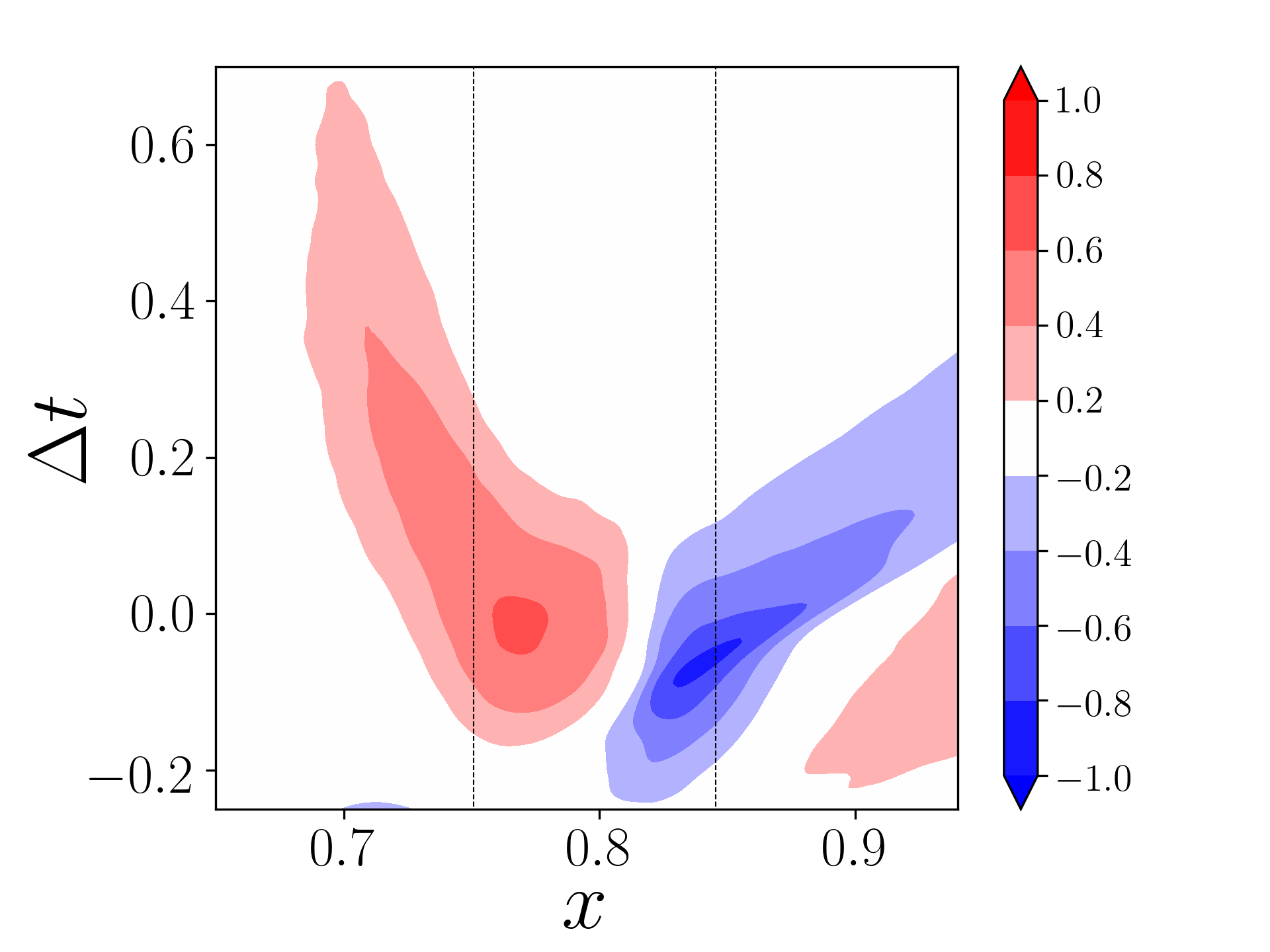}
		\put(0,65){(d)}
	\end{overpic} 
	\caption{Suction side space-time correlations computed for (a) tangential velocity at upstream location (probe 9) with skin-friction coefficient, (b) tangential velocity at probe 9 with wall pressure, (c) tangential velocity at the reattachment shock (probe 4) with skin-friction coefficient, and (d) tangential velocity at probe 4 with wall pressure.}
	\label{fig:correlation_SS}
\end{figure}

To further investigate the influence of upstream or downstream disturbances in the SBLI, an analysis of space-time correlations is presented in Fig. \ref{fig:correlation_SS} for the suction side. A similar analysis was also conducted for the pressure side SBLI, but lower correlations were observed. In the figure, the vertical dotted lines delimit the region of separated flow. Figure \ref{fig:correlation_SS}(a) presents the cross-correlation of the tangential velocity computed at probe 9 from Fig. \ref{fig:mean_flow_probes} with the skin-friction coefficient along the suction side. The probe is placed upstream of the separation bubble, at $y^{+} \approx 6$. In this figure, a region of high correlation is observed upstream of the separation region, near the reference location, indicating the presence of near-wall flow structures in the incoming boundary layer. The mean convective velocity $U_c$ of such structures is estimated from the plot as $U_c/U_{\infty} \approx 0.67$. A high correlation is also computed at the mean separation and reattachment points, and they occur at positive values of time delay $\Delta t$, based on the inlet velocity. This suggests that the large-scale structures in the incoming boundary layer drive the separation bubble motion. As can be seen from the plot, the tangential velocity fluctuations in the upstream boundary layer are positively correlated with the skin-friction coefficient. Therefore, the positive (negative) near-wall velocity fluctuations lead to an increase (decrease) in the skin-friction coefficient. This result confirms the previous observation from Fig. \ref{fig:cf_contour_SS}, i.e., the fast (slow) boundary layer streaks cause a contraction (expansion) of the separation bubble. 

The space-time correlation computed for the tangential velocity at probe 9 and the wall pressure along the suction side is shown in 
Fig. \ref{fig:correlation_SS}(b). The first observation from this plot is that the tangential velocity fluctuations in the upstream boundary layer are negatively correlated with the wall pressure at the bubble leading edge. However, the correlation is positive at the bubble trailing edge, indicating a phase jump of $\pi$ in the pressure fluctuations along the bubble. 
This finding has also been reported in previous studies \citep{dupont_haddad_debieve_2006,Touber2009,priebe2012,Hartmann2013,Klinner2021}. The time delay is positive, indicating that the pressure disturbances arise after the advection of the boundary layer streaks. A second observation from this figure concerns the propagation mechanisms of pressure disturbances. A closer look at the cross-correlation reveals that the bubble leading edge  responds to its trailing edge due to the positive time delay. After the jump, a downstream propagating pressure disturbance is seen, while before the jump, we observe an upstream propagating pressure disturbance. The propagation speeds of such disturbances can be estimated from the plot as $U_{-}/U_{\infty} \approx -0.21$ and $U_{+}/U_{\infty} \approx 0.25$, where $U_{-}$ and $U_{+}$ represent the speed of propagation of upstream and downstream pressure disturbances, respectively. 

The correlation of the tangential velocity at the reattachment shock (probe 4) with the skin-friction coefficient is shown in Fig. \ref{fig:correlation_SS}(c). It can be seen that the tangential velocity fluctuations at the reattachment shock are negatively correlated with the skin-friction coefficient. 
Similar to Fig. \ref{fig:correlation_SS}(a), three regions of high correlation are observed, indicating that the motion of the reattachment shock is associated with the skin-friction and, hence, with the streaks in the incoming boundary layer as well as the separation bubble. Results suggest that negative (positive) tangential velocity fluctuations in the incoming near-wall flow lead to an expansion (contraction) of the bubble, which in turn result in the downstream (upstream) motion of the reattachment shock. When the bubble expands (contracts), the skin-friction is reduced (increased) and the shock moves downstream (upstream) leading to positive (negative) velocity fluctuations.

Finally, Fig. \ref{fig:correlation_SS}(d) presents the space-time correlation of tangential velocity at probe 4 from Fig. \ref{fig:mean_flow_probes} and wall pressure. As can be seen from the figure, the tangential velocity fluctuations at the reattachment shock are positively correlated with the wall pressure on the bubble leading edge. The opposite is true for the bubble trailing edge due to the phase jump previously observed. From this figure, higher correlation values occur at ``negative time delays'' near the mean reattachment point, indicating that the reattachment shock is preceded by the pressure disturbances at the wall. This observation suggests that the bubble drives the shock motion.

\subsection{Analysis of coherent structures}
\label{section:POD}

\begin{figure}
	\centering
	\begin{overpic}[trim = 50mm 32mm 60mm 10mm, clip,width=.48\linewidth]{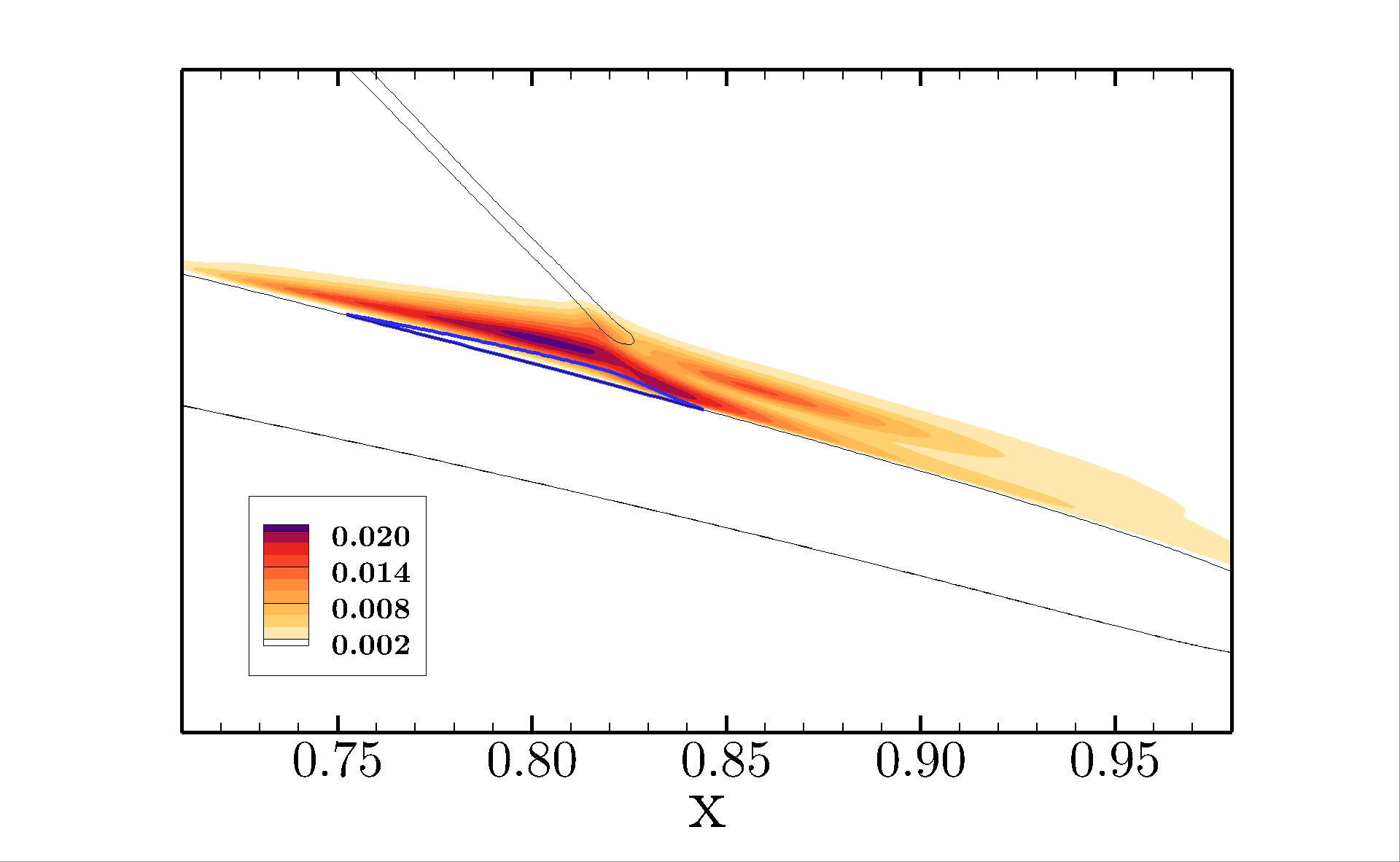}
		\put(10,55){(a)}
	\end{overpic}
	\begin{overpic}[trim = 50mm 32mm 60mm 10mm, clip,width=.48\linewidth]{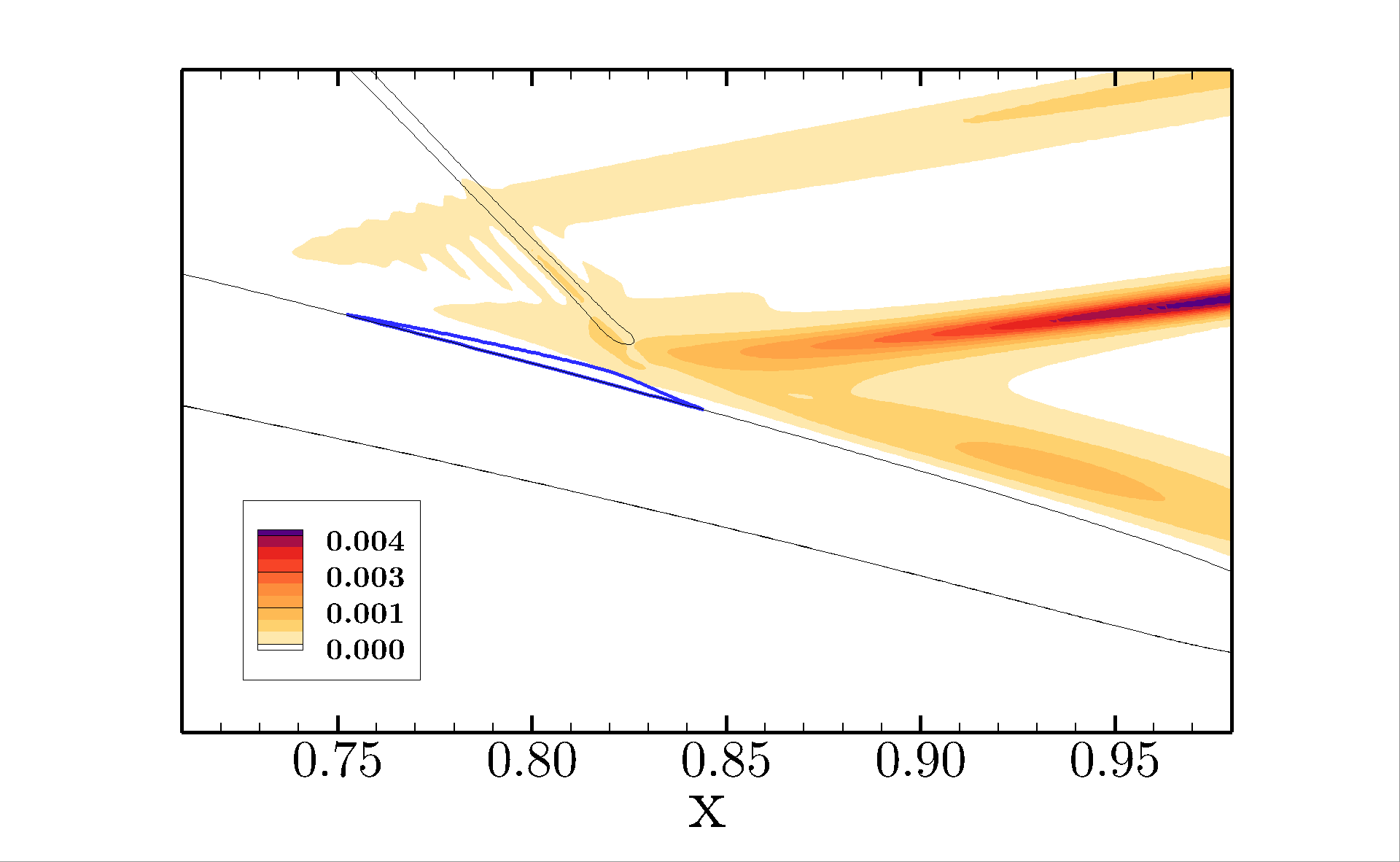}
		\put(10,55){(b)}
	\end{overpic}
	
	\begin{overpic}[trim = 50mm 5mm 60mm 10mm, clip,width=.48\linewidth]{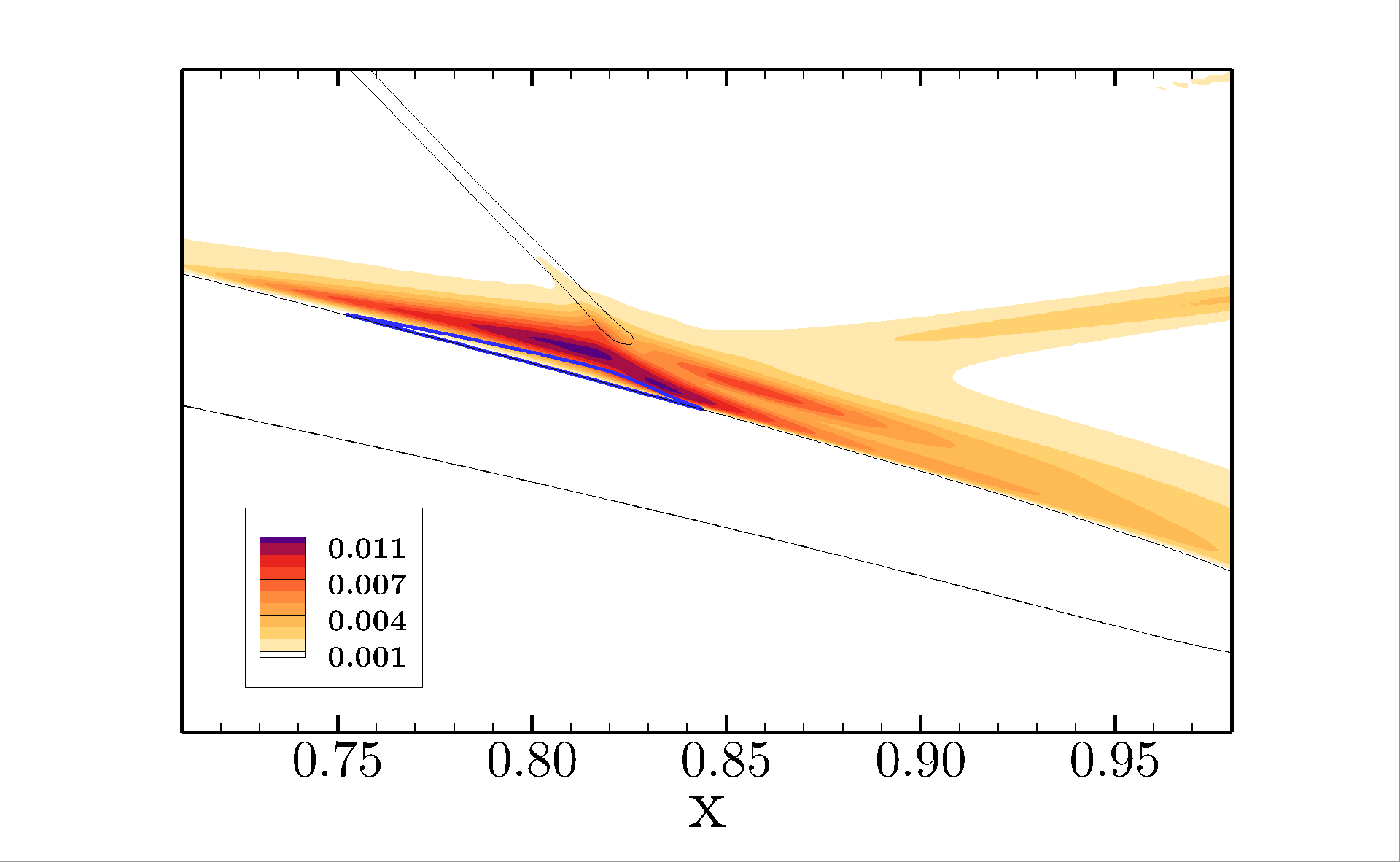}
		\put(10,60){(c)}
	\end{overpic}
	\begin{overpic}[trim = 50mm 5mm 60mm 10mm, clip,width=.48\linewidth]{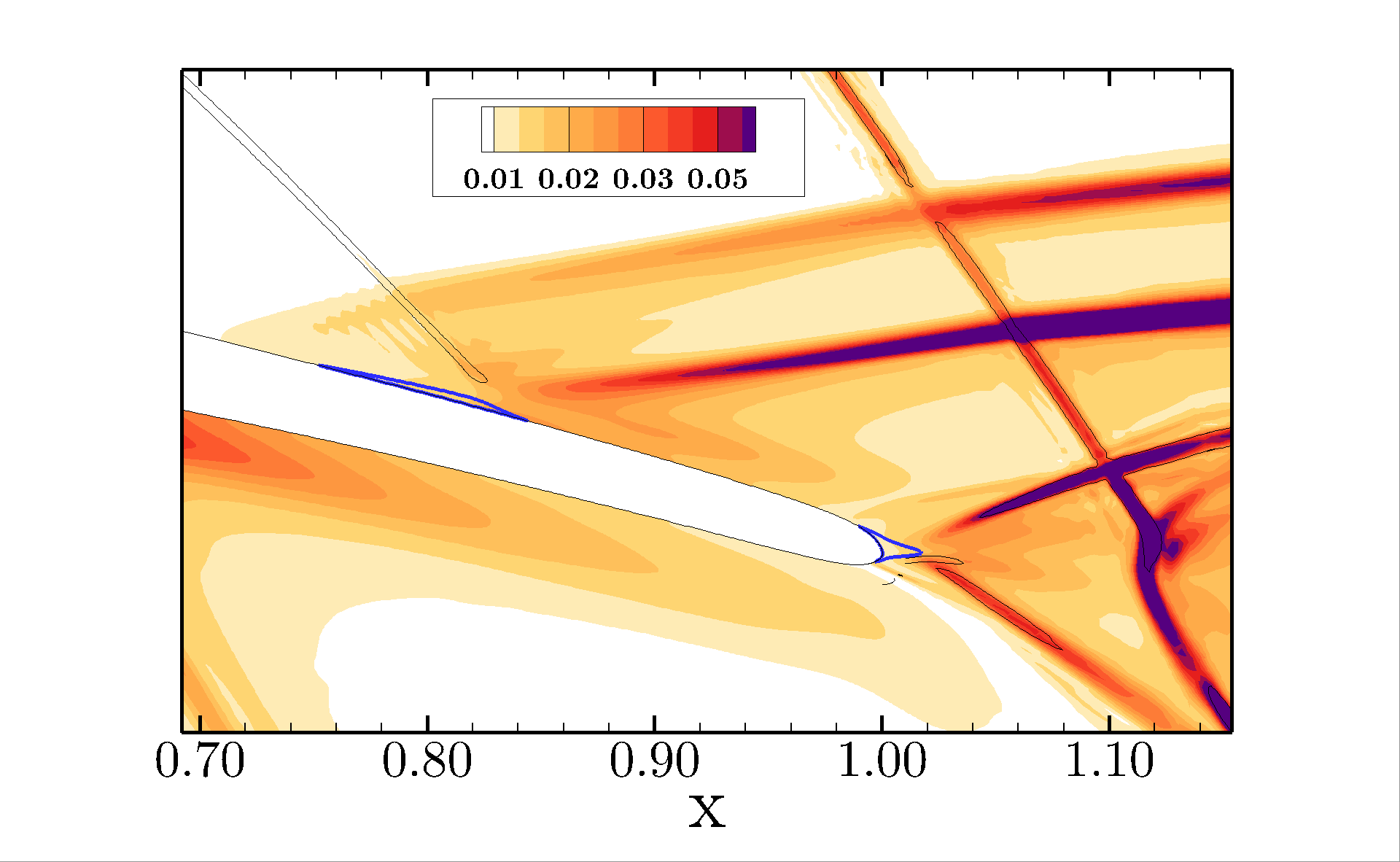}
		\put(17.5,60){(d)}
	\end{overpic}
    \caption{Spanwise and time averaged turbulence quantities on the suction side with (a) $\langle u_t u_t \rangle$, (b) $\langle u_n u_n \rangle$, (c) turbulent kinetic energy, and (d) pressure RMS.}
	\label{fig:reynolds_stresses_SS_2D}
\end{figure}

In previous sections, it was observed that the large-scale boundary layer structures impinge on the separation bubble, playing an important role on the bubble and shock motions. To better understand the bubble-shock interactions on both sides of the airfoil, including their excitation frequencies, an analysis is presented in this section via the filtered POD method. The sum of all the POD modes should provide a reconstruction of the flow fluctuation field. Hence, before the POD analysis is performed, some turbulent quantities are presented in Fig. \ref{fig:reynolds_stresses_SS_2D} for the suction side, namely, the tangential and wall-normal Reynolds stresses, $\langle u_t u_t \rangle$ and $\langle u_n u_n \rangle$, respectively, the turbulent kinetic energy (TKE) and the pressure root mean square (RMS). These quantities allow an assessment of the flow regions where the fluctuations are more pronounced.

Intense values of the tangential Reynolds stress component $\langle u_t u_t \rangle$ are found along the shear layer, over the separation bubble, as can be seen in Fig. \ref{fig:reynolds_stresses_SS_2D}(a). Downstream the incident shock, the maximum values of $\langle u_t u_t \rangle$ present a bifurcation with one branch along the free shear layer and another one along the wall. 
In Fig. \ref{fig:reynolds_stresses_SS_2D}(b), high values of $\langle u_n u_n \rangle$ are observed upstream of the bubble. The highest fluctuations of the wall-normal stress component are observed at the reattachment shock characterizing its unsteady motion. In addition, a region with high levels of $\langle u_n u_n \rangle$ is also found along the free shear layer downstream the bubble. As expected, the distribution of turbulent kinetic energy captures the fluctuations of $\langle u_t u_t \rangle$ and $\langle u_n u_n \rangle$ combined, as can be seen in Fig. \ref{fig:reynolds_stresses_SS_2D}(c). The pressure root mean square contours plotted in Fig. \ref{fig:reynolds_stresses_SS_2D}(d) show the unsteadiness of the upstream compression waves at the bubble leading edge and the reattachment shock, as well as the trailing edge shocks.

As also observed by \citet{DELERY1985}, the wall curvature leads to weaker compression waves that do not coalesce into a separation shock. The low frequency pressure fluctuations observed in Fig. \ref{fig:PSD_maps}(a) upstream of the bubble are caused by these compression waves which have their intensity affected by the bubble breathing. A similar mechanism is reported by \citet{Nichols_2017} for the interaction between a bubble and a separation shock. When the bubble expands and achieves a larger size, as shown in Fig. \ref{fig:z_vorticity}(b), the compression waves become stronger due to the more pronounced flow deflection. This, in turn, leads to stronger pressure fluctuations upstream of the mean separation location as can be seen in Fig. \ref{fig:reynolds_stresses_SS_2D}(d).

\begin{figure}
	\centering
	\begin{overpic}[trim = 50mm 34mm 60mm 10mm, clip,width=.48\linewidth]{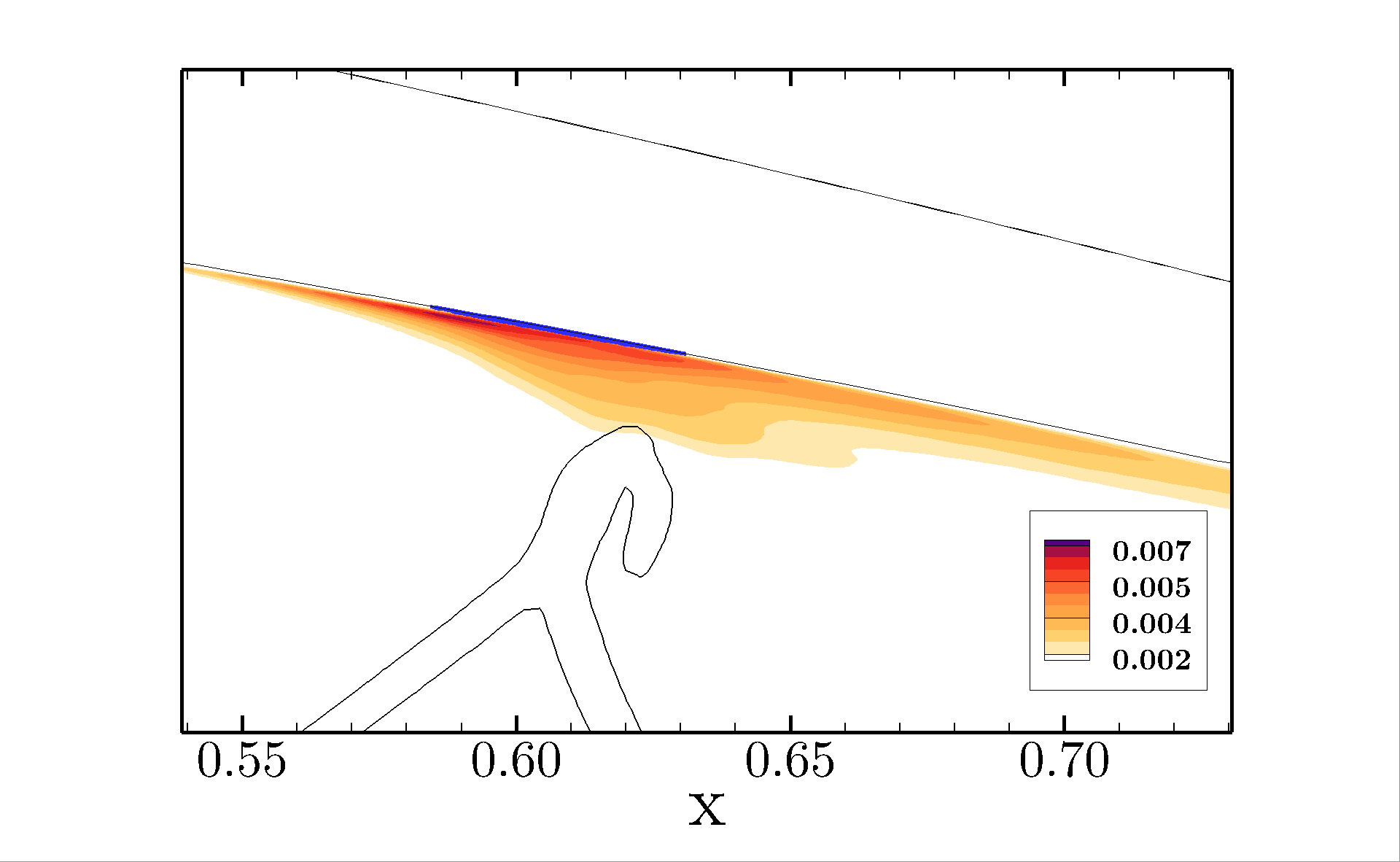}
		\put(12,52){(a)}
	\end{overpic}
	\begin{overpic}[trim = 50mm 34mm 60mm 10mm, clip,width=.48\linewidth]{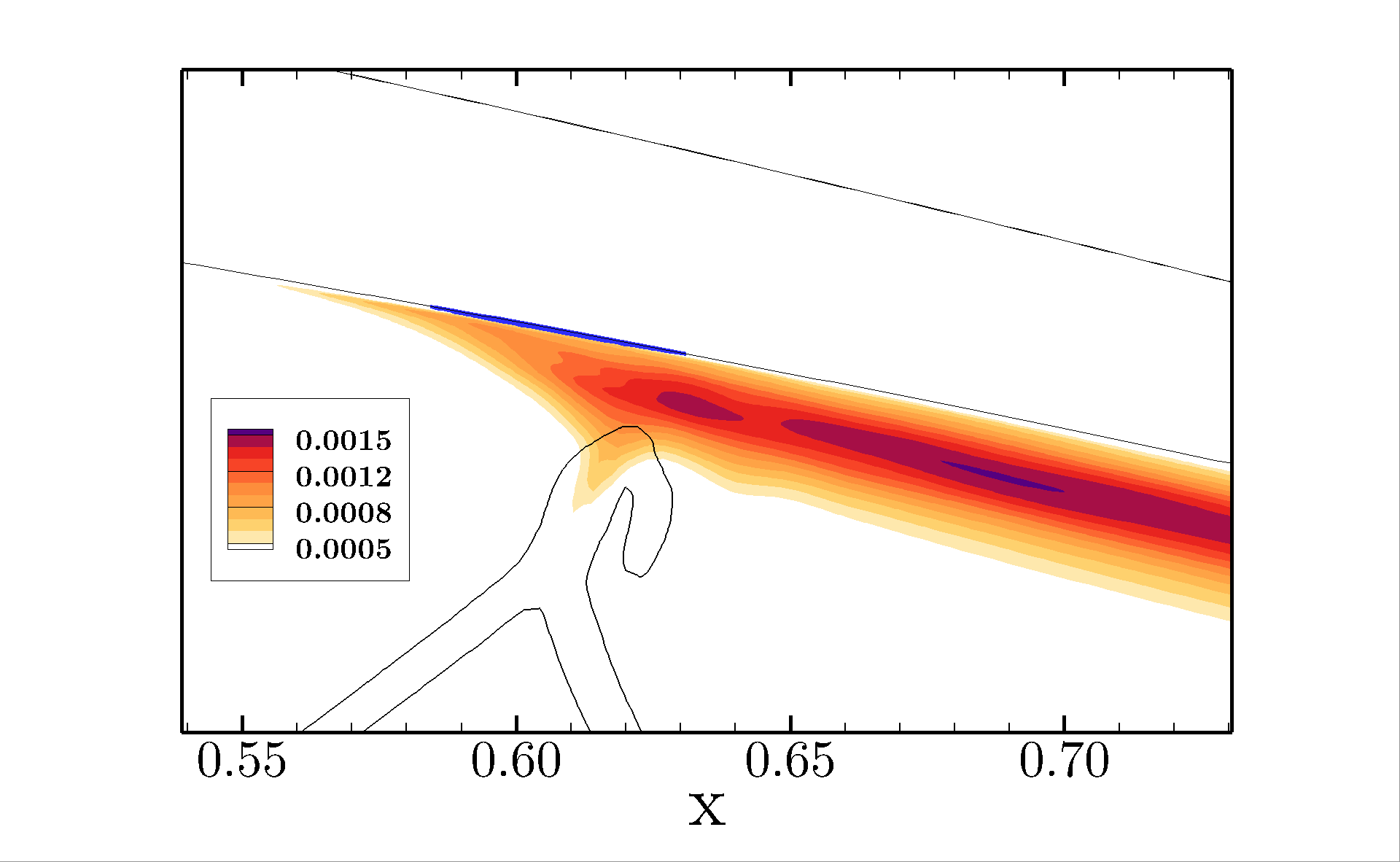}
		\put(12,52){(b)}
	\end{overpic}
	\begin{overpic}[trim = 50mm 2mm 60mm 10mm, clip,width=.48\linewidth]{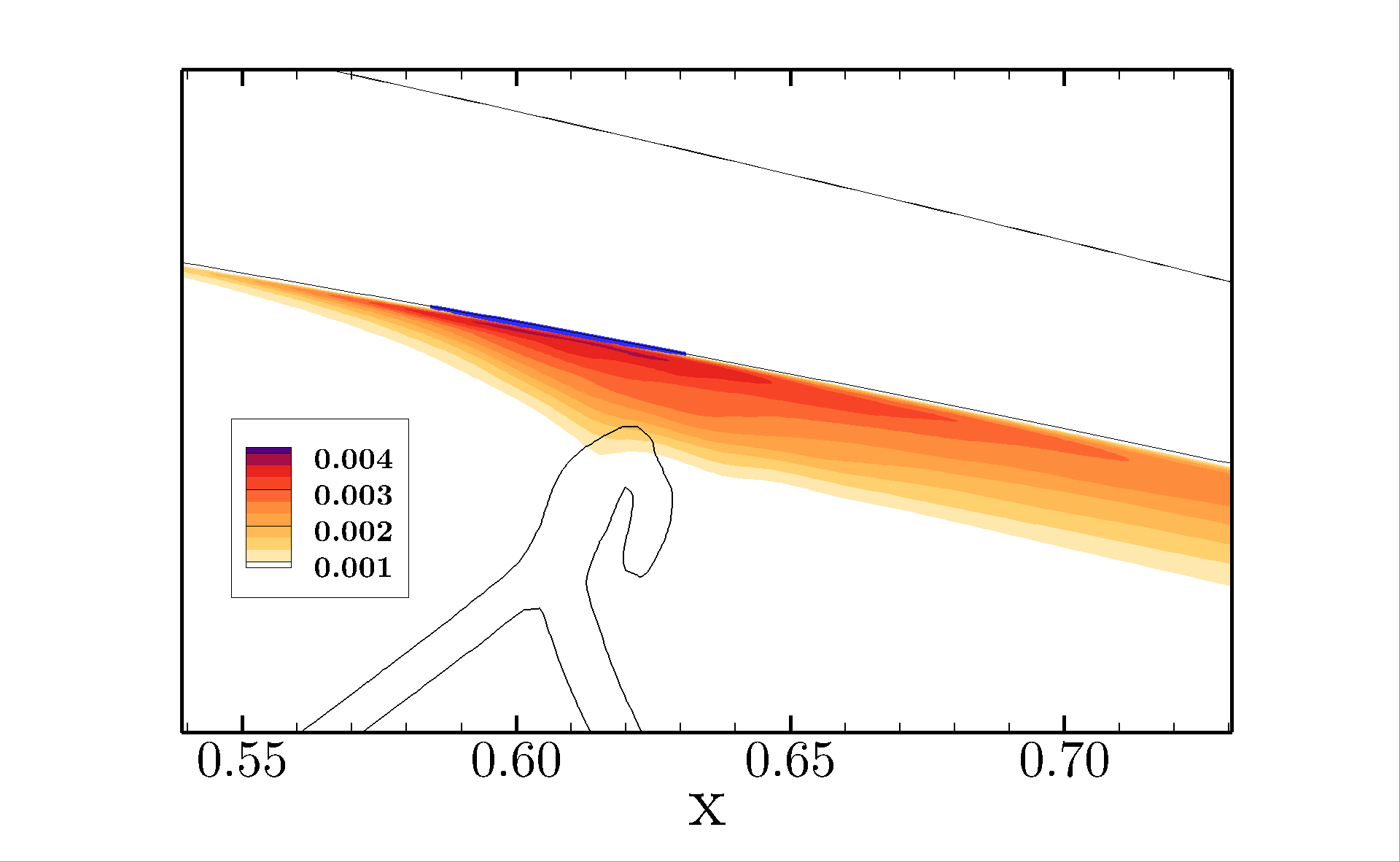}
		\put(12,57){(c)}
	\end{overpic}
	\begin{overpic}[trim = 50mm 2mm 60mm 10mm, clip,width=.48\linewidth]{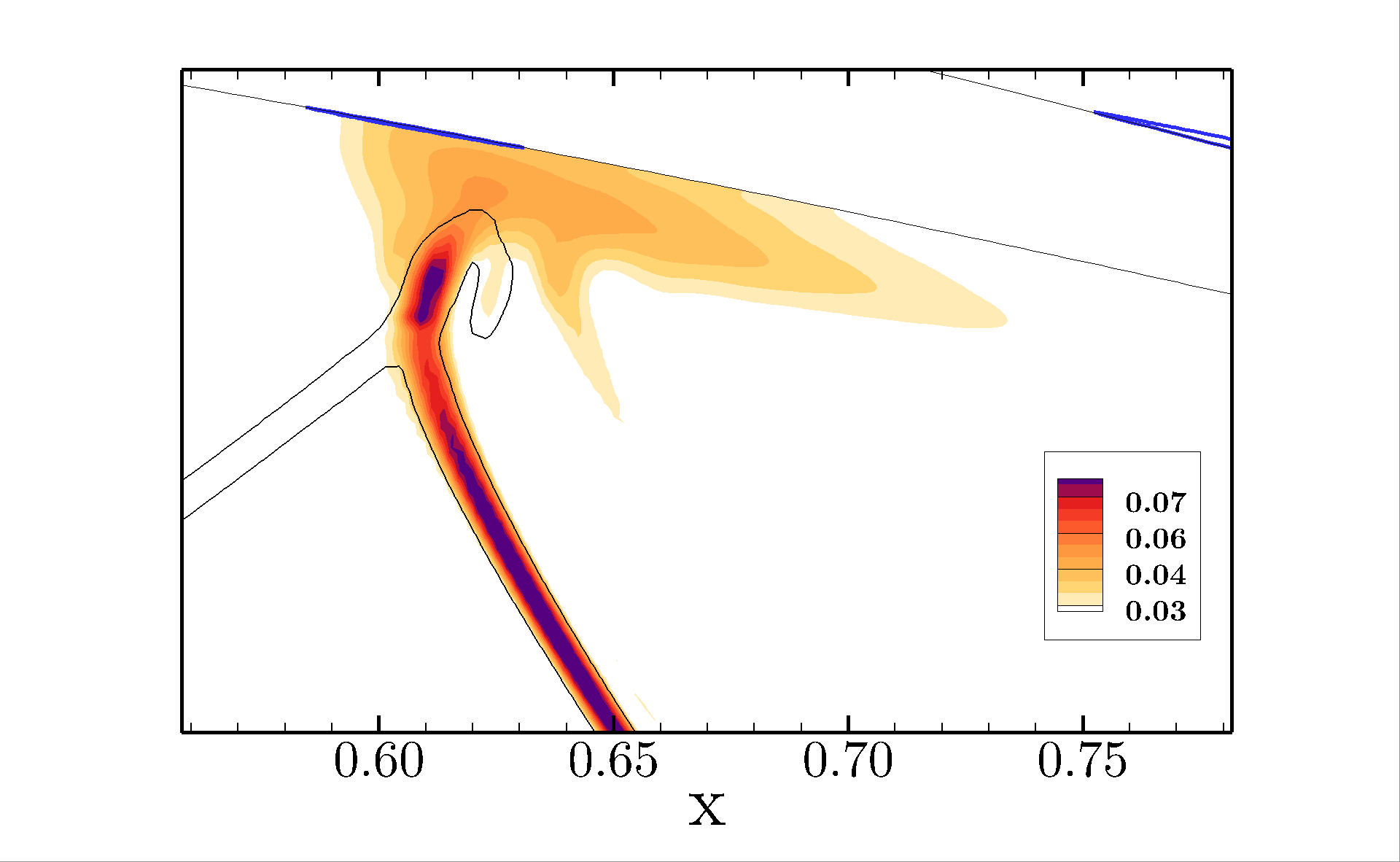}
		\put(12,15){(d)}
	\end{overpic}
	\caption{Spanwise and time averaged turbulence quantities on the pressure side with (a) $\langle u_t u_t \rangle$, (b) $\langle u_n u_n \rangle$, (c) turbulent kinetic energy, and (d) pressure RMS.}
	\label{fig:reynolds_stresses_PS_2D}
\end{figure}

Figure \ref{fig:reynolds_stresses_PS_2D} exhibits the Reynolds stress components, the TKE and the pressure RMS contours on the pressure side. Similarly to the suction side, Fig. \ref{fig:reynolds_stresses_PS_2D}(a) shows that high intensity levels of $\langle u_t u_t \rangle$ are present along the shear layer over the bubble and slightly upstream, possibly being connected to a local flow deceleration as suggested by \citet{fang2020}. A region of high $\langle u_t u_t \rangle$ values is also found downstream of the bubble, near the wall. 
Contrary to the suction side, no sign of the bifurcation is evident in the $\langle u_t u_t \rangle$ distribution after the bubble.
In Fig. \ref{fig:reynolds_stresses_PS_2D}(b), one can observe intense values of $\langle u_n u_n \rangle$ along the free shear layer downstream the shock, indicating the presence of strong wall-normal velocity fluctuations on that region. The turbulent kinetic energy combines the trends observed from the $\langle u_t u_t \rangle$ and $\langle u_n u_n \rangle$ components, as can be visualized in Fig. \ref{fig:reynolds_stresses_PS_2D}(c). Finally, the pressure RMS plot shows strong fluctuations along the reflected shock, revealing its unsteadiness.

\begin{figure}
	\begin{overpic}[trim = 1mm 1mm 1mm 1mm, clip,width=.49\linewidth]{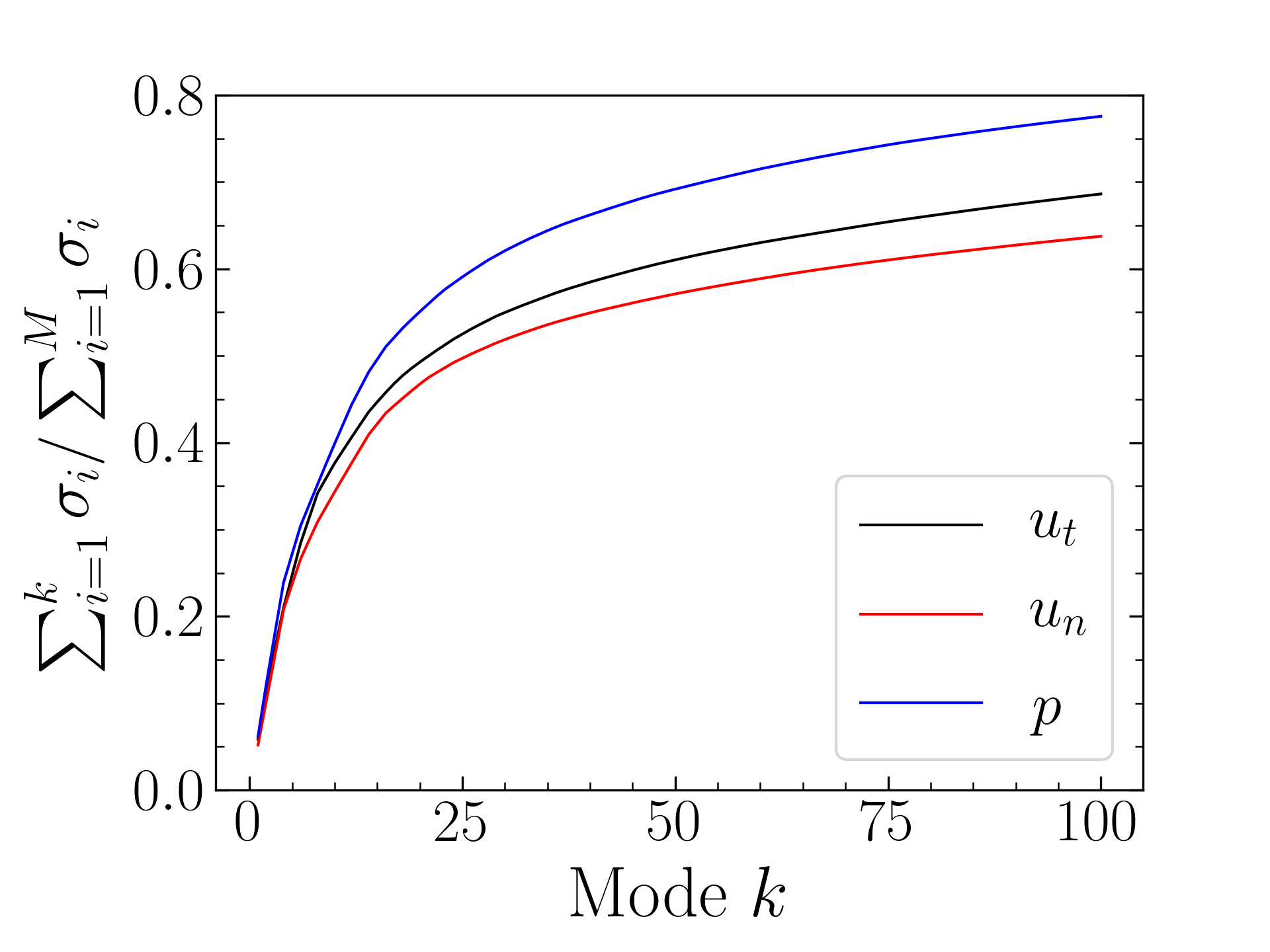}
		\put(1,65){(a)}
	\end{overpic}
	\begin{overpic}[trim = 1mm 1mm 1mm 1mm, clip,width=.49\linewidth]{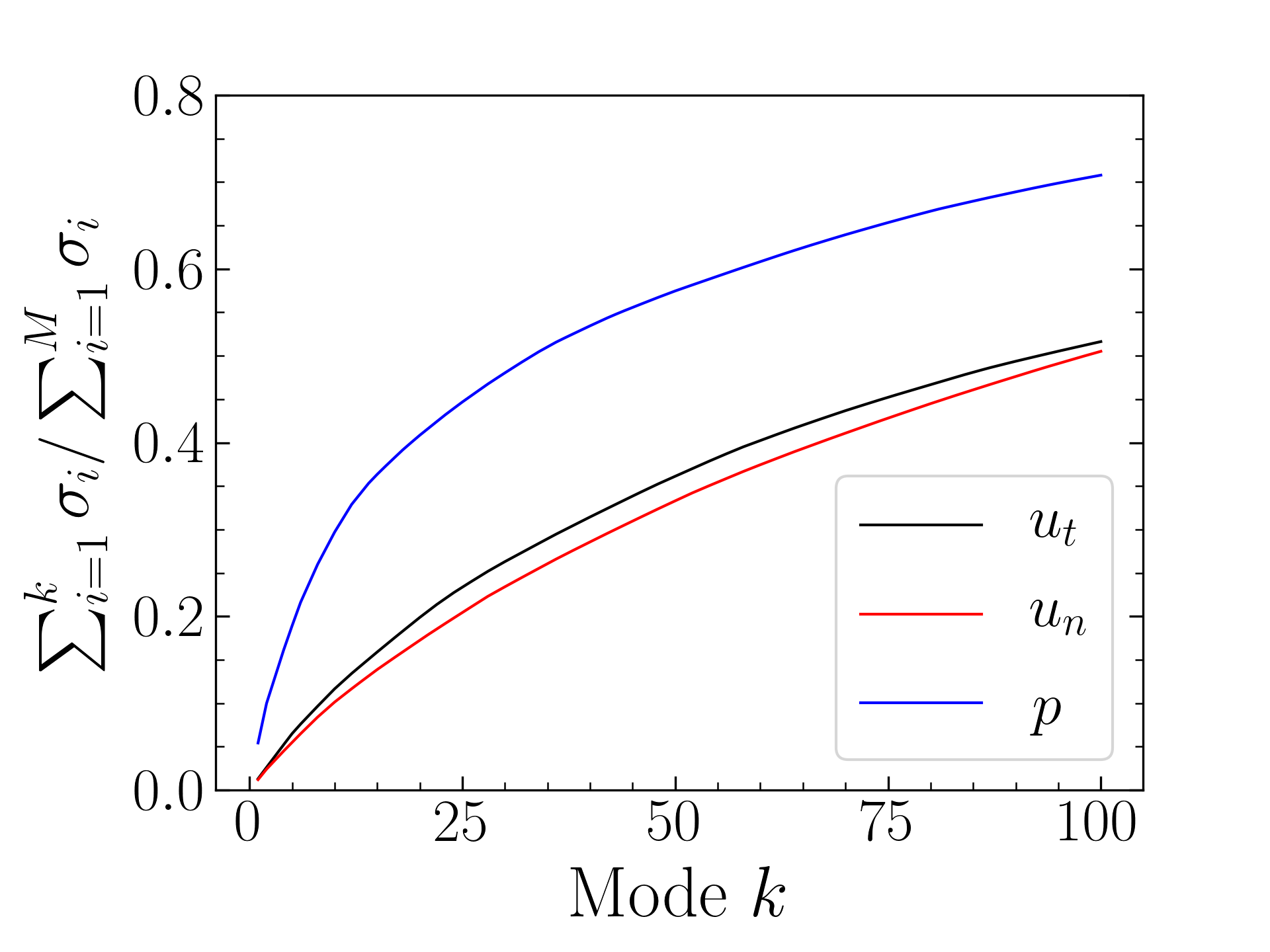}
		\put(1,65){(b)}
	\end{overpic}
	\caption{Cumulative modal energy of the $k$th mode cumulative sum for (a) suction side and (b) pressure side.}
	\label{fig:cumu_energy}
\end{figure}

\begin{figure}
	\centering
	\begin{overpic}[trim = 1mm 40mm 2mm 2mm, clip,width=.24\linewidth]{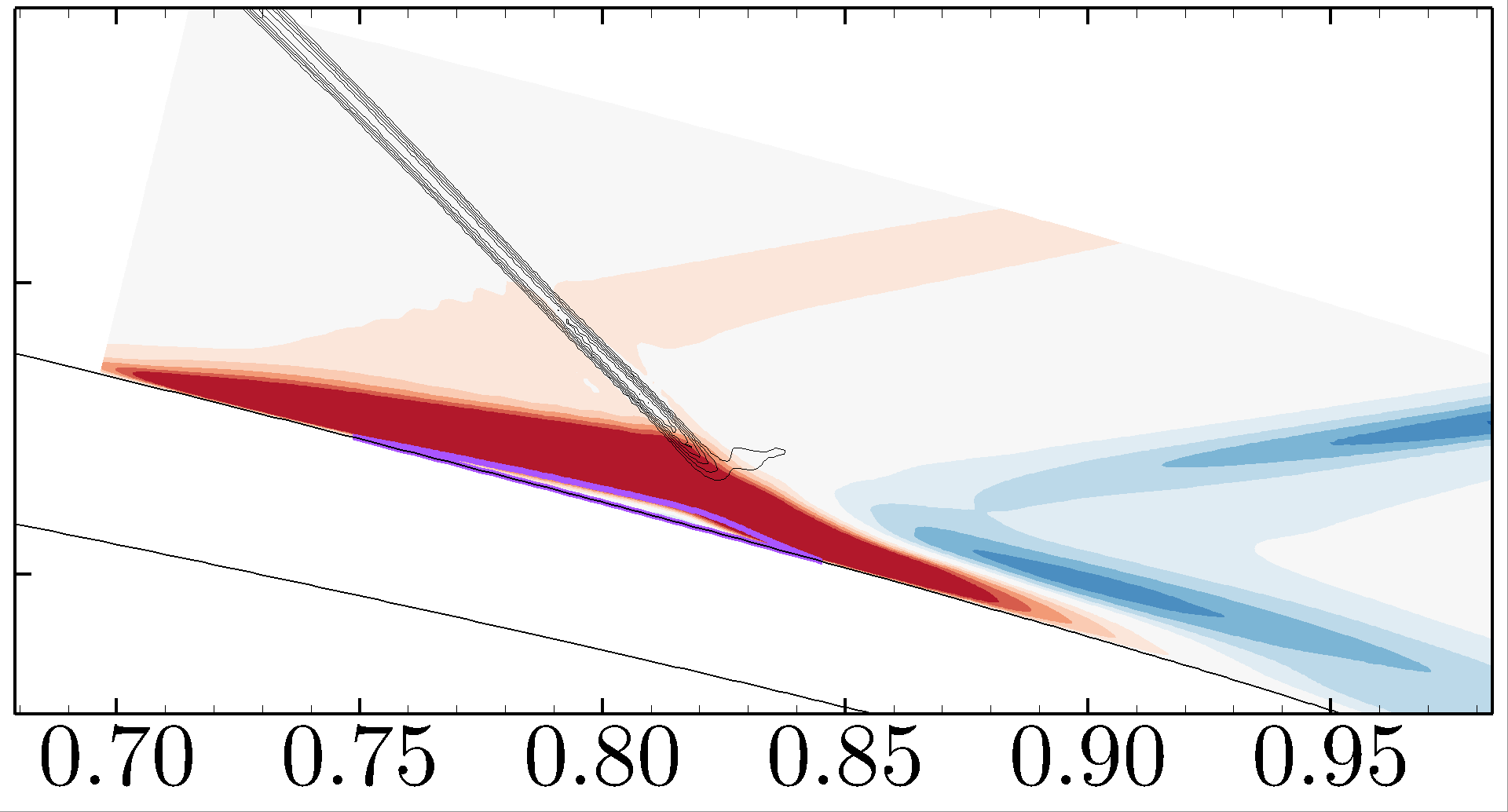}
		\put(85,37){(a)}
	\end{overpic}
	\begin{overpic}[trim = 1mm 40mm 2mm 2mm, clip,width=.24\linewidth]{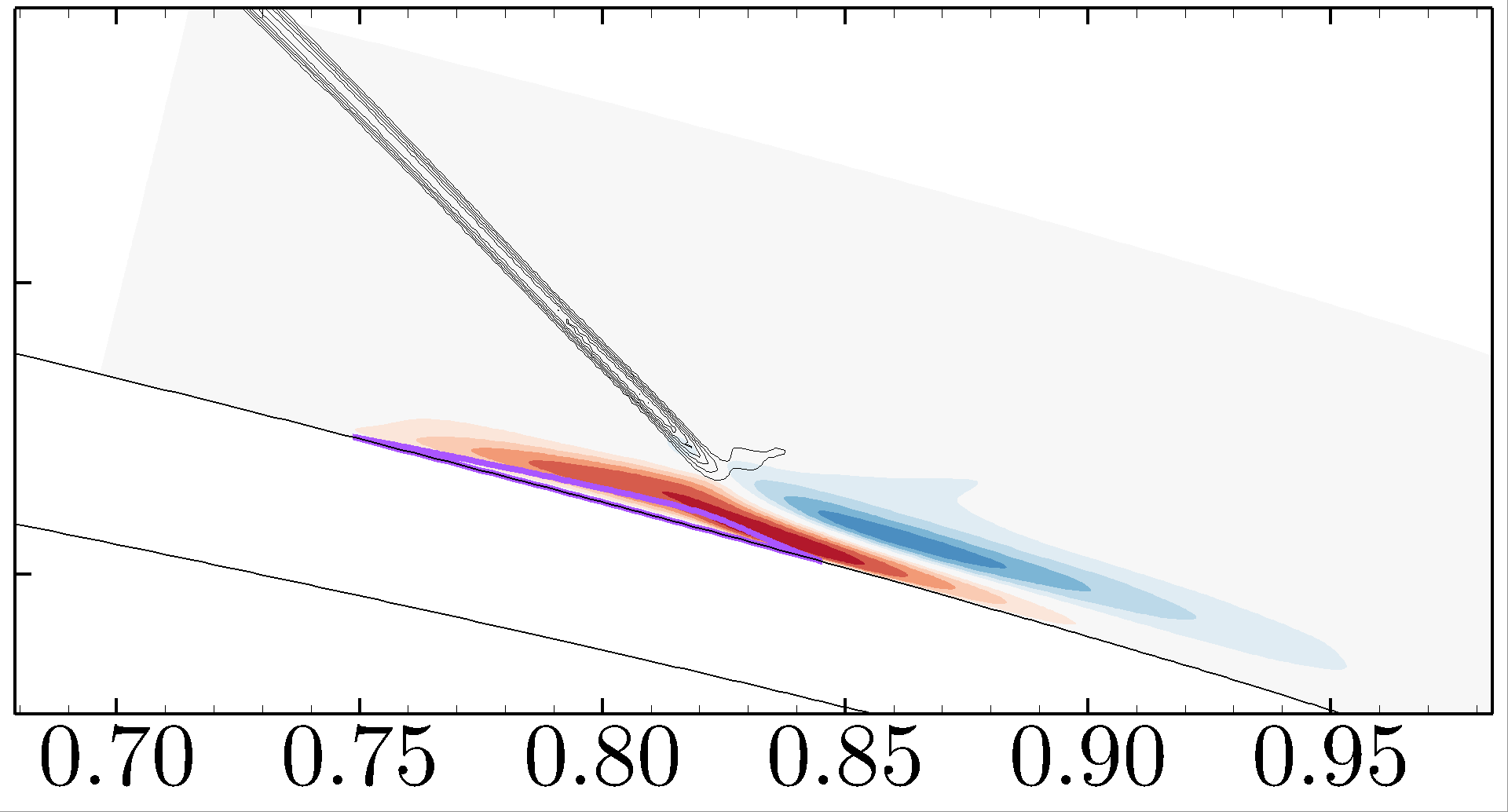}
		\put(85,37){(e)}
	\end{overpic}
	\begin{overpic}[trim = 1mm 40mm 2mm 2mm, clip,width=.24\linewidth]{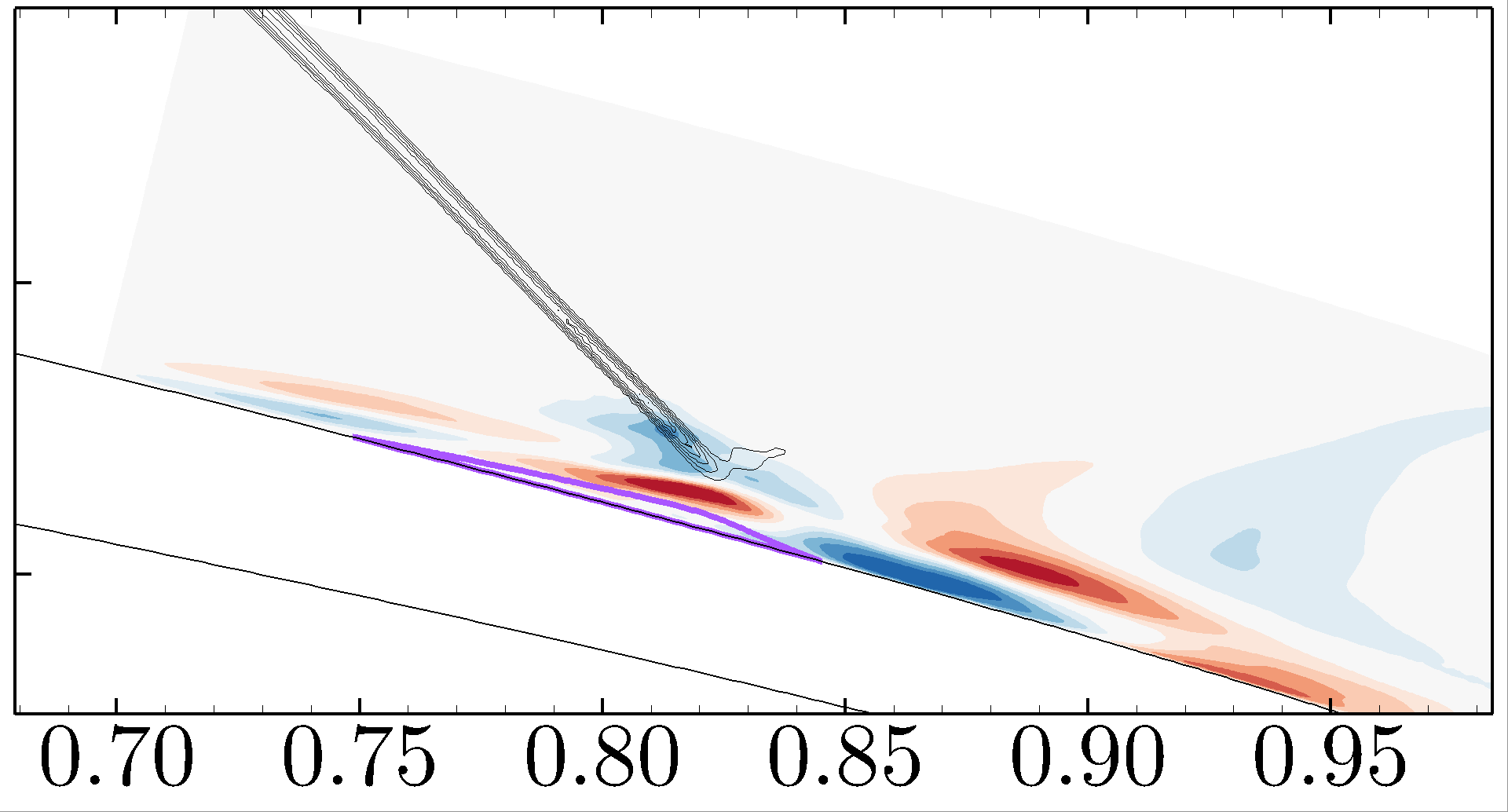}
		\put(85,37){(i)}
	\end{overpic}
	\begin{overpic}[trim = 1mm 40mm 2mm 2mm, clip,width=.24\linewidth]{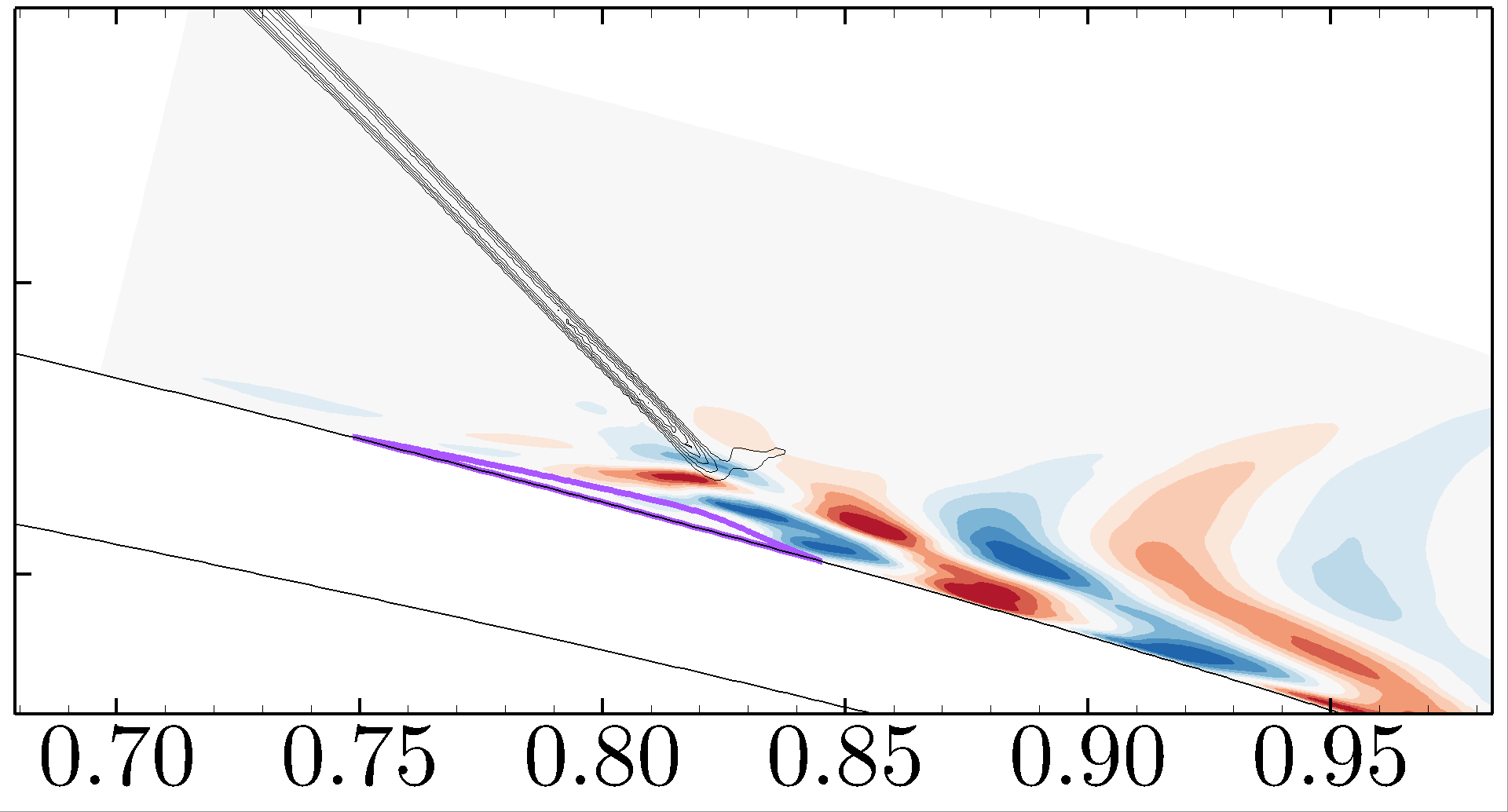}
		\put(83,37){(m)}
	\end{overpic}
	\begin{overpic}[trim = 1mm 40mm 2mm 2mm, clip,width=.24\linewidth]{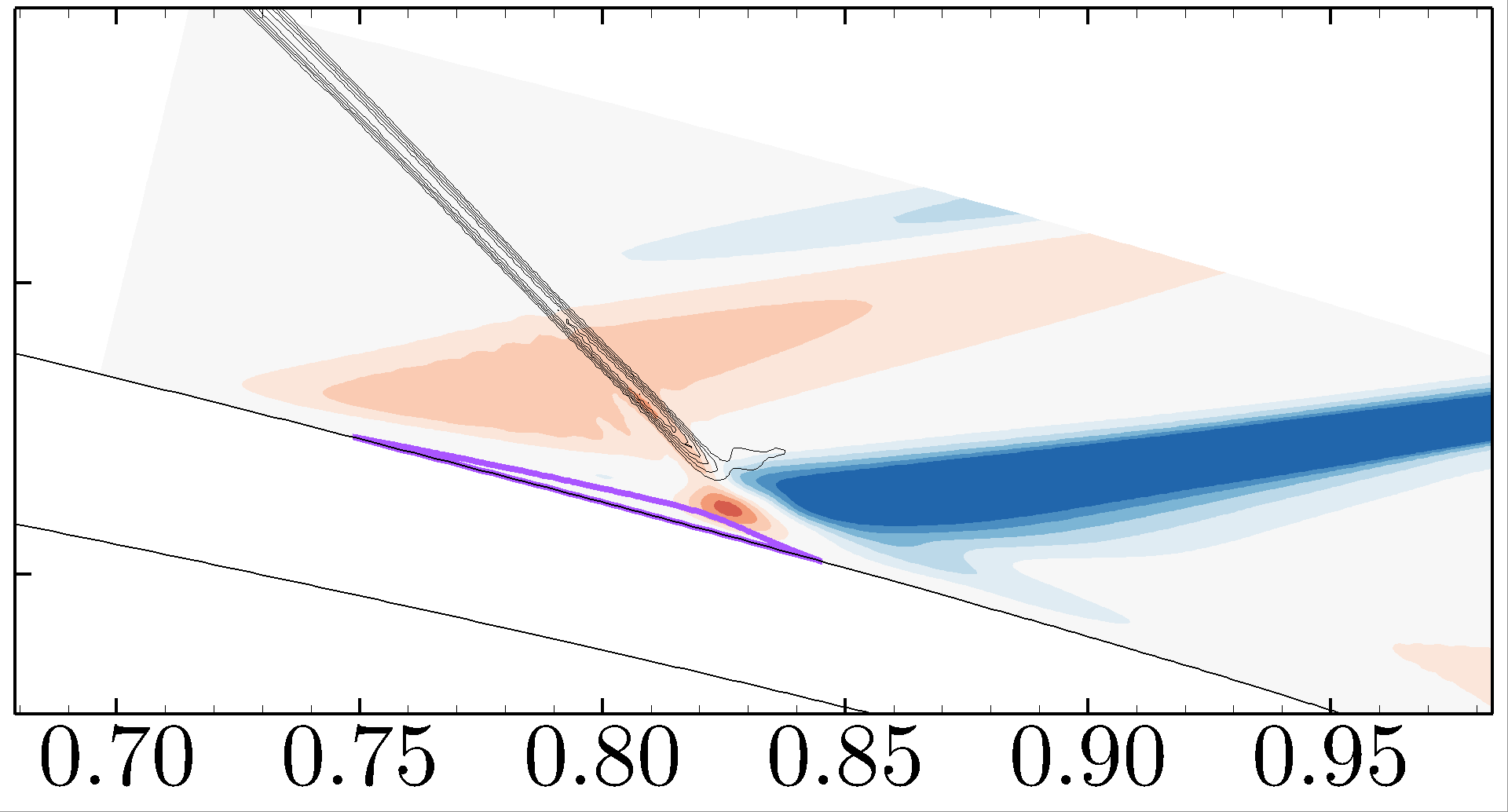}
		\put(85,37){(b)}
	\end{overpic}
	\begin{overpic}[trim = 1mm 40mm 2mm 2mm, clip,width=.24\linewidth]{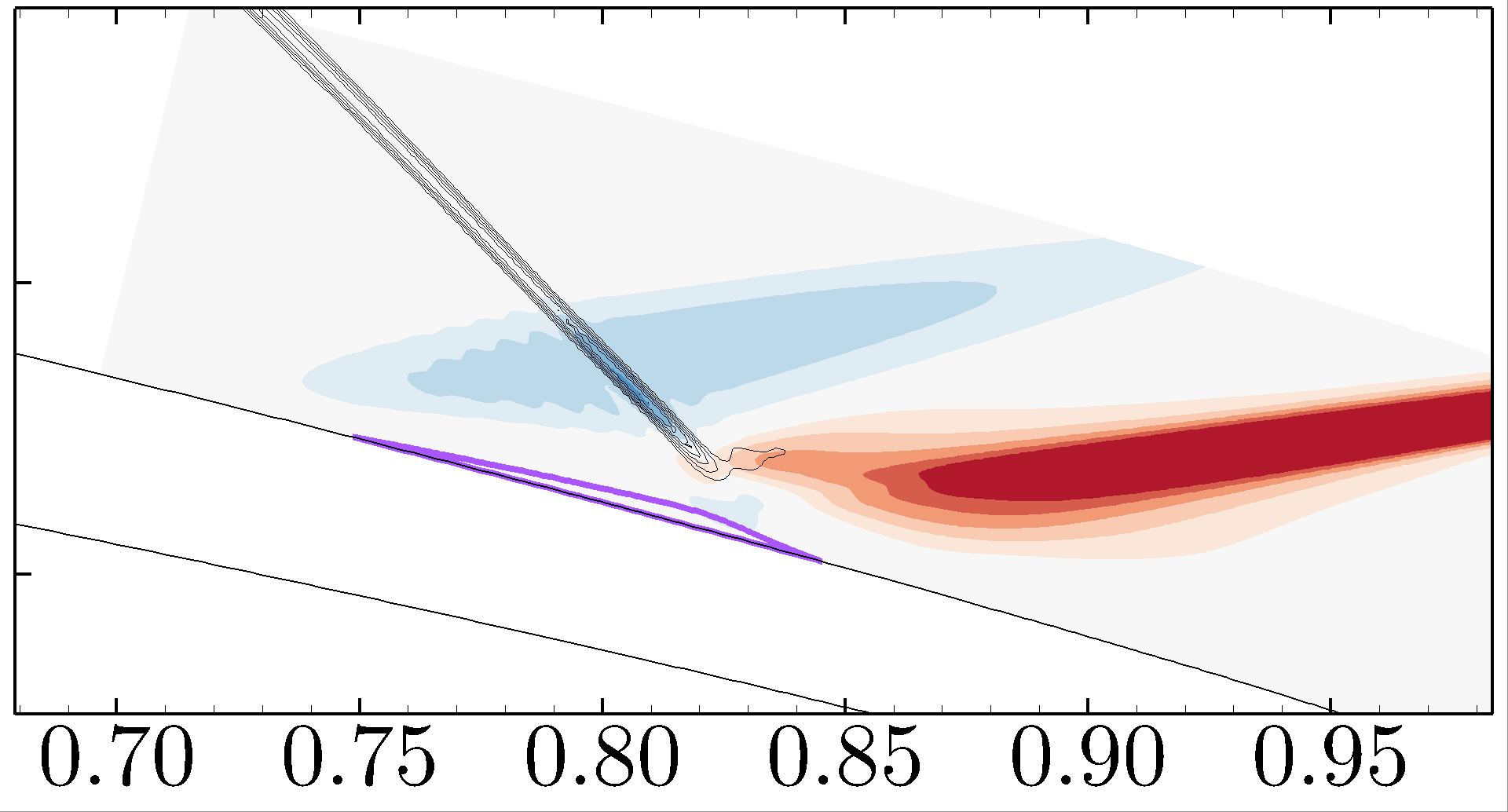}
		\put(85,37){(f)}
	\end{overpic}
	\begin{overpic}[trim = 1mm 40mm 2mm 2mm, clip,width=.24\linewidth]{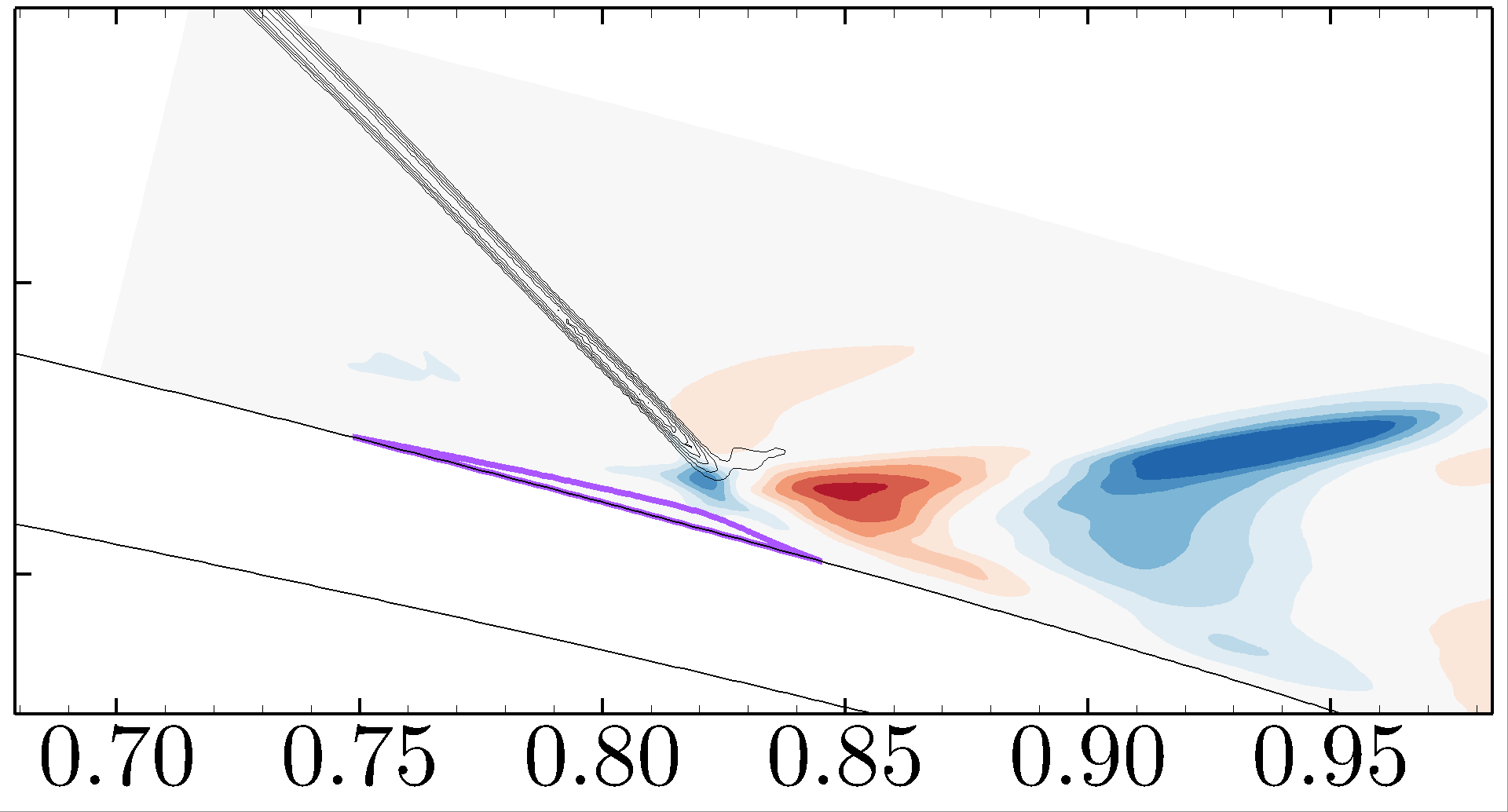}
		\put(85,37){(j)}
	\end{overpic}
	\begin{overpic}[trim = 1mm 40mm 2mm 2mm, clip,width=.24\linewidth]{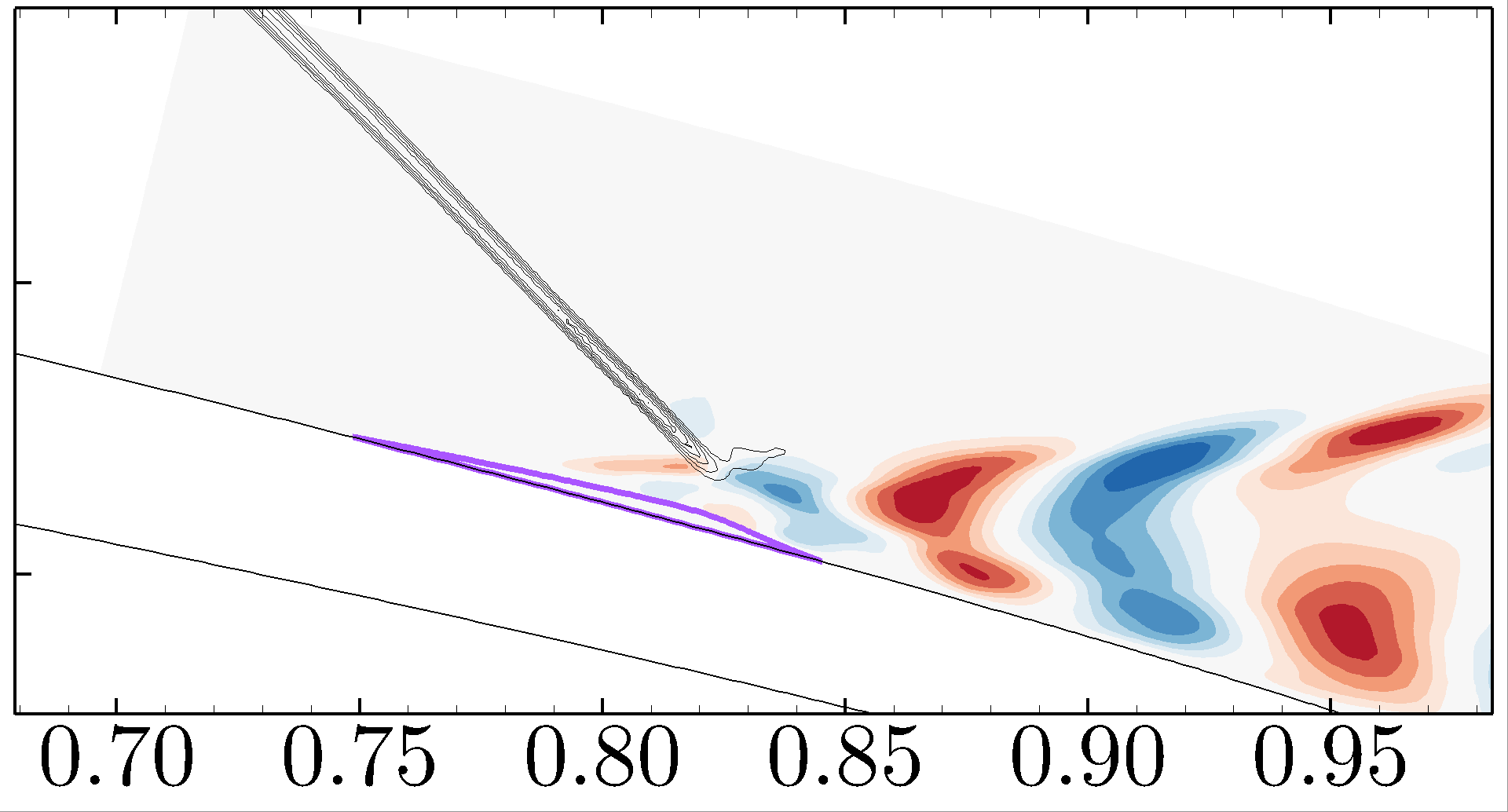}
		\put(85,37){(n)}
    \end{overpic}
    \begin{overpic}[trim = 1mm 2mm 2mm 2mm, clip,width=.24\linewidth]{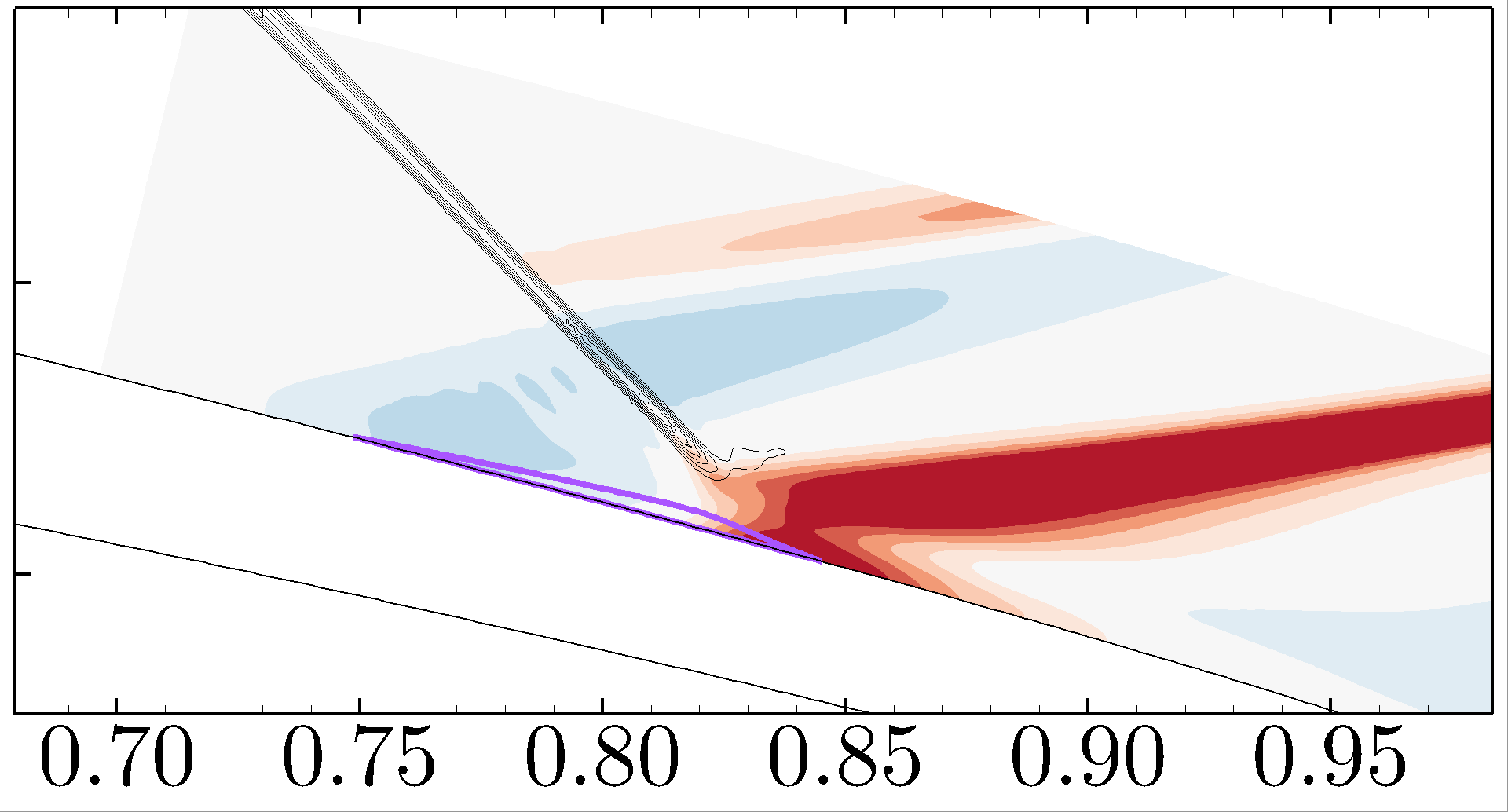}
		\put(85,43){(c)}
	\end{overpic}
	\begin{overpic}[trim = 1mm 2mm 2mm 2mm, clip,width=.24\linewidth]{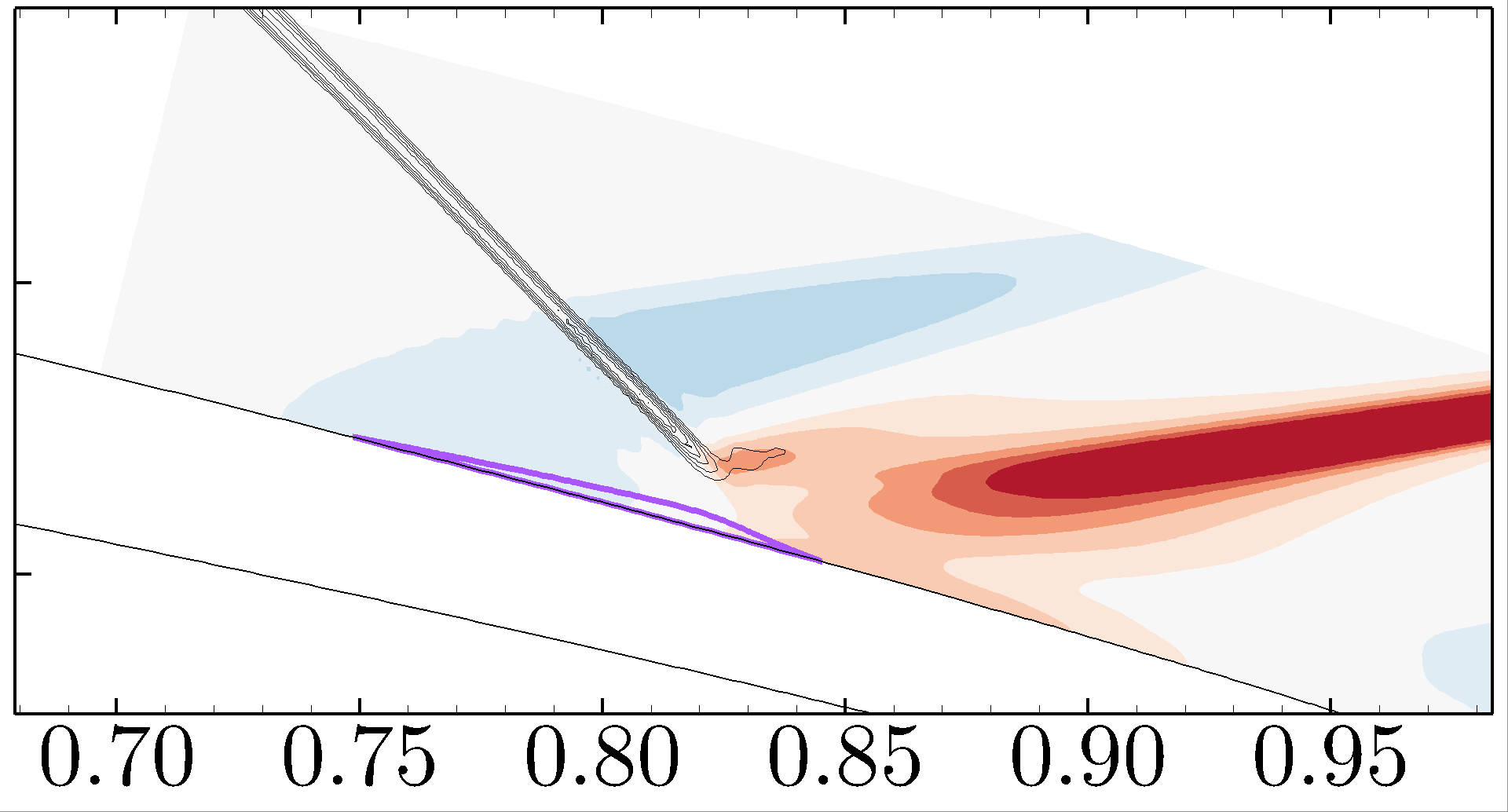}
		\put(85,43){(g)}
	\end{overpic}
	\begin{overpic}[trim = 1mm 2mm 2mm 2mm, clip,width=.24\linewidth]{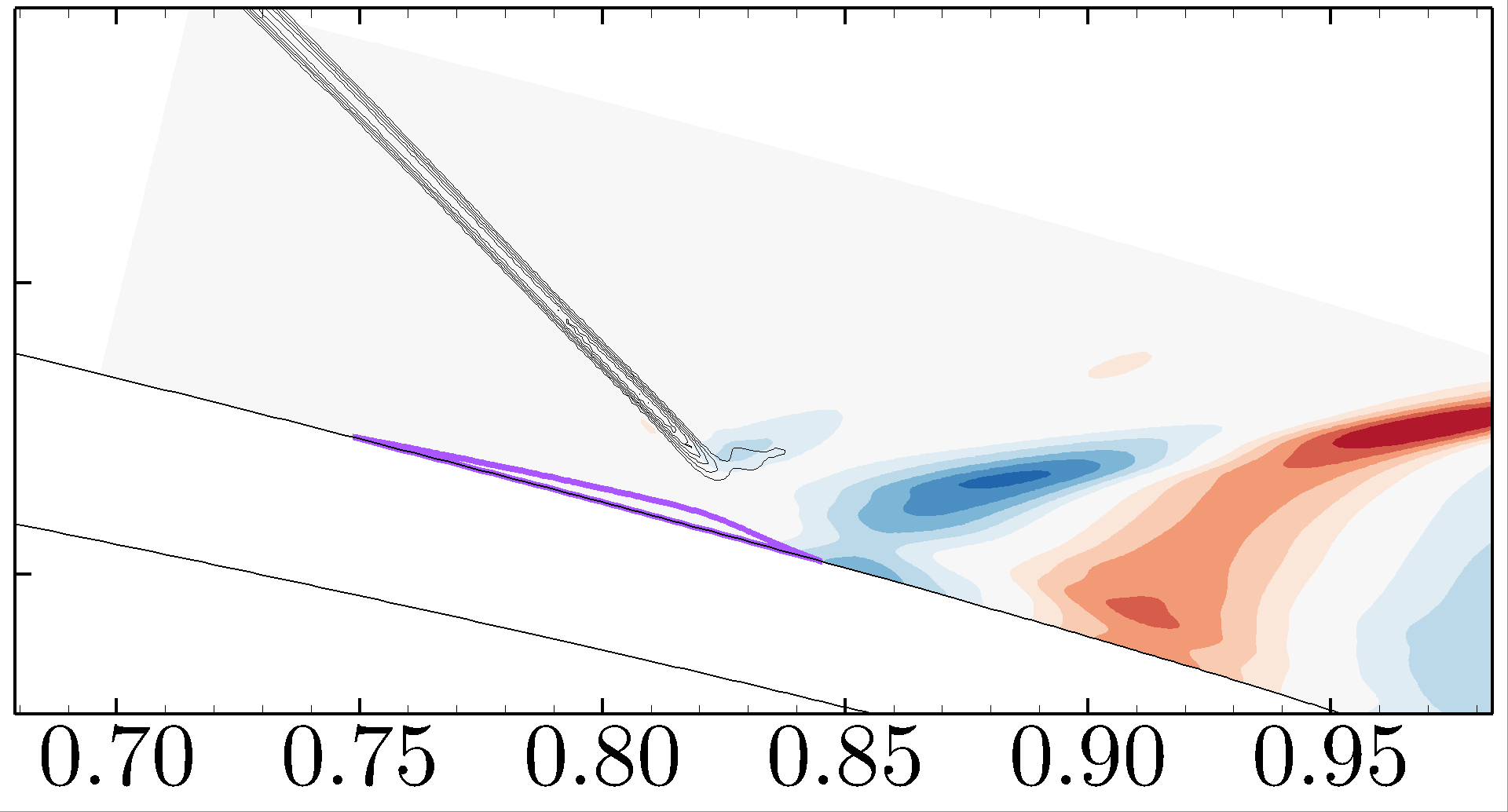}
		\put(85,43){(k)}
	\end{overpic}
	\begin{overpic}[trim = 1mm 2mm 2mm 2mm, clip,width=.24\linewidth]{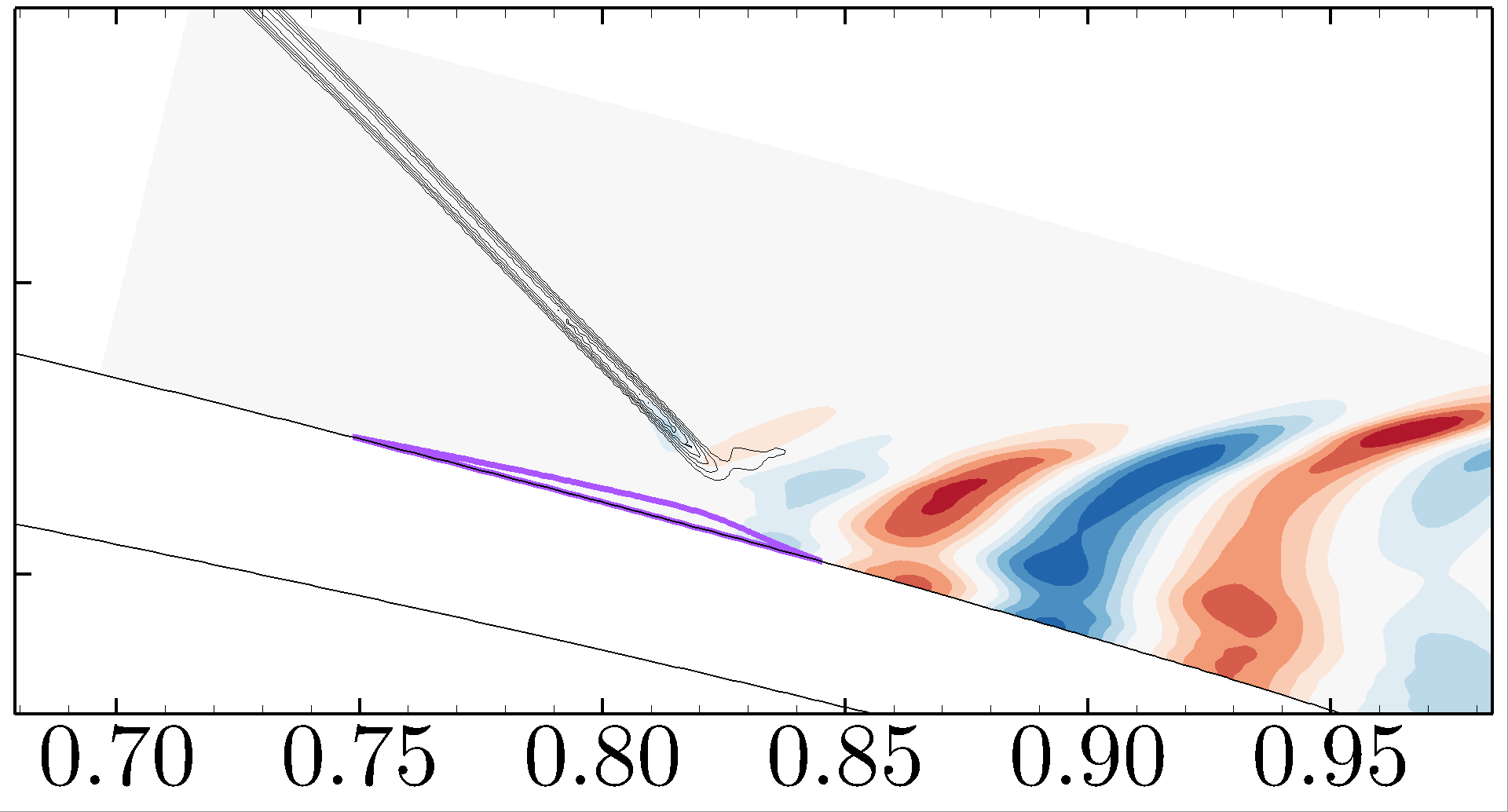}
		\put(85,43){(o)}
    \end{overpic}
    \vskip 0.1cm
     \begin{overpic}[trim = 1mm 1mm 2mm 2mm, clip,width=.24\linewidth]{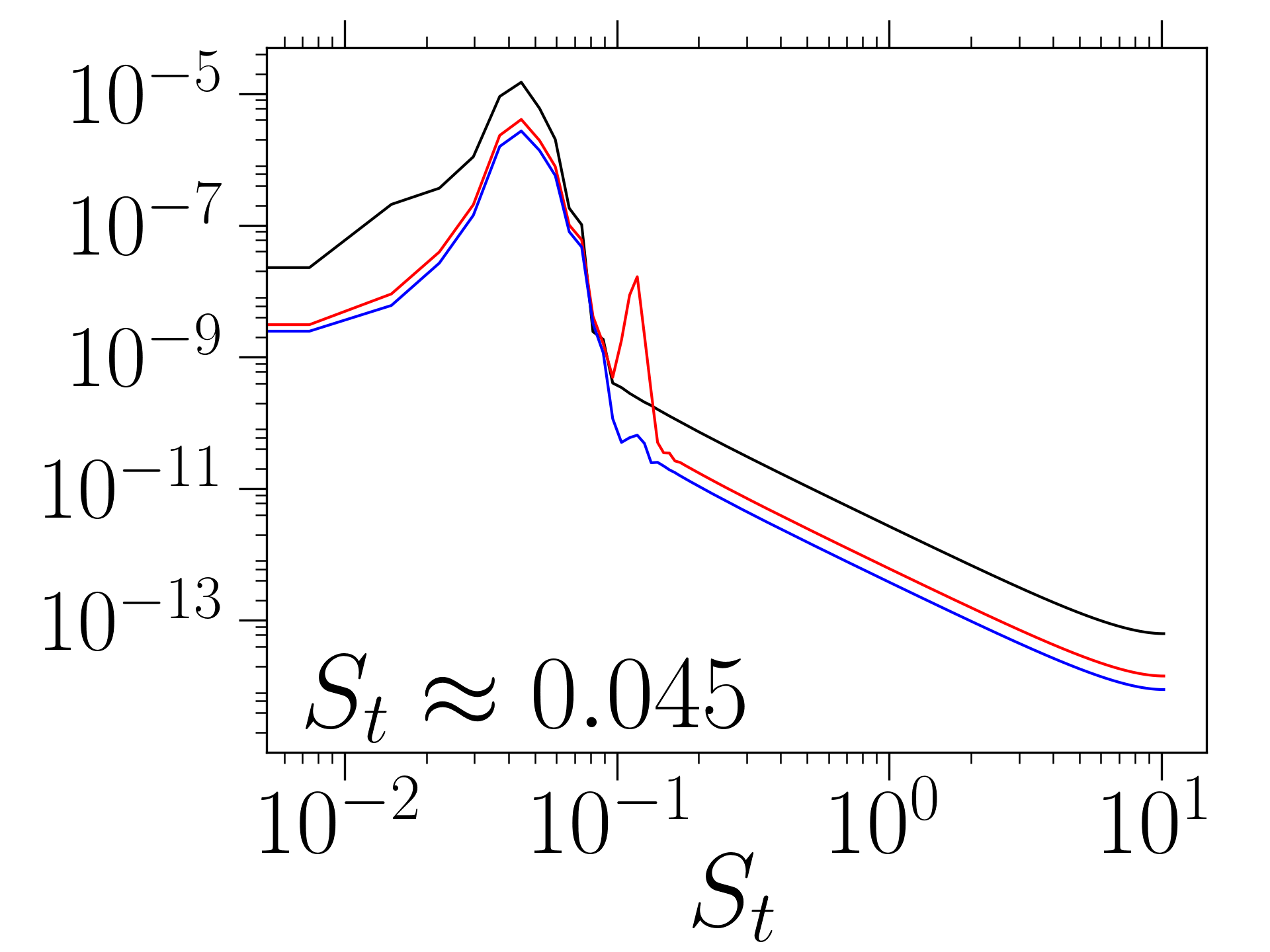}
		\put(82,60){(d)}
	\end{overpic}
	\begin{overpic}[trim = 1mm 1mm 2mm 2mm, clip,width=.24\linewidth]{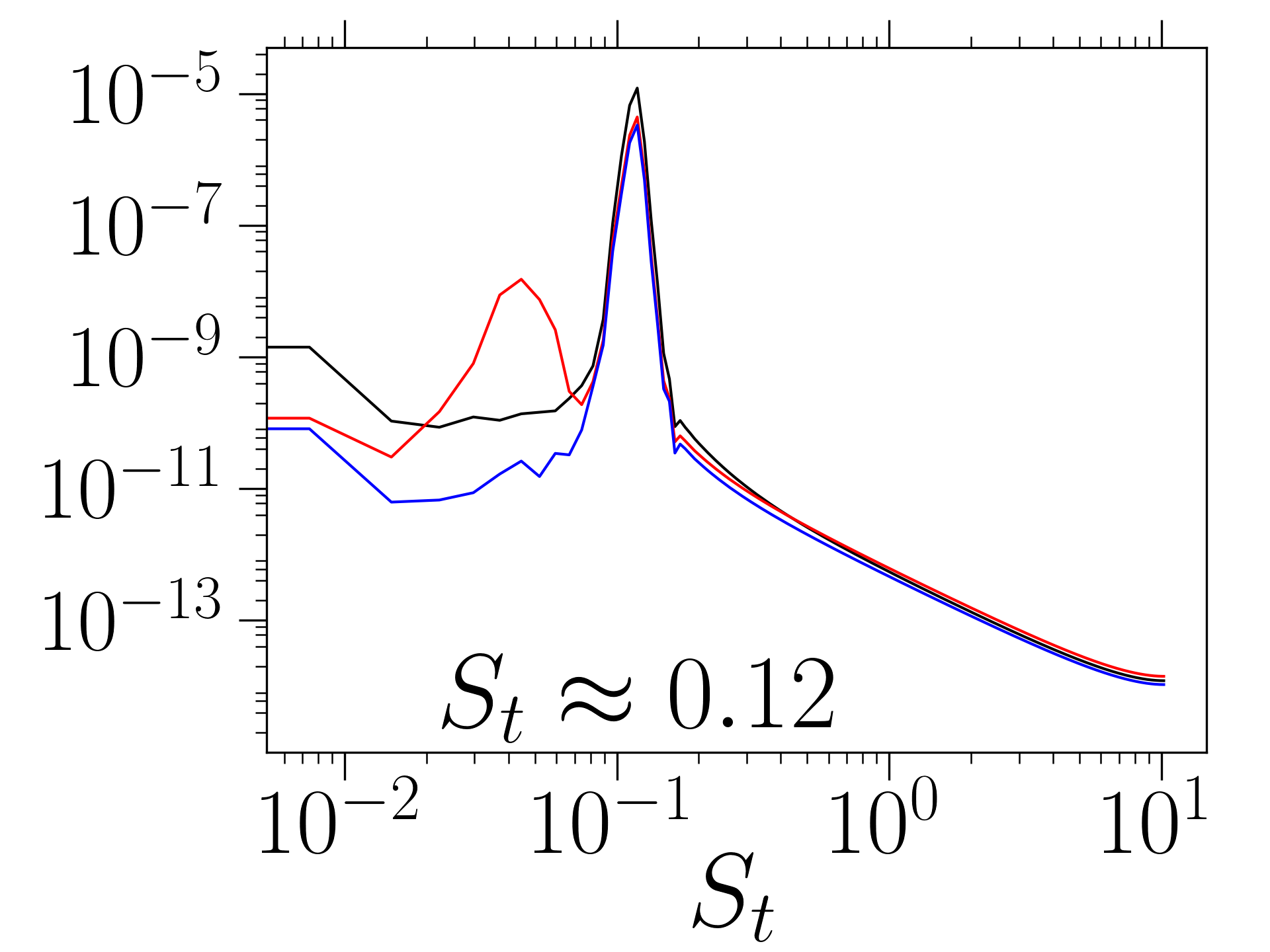}
		\put(82,60){(h)}
	\end{overpic}
	\begin{overpic}[trim = 1mm 1mm 2mm 2mm, clip,width=.24\linewidth]{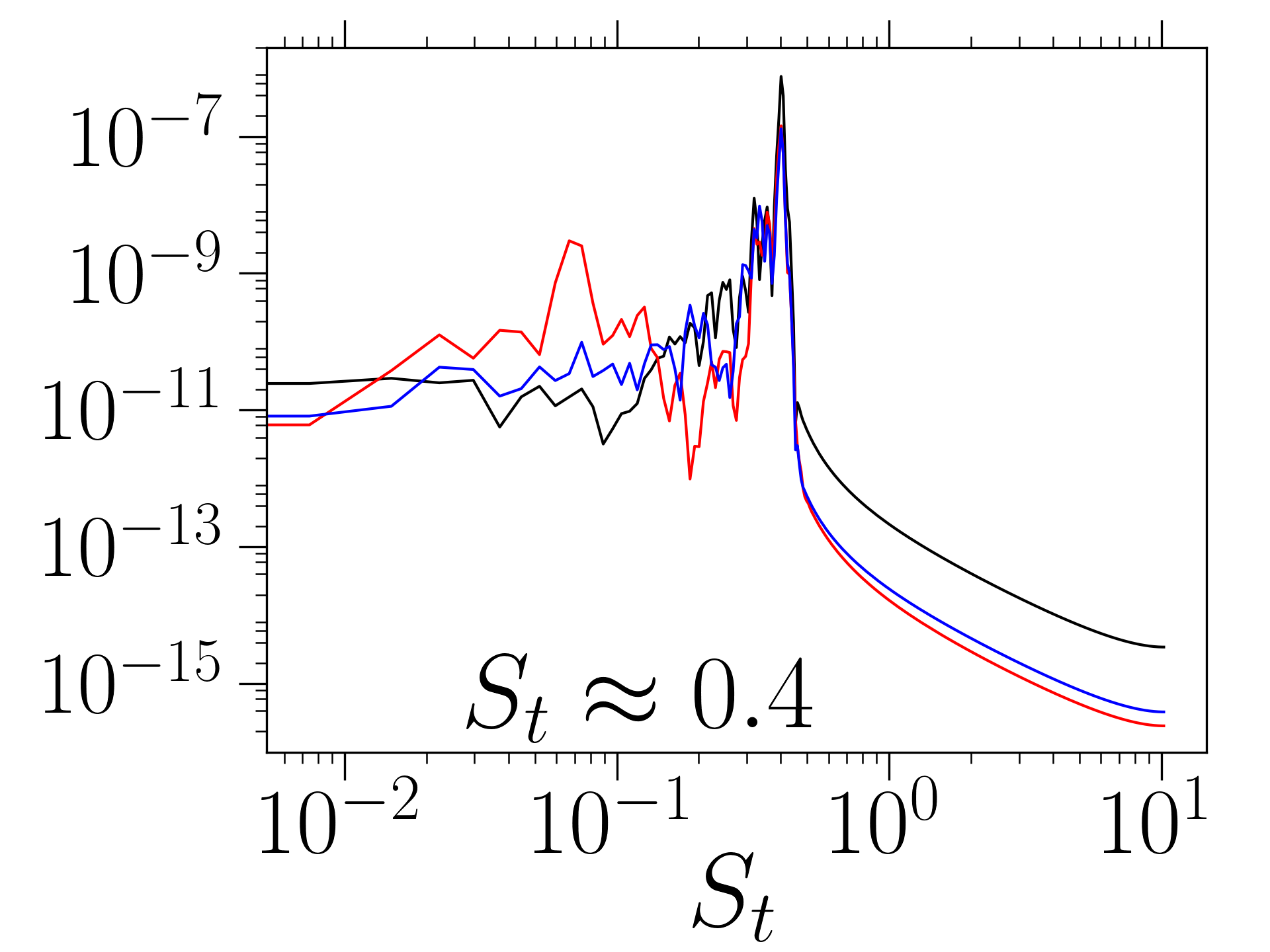}
		\put(82,60){(l)}
	\end{overpic}
	\begin{overpic}[trim = 1mm 1mm 2mm 2mm, clip,width=.24\linewidth]{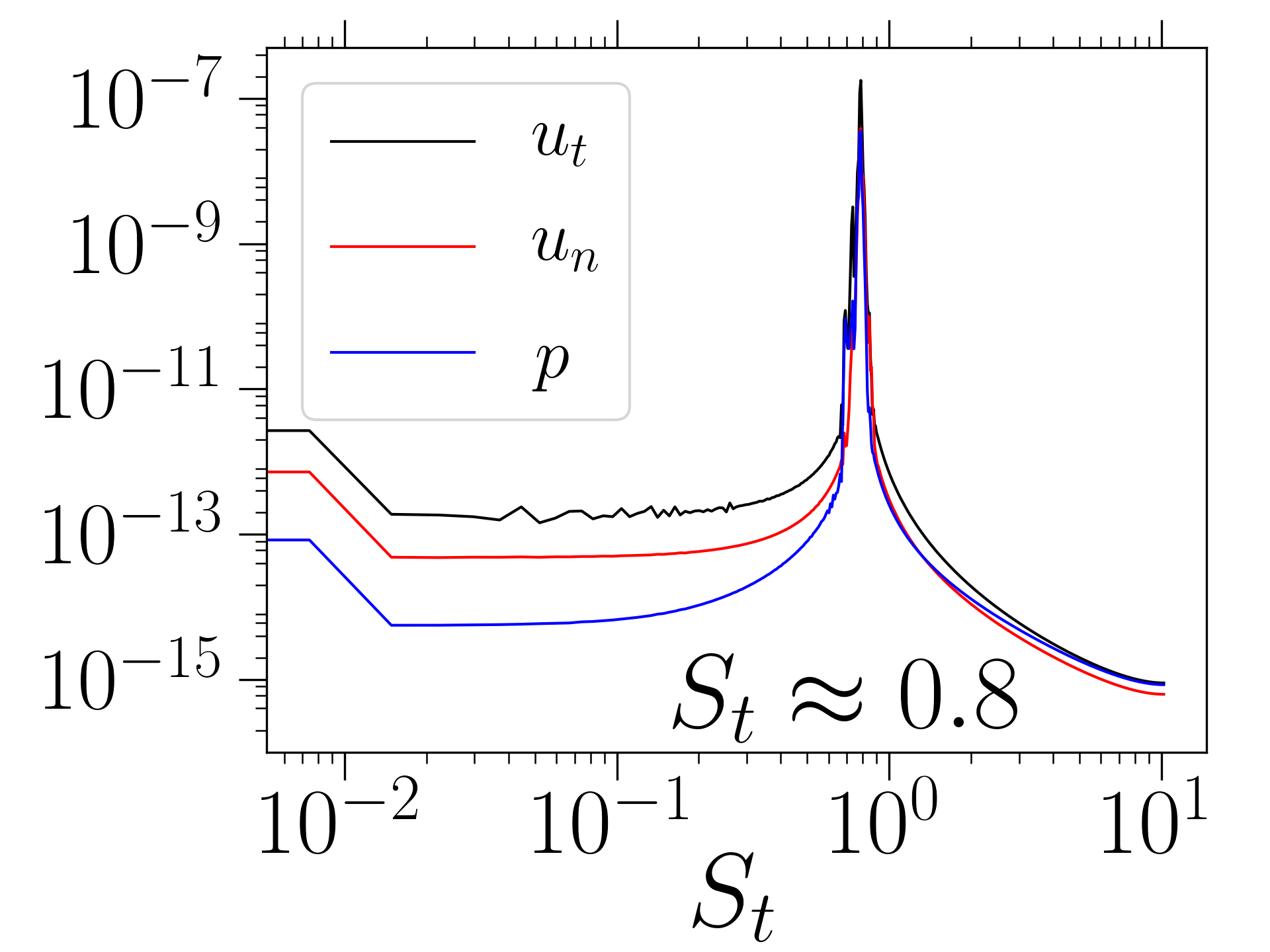}
		\put(82,60){(p)}
    \end{overpic}
	
	\caption{Filtered POD modes selected based on the frequency content along the suction side. The first, second and third rows show the spatial modes of $u_{t}$, $u_{n}$, and $p$, respectively. The fourth row presents the PSD of the temporal modes.}
	\label{fig:pod_spatial_modes_SS}
\end{figure}

\begin{figure}
	\centering
	\begin{overpic}[trim = 1mm 38mm 2mm 2mm, clip,width=.32\linewidth]{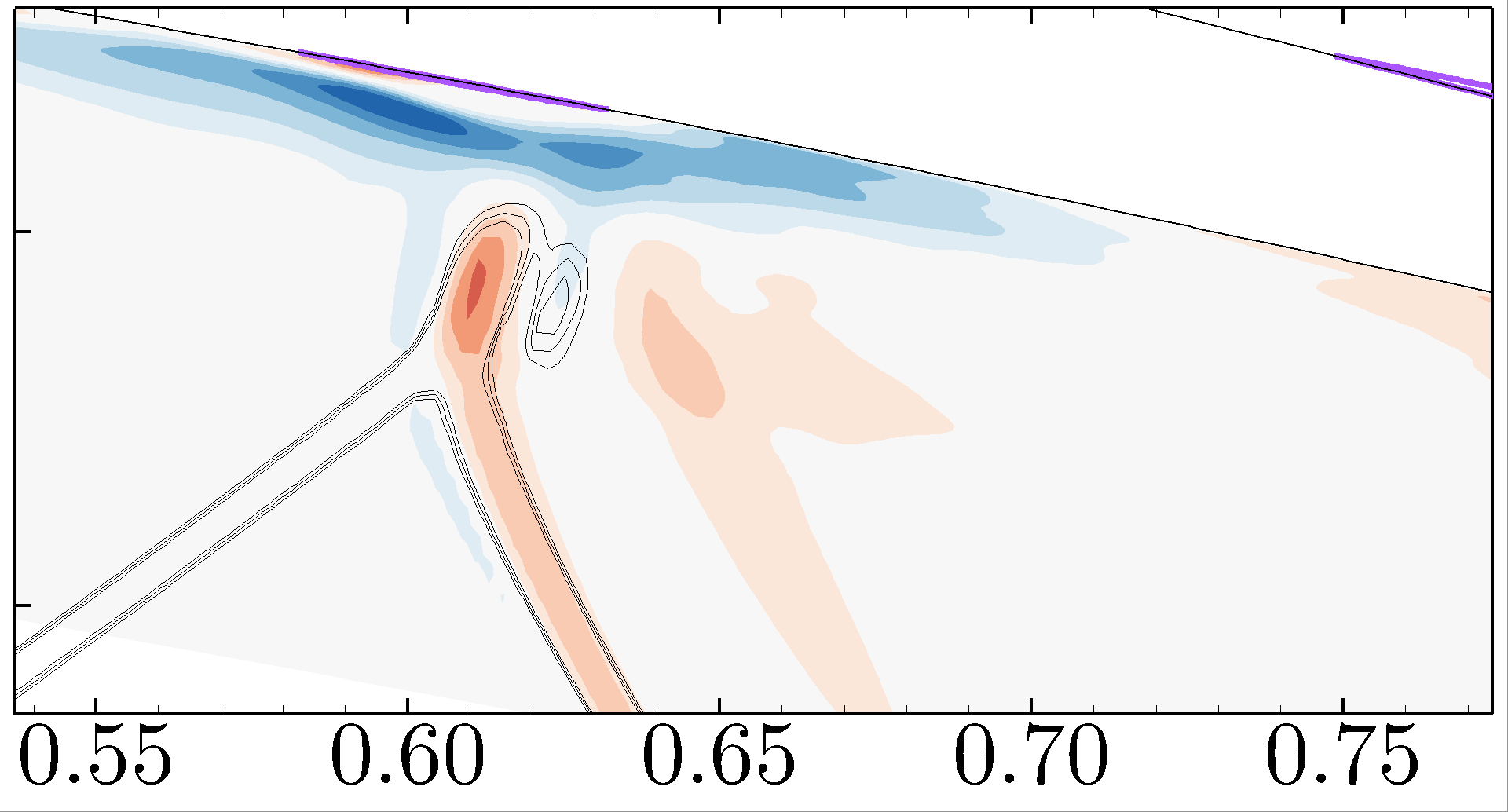}
		\put(84,35){(a)}
	\end{overpic}
	 \begin{overpic}[trim = 1mm 38mm 2mm 2mm, clip,width=.32\linewidth]{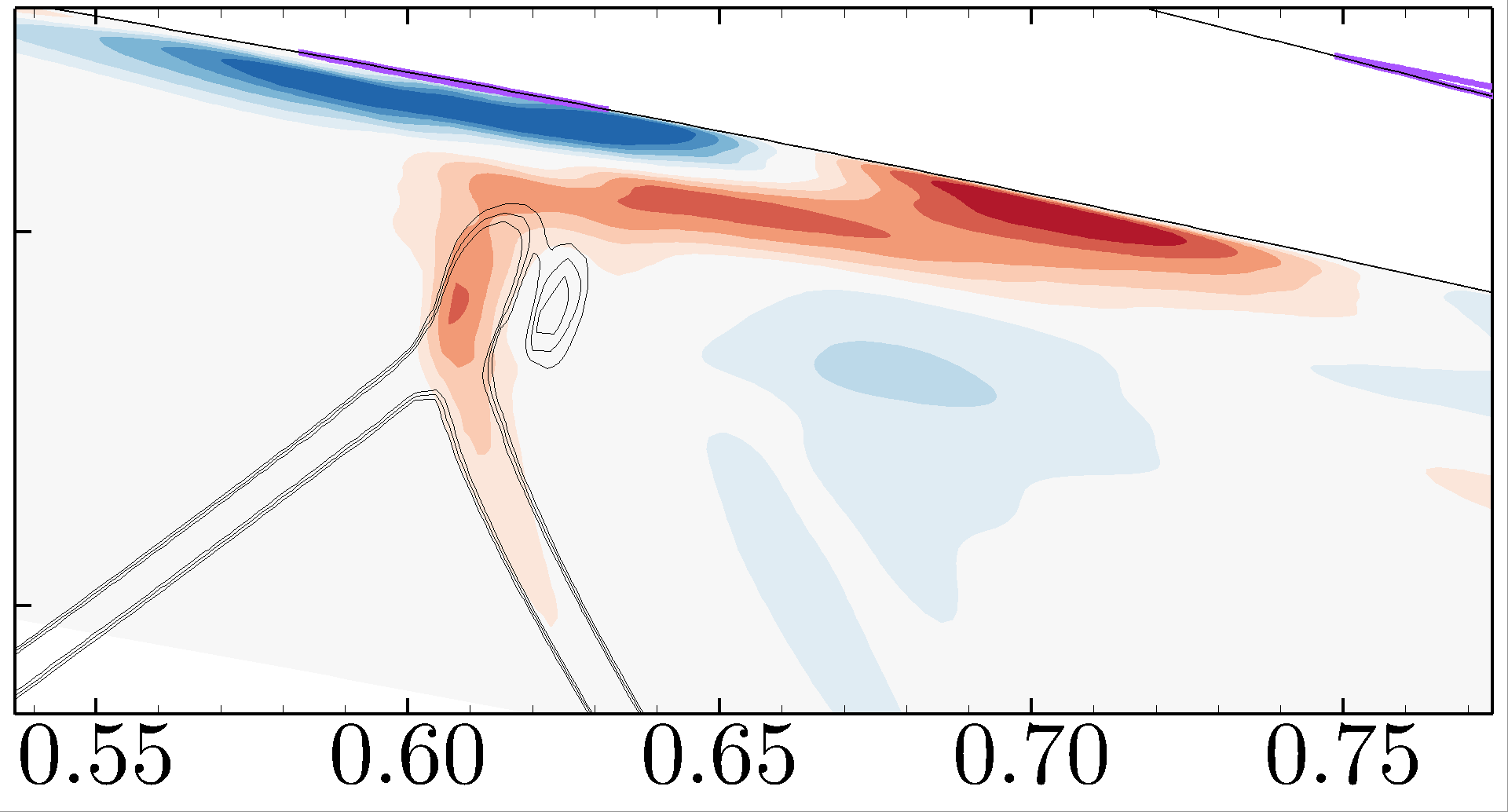}
 		\put(84,35){(e)}
 	\end{overpic}
 	\begin{overpic}[trim = 1mm 38mm 2mm 2mm, clip,width=.32\linewidth]{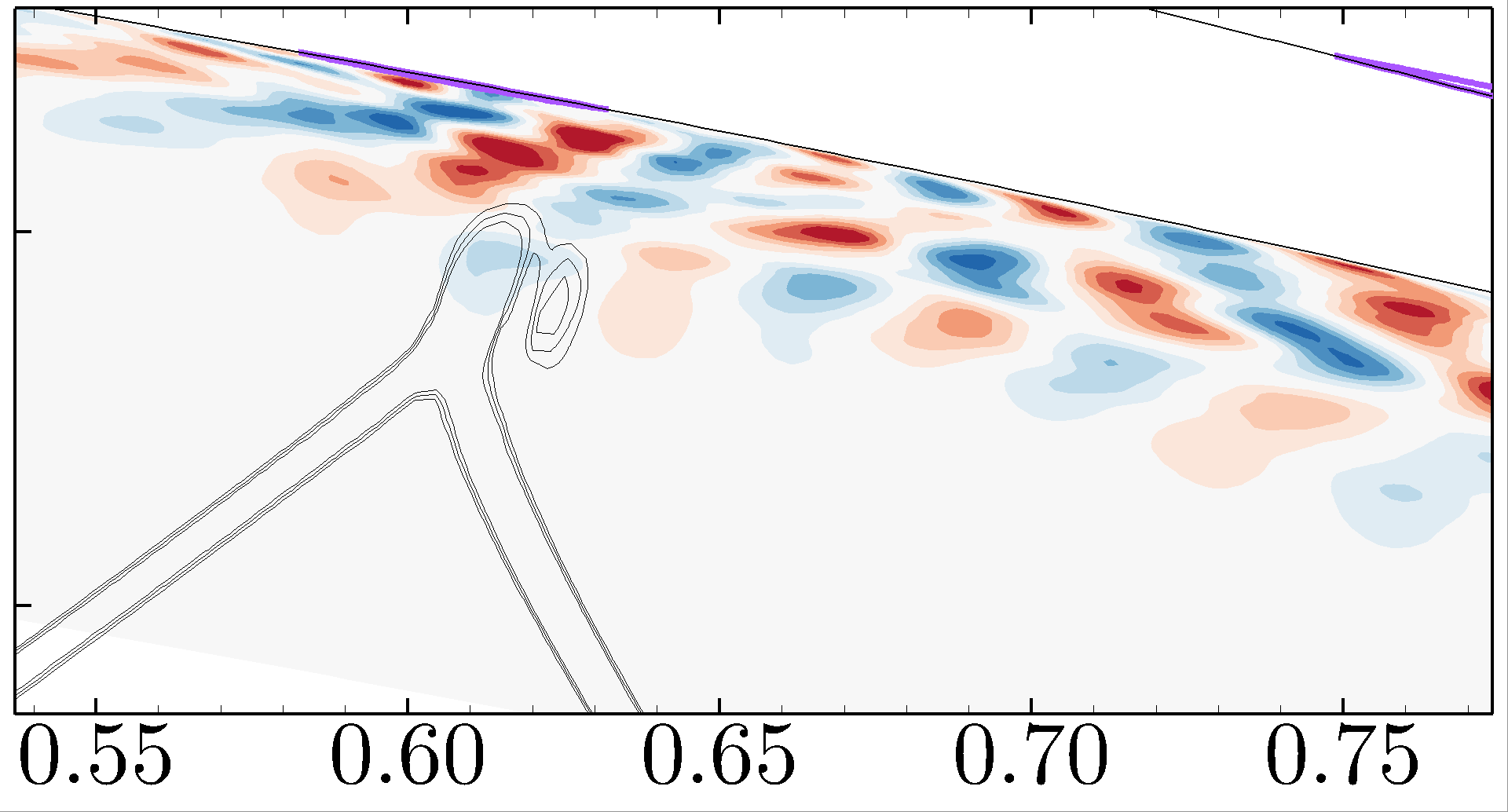}
 		\put(84,35){(i)}
 	\end{overpic}
 	\begin{overpic}[trim = 1mm 40mm 2mm 2mm, clip,width=.32\linewidth]{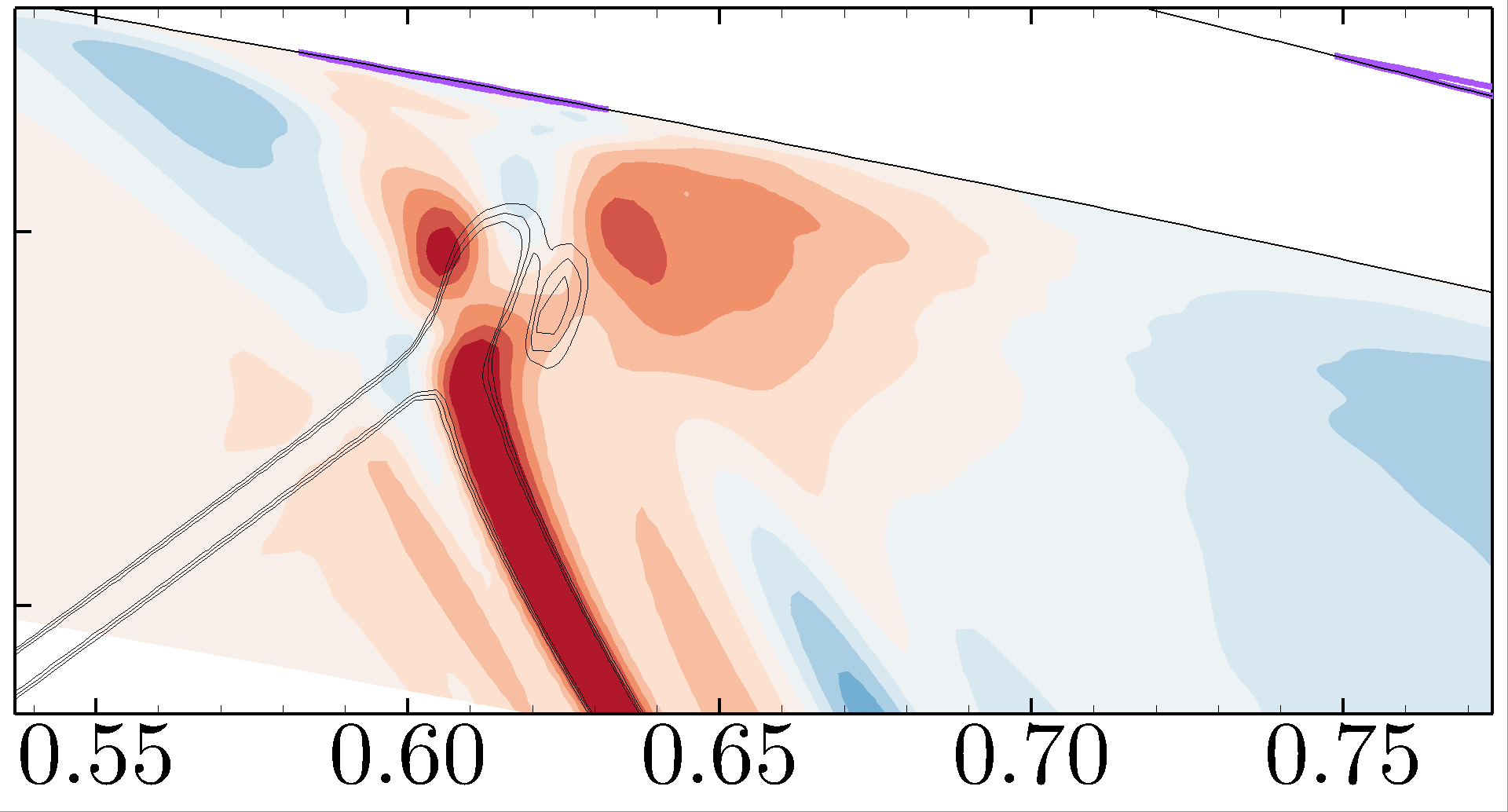}
		\put(84,35){(b)}
	\end{overpic}
	 \begin{overpic}[trim = 1mm 40mm 2mm 2mm, clip,width=.32\linewidth]{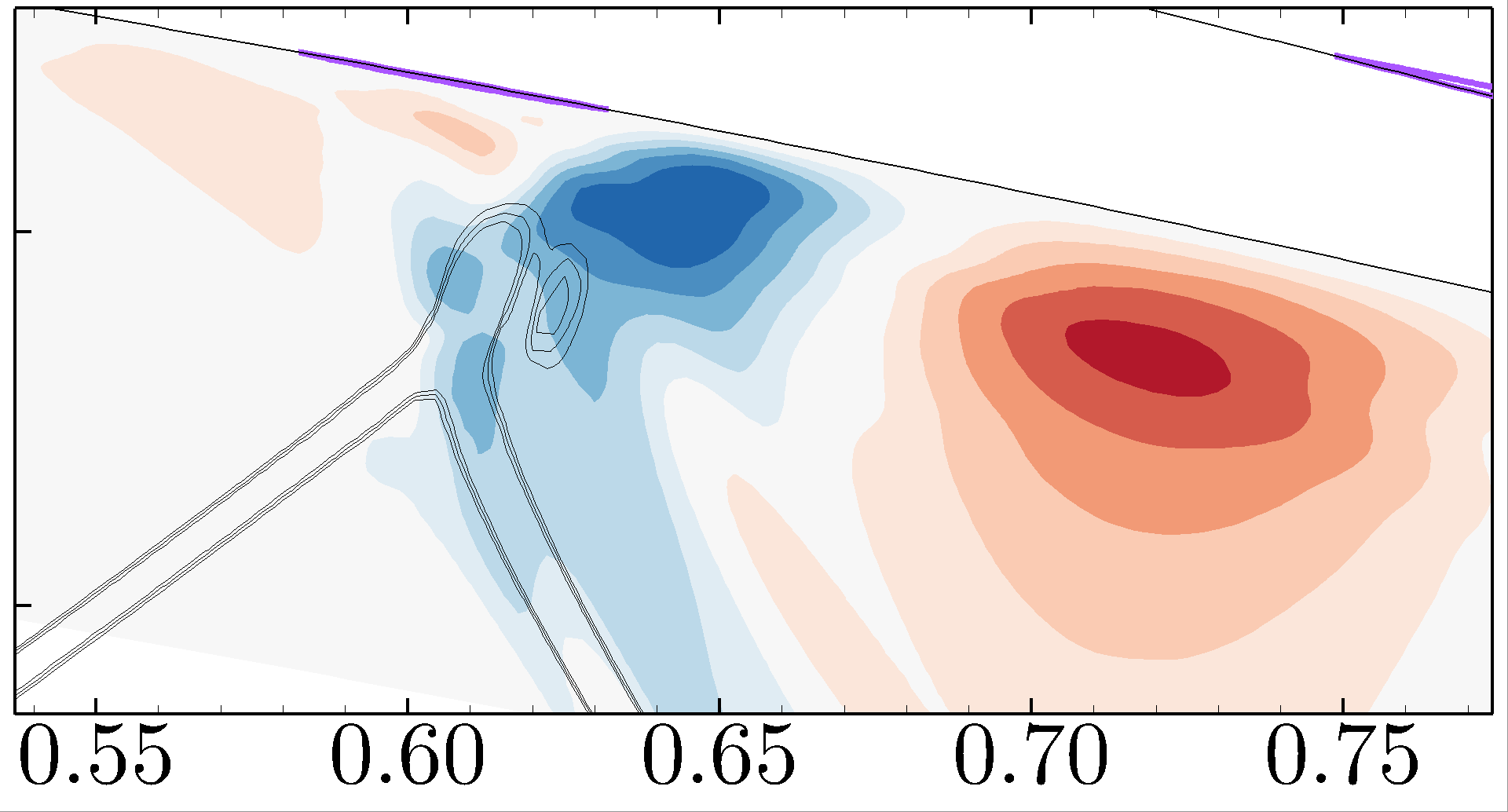}
 		\put(84,35){(f)}
 	\end{overpic}
 	\begin{overpic}[trim = 1mm 40mm 2mm 2mm, clip,width=.32\linewidth]{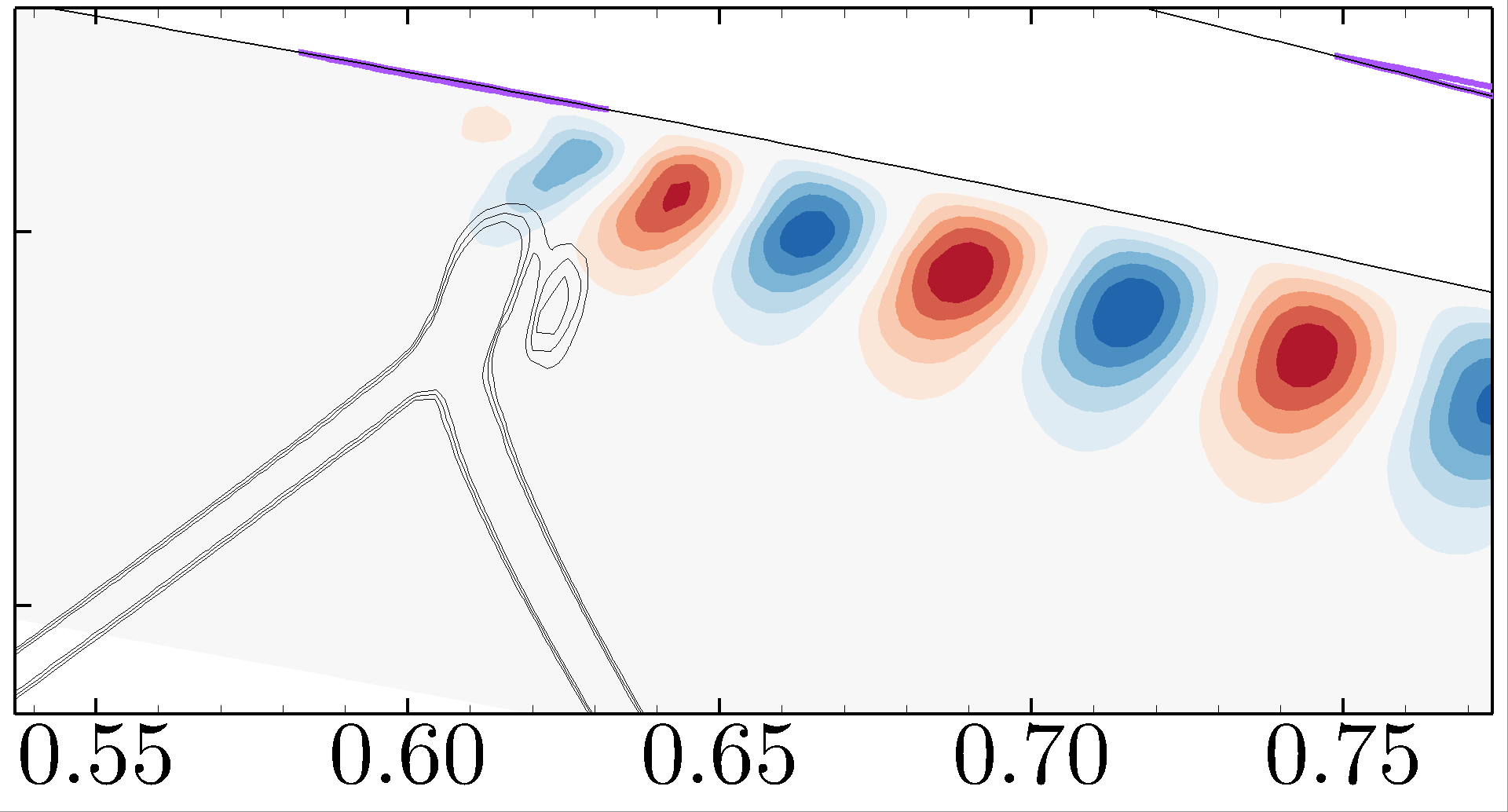}
 		\put(84,35){(j)}
 	\end{overpic}
 	 \begin{overpic}[trim = 1mm 2mm 2mm 2mm, clip,width=.32\linewidth]{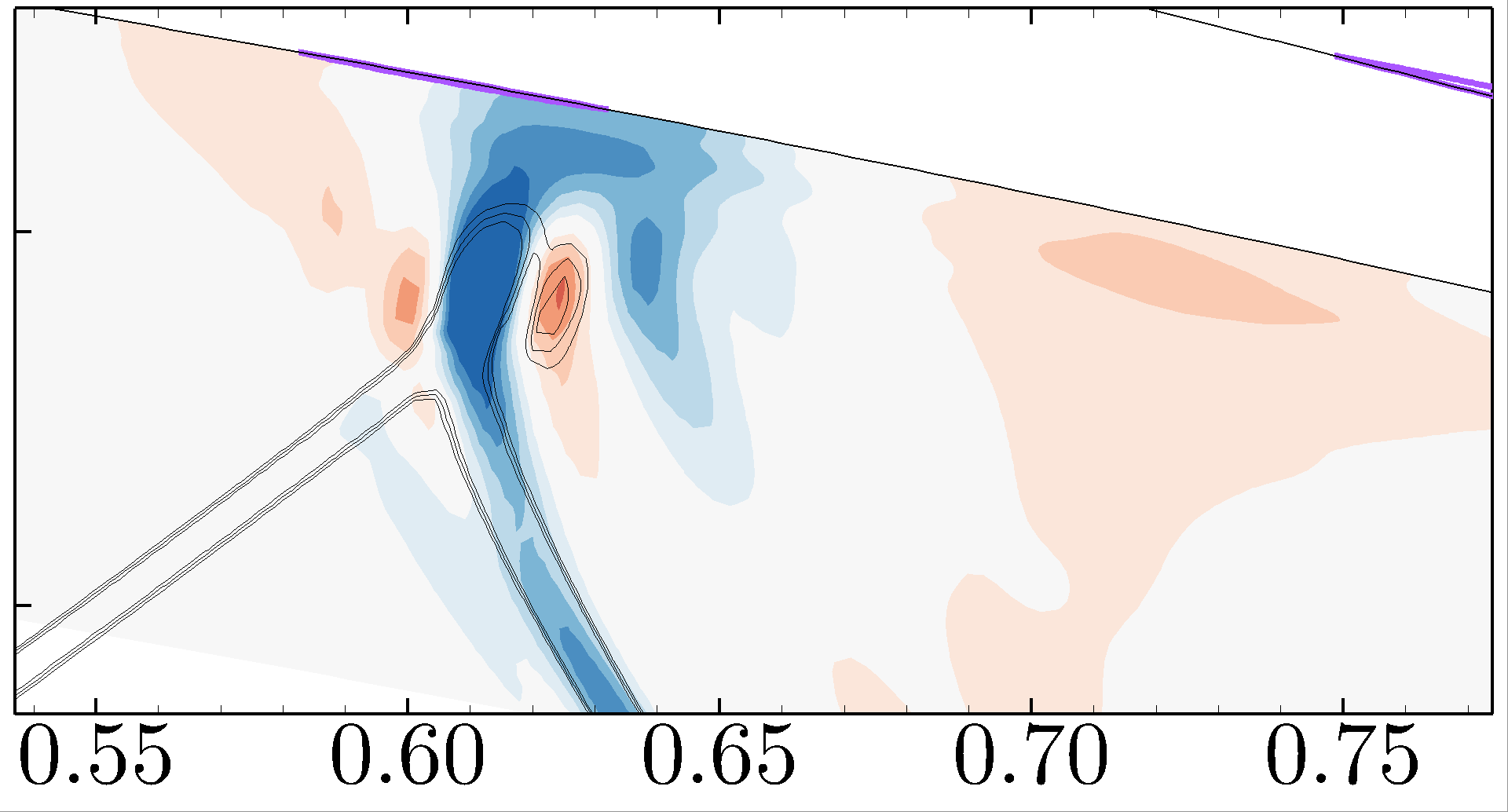}
		\put(84,42){(c)}
	\end{overpic}
	 \begin{overpic}[trim = 1mm 2mm 2mm 2mm, clip,width=.32\linewidth]{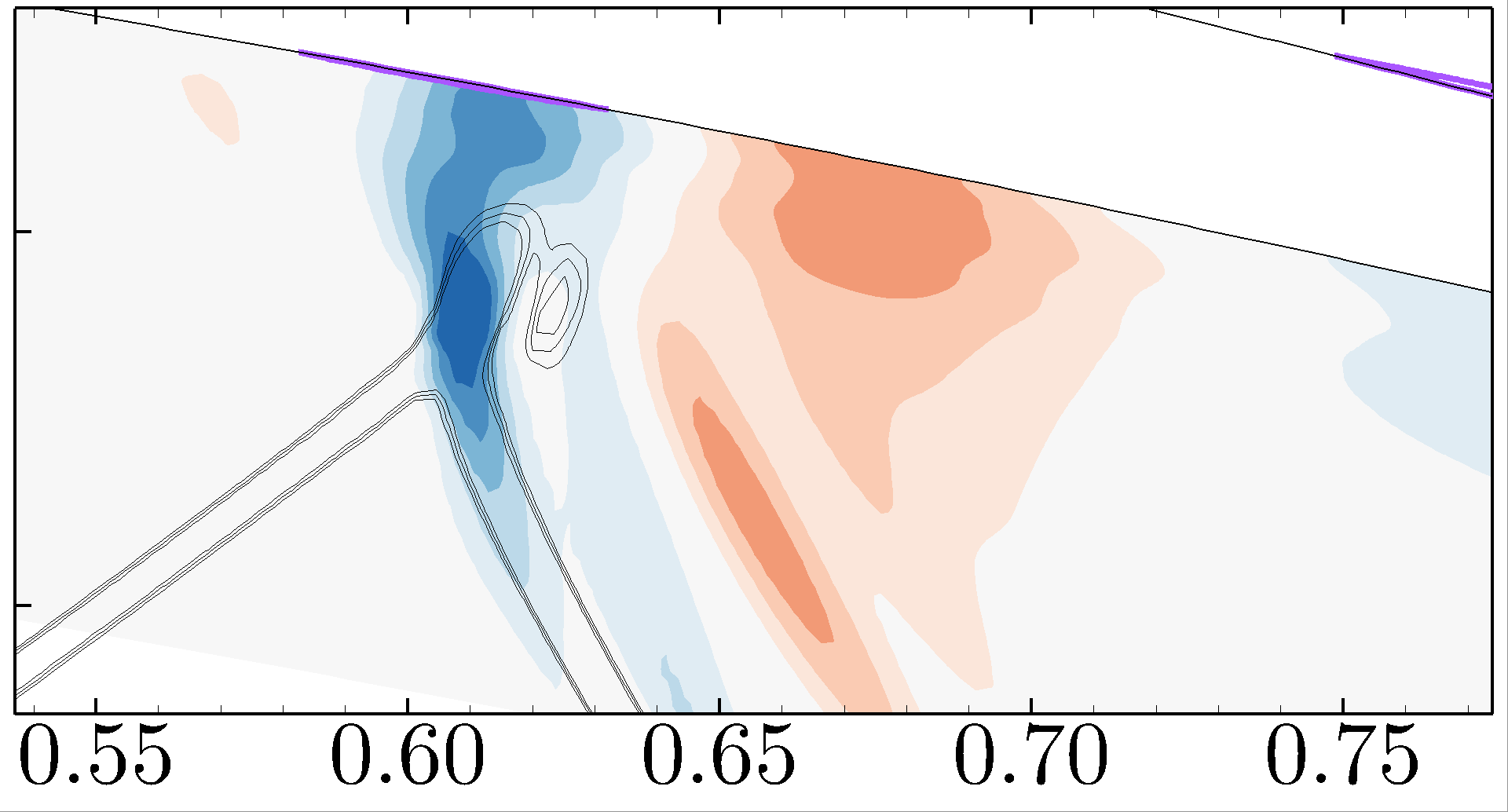}
 		\put(84,42){(g)}
 	\end{overpic}
 	\begin{overpic}[trim = 1mm 2mm 2mm 2mm, clip,width=.32\linewidth]{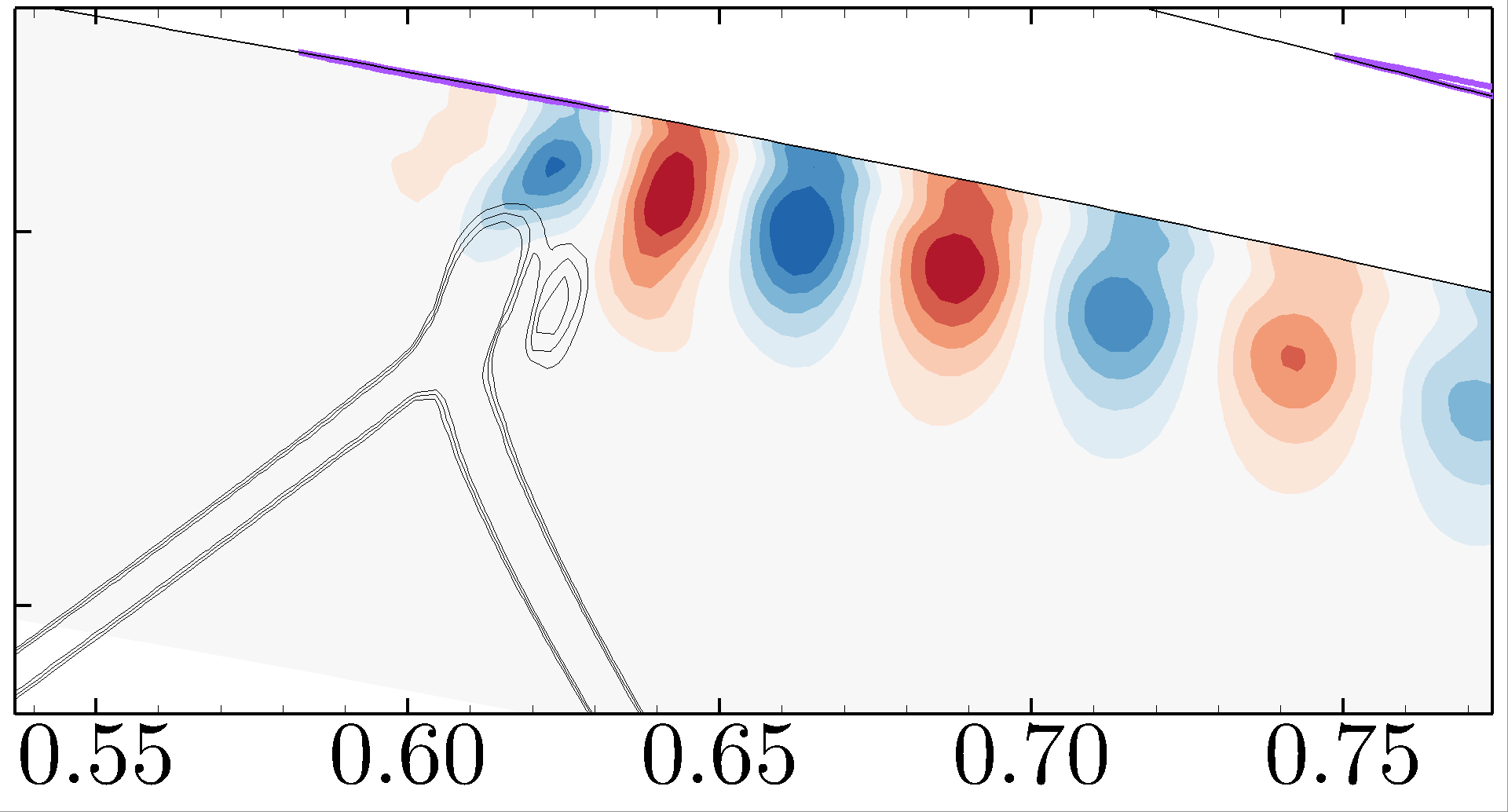}
 		\put(84,42){(k)}
 	\end{overpic}
 	\vskip 0.1cm
 	\begin{overpic}[trim = 1mm 1mm 1mm 1mm, clip,width=.32\linewidth]{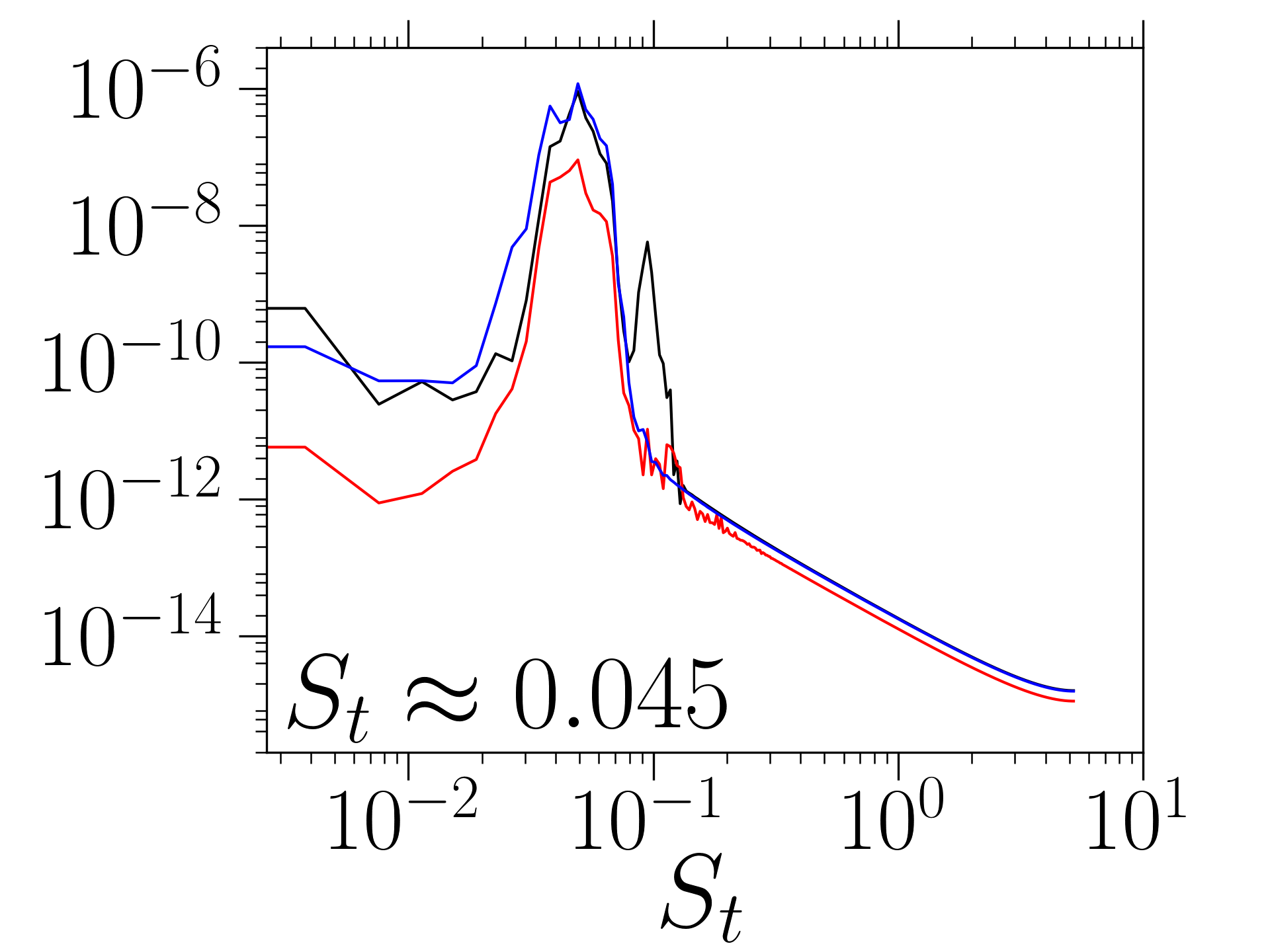}
		\put(80,62){(d)}
	\end{overpic}
	 \begin{overpic}[trim = 1mm 1mm 1mm 1mm, clip,width=.32\linewidth]{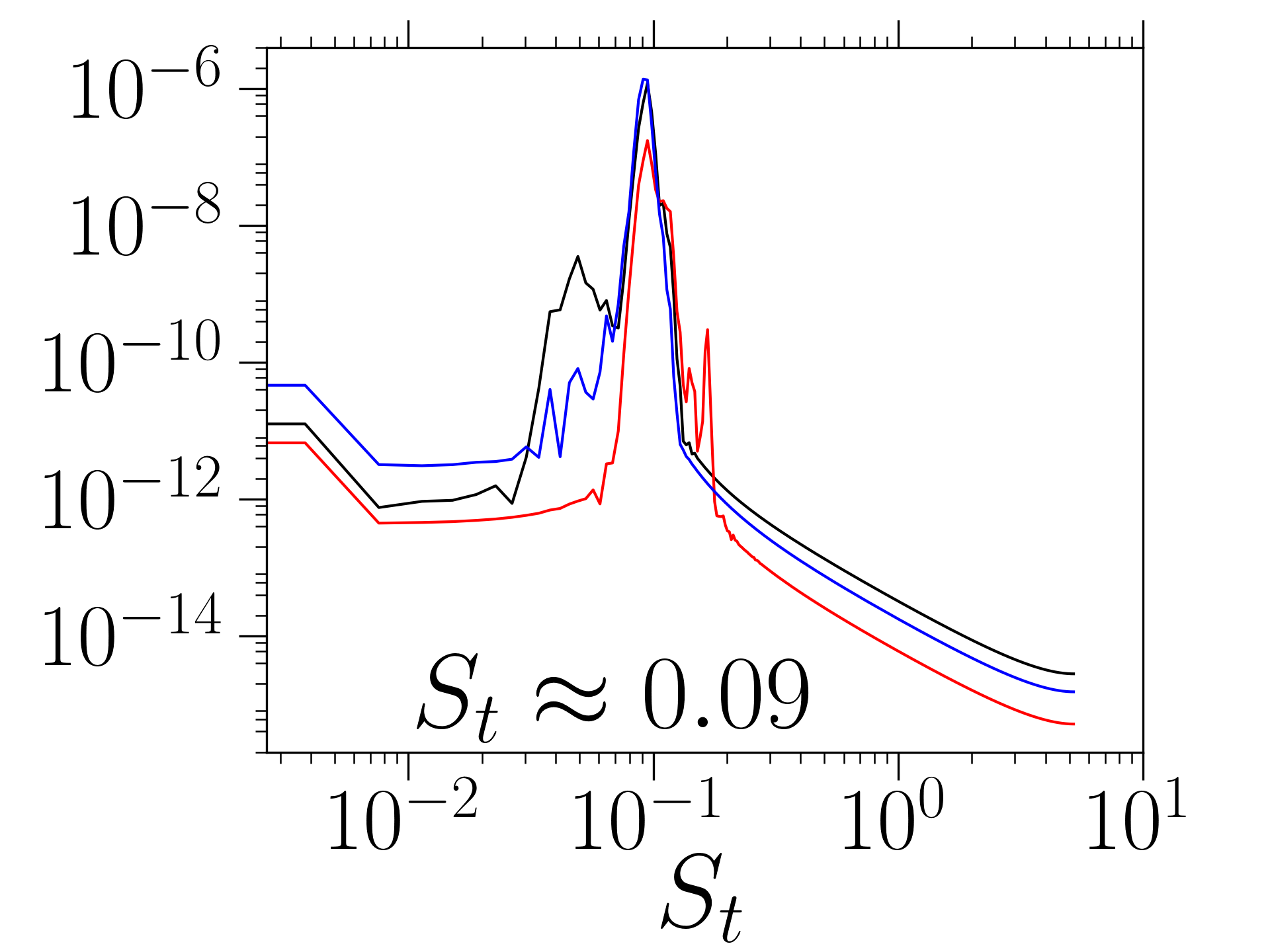}
		\put(80,62){(h)}
	\end{overpic}
	 \begin{overpic}[trim = 1mm 1mm 1mm 1mm, clip,width=.32\linewidth]{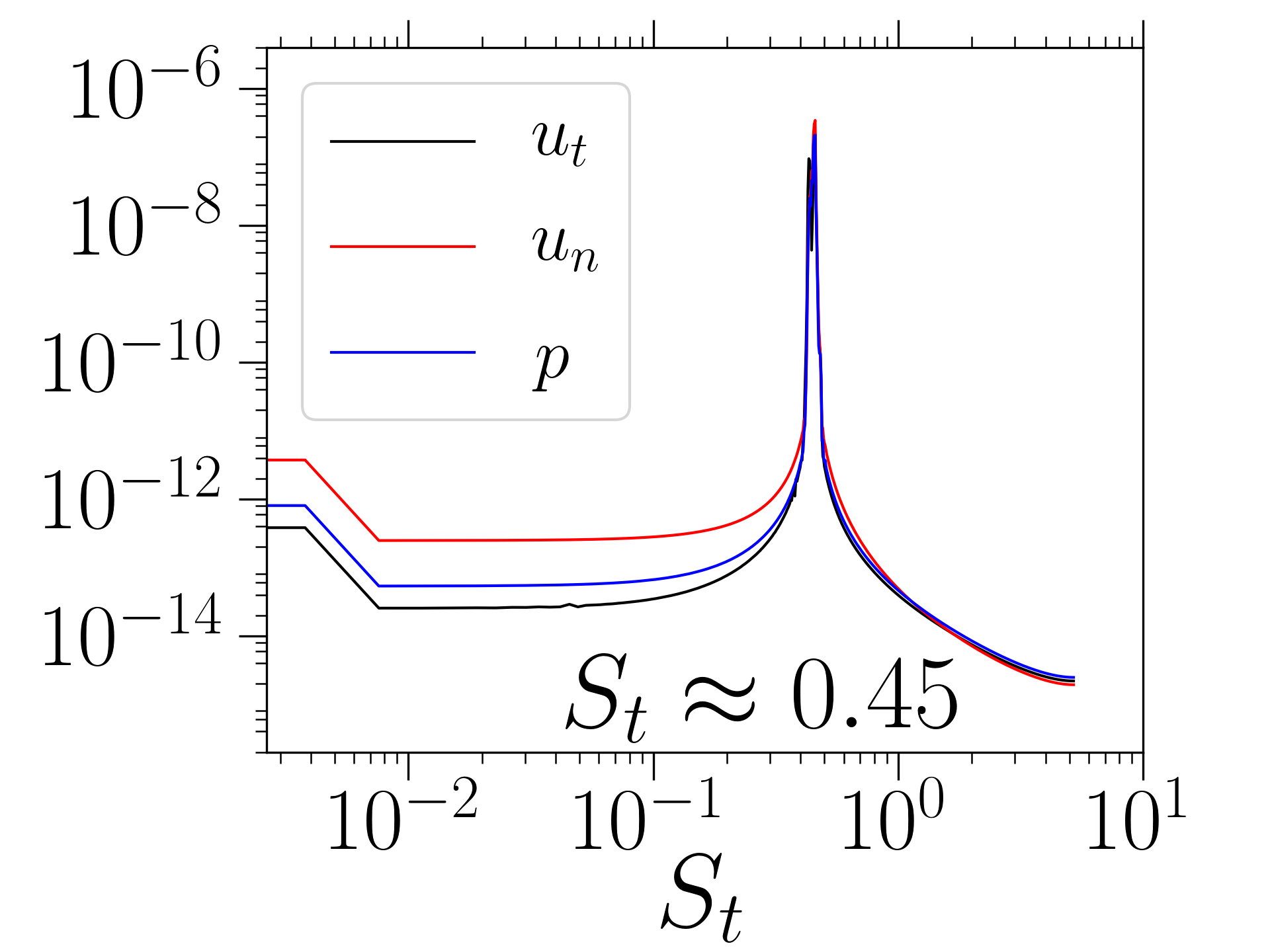}
		\put(80,62){(l)}
	\end{overpic}
	
	 \caption{Filtered POD modes selected based on frequency content along the pressure side. The first, second and third rows show the spatial modes of $u_{t}$, $u_{n}$, and $p$, respectively. The fourth row presents the PSD of the temporal modes.}
	\label{fig:pod_spatial_modes_PS}
\end{figure}

The filtered (spectral) proper orthogonal decomposition is applied to identify organized motions in the flow and their corresponding characteristic frequencies. This technique will enable associating the low-frequency unsteady motions observed in the SBLIs to particular flow features. The current modal analysis is conducted using 5610 snapshots with a time interval of 0.0048. A Gaussian filter is applied to 50 $\%$ of the POD correlation matrix. If a lower filter size is employed, noise can still be present in the low-frequency dynamics of the POD temporal modes. 
The first 100 POD modes are analyzed and we select specific POD modes based on the characteristic frequencies identified by the spectral analysis presented throughout this paper. 
The cumulative energy of the first 100 modes, for each independent quantity, is shown in Figs. \ref{fig:cumu_energy}(a) and \ref{fig:cumu_energy}(b) for the suction and pressure sides, respectively.
Figures \ref{fig:pod_spatial_modes_SS} and \ref{fig:pod_spatial_modes_PS} show the POD modes computed for a region covering the SBLIs on the suction and pressure sides of the airfoil, respectively. The first, second and third rows represent the spatial modes of tangential velocity $u_t$, normal velocity $u_n$, and pressure, respectively, and the last row depicts the PSD of the POD temporal modes.

On the suction side, the POD spatial modes associated with $St \approx 0.045$ are shown in Figs. \ref{fig:pod_spatial_modes_SS}(a) - \ref{fig:pod_spatial_modes_SS}(c). 
The POD modes of $u_t$ and $u_n$, from Figs. \ref{fig:pod_spatial_modes_SS}(a) and \ref{fig:pod_spatial_modes_SS}(b), highlight the tangential velocity fluctuations along the shear layer over the separation bubble and the wall-normal velocity fluctuations of the upstream compression waves and reattachment shock. These plots confirm that the bubble and shock motions are interrelated.
Pressure fluctuations are observed over the recirculation region in Fig. \ref{fig:pod_spatial_modes_SS}(c), indicating the bubble breathing. The POD pressure mode displays two major regions of fluctuations: a blue one related to the bubble leading edge, and a red one connected to its trailing edge. This result shows that these regions tend to fluctuate out of phase, confirming the findings from Fig. \ref{fig:correlation_SS}(b). Moreover, these regions of intense pressure fluctuations are associated with those of high spectral energy at $St \approx 0.045$, displayed in Fig. \ref{fig:PSD_maps}(a). The pressure fluctuations upstream and downstream of the bubble represent the compression waves and reattachment shock, respectively, as observed in the pressure RMS plot of Fig. \ref{fig:reynolds_stresses_SS_2D}(c).

Figures \ref{fig:pod_spatial_modes_SS}(e) - \ref{fig:pod_spatial_modes_SS}(g) show the POD spatial modes associated with $St \approx 0.12$. The characteristic frequency at $St \approx 0.12$ is most likely to be related to a low/mid frequency motion of the separation bubble, the subsequent shock oscillations and the shear layer flapping. The present frequency is also observed in the wall pressure PSDs of previous studies \citep{dupont_haddad_debieve_2006,Touber2009, priebe2012}, and \citet{adler_gaitonde_2018} reports that this frequency should correspond to bubble oscillations. Their results indicated the presence of high spectral energy associated to this frequency near the separation and reattachment points, as well as downstream of the SBLI. The pressure fluctuations at this frequency are also observed in the regions of high spectral energy displayed in Fig. \ref{fig:PSD_maps}(a).

The POD modes associated with $St \approx 0.4$ are presented in Figs. \ref{fig:pod_spatial_modes_SS}(i) - \ref{fig:pod_spatial_modes_SS}(k). This frequency was not clearly
identified by the previous spectral analysis due to the broadband content present around this particular frequency. With the spectral POD approach, we can investigate these mid-frequencies since the present modal decomposition technique works as a band-pass filter of the POD temporal modes. In the figures, the POD spatial modes illustrate the vortical structures of a K-H instability produced on the shear layer downstream the incident shock. Velocity and pressure fluctuations are also propagated along the reattachment shock which transports tangential disturbances once the flow is supersonic. Similar results have been reported in Refs. \citep{pasquariello_2017,Nichols_2017,Hu_2019,hu_2021}. 
The POD spatial modes for the first harmonic of the K-H instability at $St \approx 0.8$ are shown in Figs. \ref{fig:pod_spatial_modes_SS}(m) - \ref{fig:pod_spatial_modes_SS}(o). These modes are similar to the previous ones and display the shedding of smaller vortices, as well as the propagation of waves along the reattachment shock.  

On the pressure side, the POD modes associated with $St \approx 0.045$ are shown in Figs. \ref{fig:pod_spatial_modes_PS}(a) - \ref{fig:pod_spatial_modes_PS}(c). Similarly to the suction side, high fluctuations are found along the flow recirculation and shear layer regions, indicating the bubble motion. The pressure mode displays two regions of high fluctuations, where the red one is related with the bubble leading edge and upstream compression waves, while the blue region is connected to the bubble trailing edge and the reflected shock. The flow locations where these pressure fluctuations appear are associated with those of high spectral energy at $St \approx 0.045$ displayed in Fig. \ref{fig:PSD_maps}(c). The POD spatial mode of tangential velocity shows strong fluctuations along the shear layer, indicating its flapping. Moreover, high fluctuations of tangential and normal velocities, and pressure, are also observed downstream of the shock, suggesting the presence of large-scale structures on that region. This suggests that the frequency of $St \approx 0.045$ is associated with the separation bubble motion,  the subsequent reflected shock oscillation, shear layer flapping, and the convection of large-scale structures.

Figures \ref{fig:pod_spatial_modes_PS}(e) - \ref{fig:pod_spatial_modes_PS}(g) display the POD modes associated with $St \approx 0.09$, which is the first harmonic of $St \approx 0.045$. One can observe that the POD spatial modes are similar to those previously analyzed, and this frequency is also associated with a low/mid frequency motion of the separation bubble and the reflected shock. The transport of large-scale structures is also observed in the spatial modes for this frequency. High spectral energy associated to this frequency is present near the flow separation and reattachment points as can be seen in Fig. \ref{fig:PSD_maps}(c).
The POD modes at $St \approx 0.45$ are presented in Figs. \ref{fig:pod_spatial_modes_PS}(i) - \ref{fig:pod_spatial_modes_PS}(k). These modes display the shedding of K-H vortices along the shear layer, downstream of the shock. Differently from the suction side, high fluctuations are not transported along the reflected shock since these structures appear closer to the wall, where the flow is locally subsonic.

\section{Conclusions}

In the this work, an investigation of SBLIs is presented for a supersonic turbine cascade at Mach 2.0 and Reynolds number 395,000 using a wall-resolved large eddy simulation. Different shock structures are formed on both sides of the airfoil. On the suction side, an oblique shock impinges on the boundary layer and that leads to an intermittent reattachment shock downstream of the recirculation bubble.  
On the pressure side, the incident shock interacts with the boundary layer and generates a Mach reflection. The mean flow visualization shows that the separation bubbles have different topologies, with the suction side bubble being longer and thicker while the pressure side bubble is shorter and thinner.

Instantaneous flow visualizations illustrate the unsteady dynamics of the SBLIs, which involve breathing of the separation bubbles, flapping of the shear layers and shock motions. The presence of elongated streamwise structures is noticed in the incoming boundary layers on both sides of the airfoil and the passage of high-speed (low-speed) streaks through the bubbles leads to downstream (upstream) motions of the separation points. These motions are highlighted by the skin-friction distributions along the vane surface, which show wavy spanwise variations of the bubbles. Tangential velocity fluctuations computed along a plane near the wall reveal that the separation region conforms to the presence of negative incoming velocity fluctuations, while the reattached regions are associated with positive fluctuations. These findings are consistent with previous studies. 

A spatiotemporal analysis of the suction side separation bubble indicates that the separation and reattachment points are mostly out of phase. When the bubble undergoes a contraction, the instantaneous separation point moves considerably downstream with respect to its mean position, while the reattachment point moves slightly upstream. The opposite is true for the bubble expansion. Similar analyses are performed for the pressure side bubble but, contrary to the suction side, the separation and reattachment points tend to oscillate around their mean values, without large excursions and being less correlated.

The spectral analyses of the suction side separation and reattachment points, as well as of the bubble length, show the presence of two energetic low-frequency peaks at $St \approx 0.045$ and $St \approx 0.12$. On the suction side, the influence of the incoming boundary layer on the bubble low-frequency unsteadiness is investigated by means of pressure and velocity power spectral density maps, as well as space-time correlations. Results show that velocity fluctuations in the incoming boundary layer at the specific low-frequency of $St \approx 0.045$ excite both the bubble separation and reattachment locations. The motion of the separation bubble would in turn cause the reattachment shock motion. 
On the pressure side, the PSD map of wall pressure shows two dominant peaks at $St \approx 0.045$ and $0.09$. For the first peak, high spectral energy is observed upstream of the recirculation bubble and near the reattachment position, while the second peak displays regions of high energy content along the entire separation bubble and downstream of the SBLI. Differently from the suction side, the separation bubble is not excited by tangential velocity fluctuations at low-frequencies.

The filtered (spectral) POD is employed to identify organized motions in the SBLIs and their corresponding characteristic frequencies. On the suction side, the separation bubble and shock motions, and the shear layer flapping, are associated with the frequencies of $St \approx 0.045$ and $0.12$. 
The POD analysis also shows that the shedding of K-H vortices occurs at the frequency of $St \approx 0.4$. Velocity and pressure fluctuations are also observed propagating along the reattachment shock for this frequency. On the pressure side, the POD modes associated with $St \approx 0.045$ are related to the separation bubble motion, the subsequent reflected shock oscillation, shear layer flapping, and the convection of large-scale structures. The frequency of $St \approx 0.09$, which is the first harmonic of $St \approx 0.045$, is also related to the aforementioned flow features. Similarly to the suction side, the shedding of K-H structures is associated with a frequency of $St \approx 0.45$, but in this case, flow fluctuations are not transported along the reflected shock which is located away from the wall due to the Mach reflection.

The present turbine cascade configuration depicts some flow differences when compared to the more canonical studies found in literature (impinging oblique shock on a flat plate and compression ramp) due to the curved walls of the airfoils. This leads to some distinctive features such as the absence of a separation shock upstream the suction side bubble due to the convex surface curvature. In this case, the compression waves do not coalesce to form a separation shock, resulting in a lower pressure rise compared to the more canonical cases. The lack of a separation shock allows some streaks to flow over the bubble, where they reach the reattachment shock and burst. This mechanism would be responsible for the strong low-frequency fluctuations observed in the PSD maps at the mean reattachment point. On the pressure side, the concave wall leads to a Mach reflection with a strong normal shock away from the airfoil surface. Overall, a strong SBLI should lead to a higher recirculation region. However, the pressure side bubble is thin and small, departing from the common triangular shapes observed in other studies. 
This behavior may be related to the state of the incoming flow since a concave curvature has the potential of destabilizing the boundary layer and enhancing turbulent mixing. The additional momentum added to the boundary layer by the turbulence would sustain a higher adverse pressure gradient induced by the shock without a large flow separation.

\begin{acknowledgments}

The authors acknowledge the financial support received from Funda\c{c}\~ao de Amparo \`a Pesquisa do Estado de S\~ao Paulo, FAPESP, under Grants No.\ 2013/08293-7, No.\ 2019/26196-5, and No. 2021/06448-0. Numerical simulations were performed in the SDumont supercomputer at the National Laboratory for Scientific Computing (LNCC/MCTI, Brazil) through project SimTurb, and also on the
Stampede 2 cluster at the Texas Advanced Computing Center (TACC), via allocation TG-CTS200027. For the latter, this work used the Extreme Science and Engineering Discovery Environment (XSEDE) \citep{xsede}, which is supported by the National Science Foundation Grant No. ACI-1548562.

\end{acknowledgments}

\section{Appendix}
\label{section:appendix}

In this Appendix, a validation test case is presented to demonstrate the ability of the numerical code to capture shock waves and boundary layers, including their interactions. Comparisons of the present LES against RANS solutions are also provided for different turbulence models.

\subsection{Two-dimensional laminar shock-boundary layer interaction}

The setup investigated is that from \citet{hakkinen} and \citet{katzer_1989}, where a two-dimensional oblique shock-wave impinges on a laminar boundary layer. The freestream Mach number is 2.0 and the Reynolds number based on the distance between the plate leading edge to the shock impingement point is $Re_{x_0} = 2.96 \times 10^5$. 
A wedge with angle $\theta = 3^{\circ}$ is used to create an oblique shock wave with an incident angle $\sigma =32.58^{\circ}$. The net pressure rise across the impinging shock and its reflection is equal to 1.4. The fluid is assumed to be a calorically perfect gas with $\gamma = 1.4$ and $Pr = 0.72$.

At the inflow, the conservative flow variables are imposed according to a laminar boundary layer profile obtained from a numerical solution of a supersonic flow over a flat plate. For the outflow, a boundary condition based on the Navier-Stokes characteristic boundary condition (NSCBC) \cite{Poinsot1992} is employed. A no-slip adiabatic wall is applied along the flat plate surface. In the far-field boundary opposite to the wall, a characteristic boundary condition based on Riemann invariants is employed, where the far-field conditions correspond to the Rankine-Hugoniot shock-jump conditions at $x=40$, which is the position where the shock wave is introduced at the top domain boundary. A damping sponge is also used near the inflow (outside of the boundary layer) and outflow to minimize the reflection of disturbances. The simulation is advanced in time until the nondimensional time $t=6,000$ is achieved, which is that required to obtain a converged flow \cite{Sansica}.
\begin{figure}
	\begin{center}
		{\includegraphics[trim = 2mm 2mm 2mm 2mm, clip, width=0.99\textwidth]{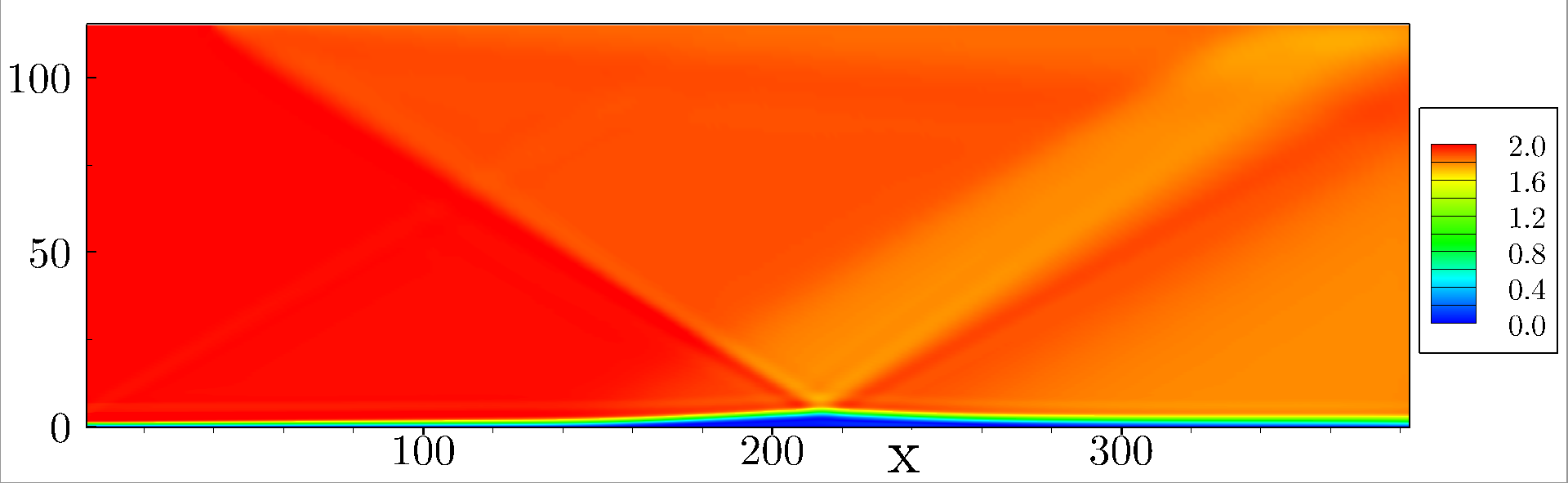}}
	\end{center}
    \caption{Contours of Mach number for the laminar shock-boundary layer interaction.}
	\label{fig:mach_validation}
\end{figure}

\begin{figure}
	\begin{overpic}[trim = 1mm 1mm 1mm 1mm, clip,width=.49\linewidth]{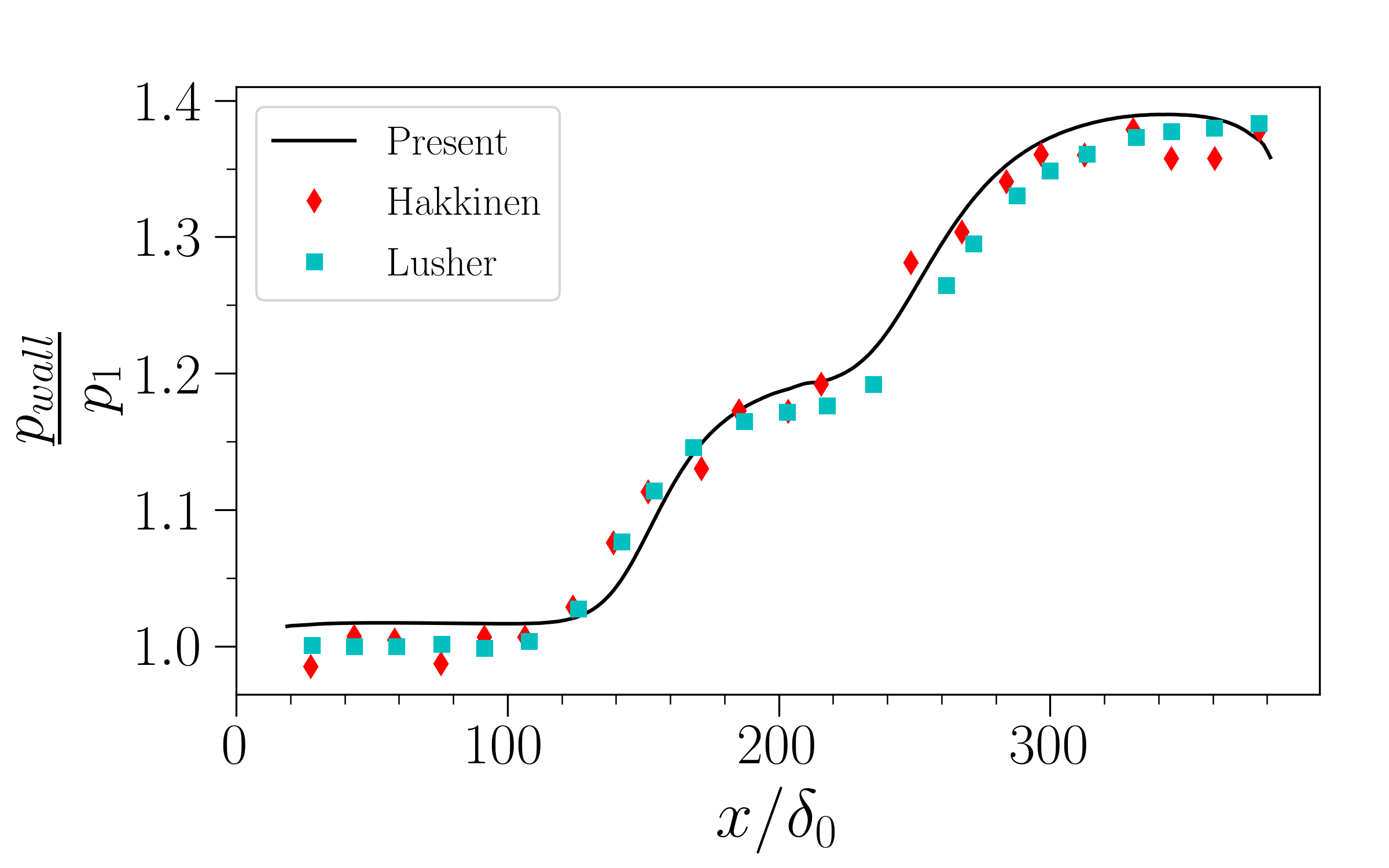}
		\put(0,55){(a)}
	\end{overpic}
	\begin{overpic}[trim = 1mm 1mm 1mm 1mm, clip,width=.49\linewidth]{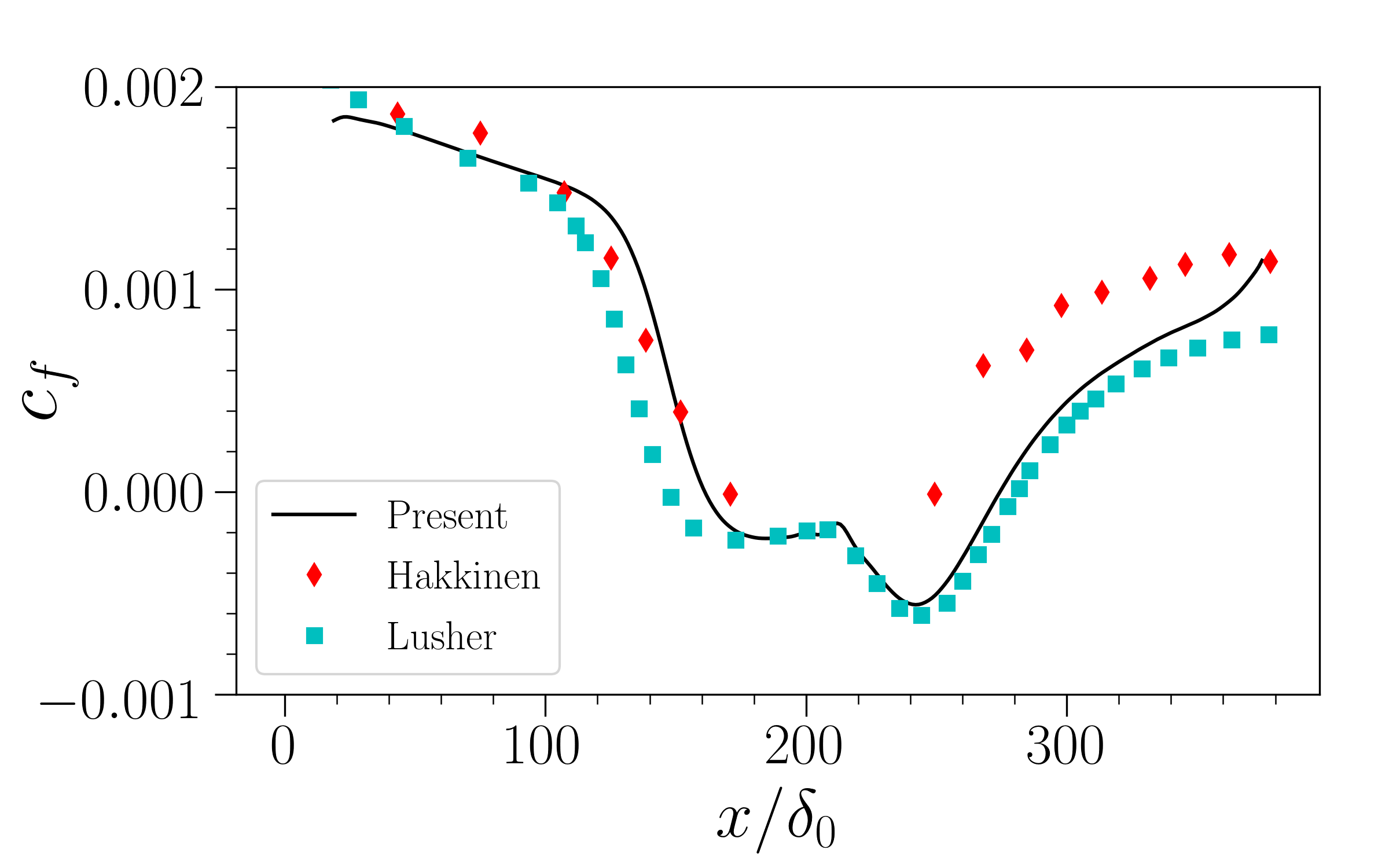}
		\put(-4,55){(b)}
	\end{overpic}
	\caption{Comparison of wall pressure (left) and skin-friction coefficient (right) distributions with the experimental data from \citet{hakkinen} and the numerical simulation from \citet{lusher}.}
	\label{fig:coeff_validation}
\end{figure}

Figure \ref{fig:mach_validation} shows contours of Mach number at time $t=6,000$. One can observe the interaction between the oblique shock impinging on the laminar boundary layer causing flow separation. The wall pressure and skin-friction coefficients along the plate are shown in Figs. \ref{fig:coeff_validation}(a) and \ref{fig:coeff_validation}(b), respectively. Results are compared to experimental \cite{hakkinen} and numerical \cite{lusher} reference solutions and good agreement is observed. Some discrepancies are noticed in the $c_f$ recovery region, but other numerical studies present similar results \cite{Sansica,Britton,lusher}.

\subsection{Comparison between LES and RANS solutions of supersonic turbine cascade}

It is worth mentioning that there are no experimental data available in the literature for this case. Hence, this section shows comparisons between LES and RANS solutions for the present supersonic turbine cascade. The flow conditions and geometrical details are those described in Sec. \ref{section:configurations}. 
The RANS calculations are obtained using the commercial solver CFD++ \cite{metacomp}. The time-marching of the equations is carried out by an implicit Euler scheme while a second-order spatial discretization is applied for convective and diffusive fluxes. Convective fluxes are calculated with the Harten-Lax-van Leer-contact (HLLC) approximate Riemann solver with a continuous TVD limiter. Three RANS turbulence models are used including the SST, cubic $k$-$\epsilon$ and the Goldberg models. For all RANS cases, a $3\%$ turbulence intensity is employed at the inlet of the domain.
\begin{figure}
	\begin{overpic}[trim = 1mm 5mm 2mm 15mm, clip,width=.49\linewidth]{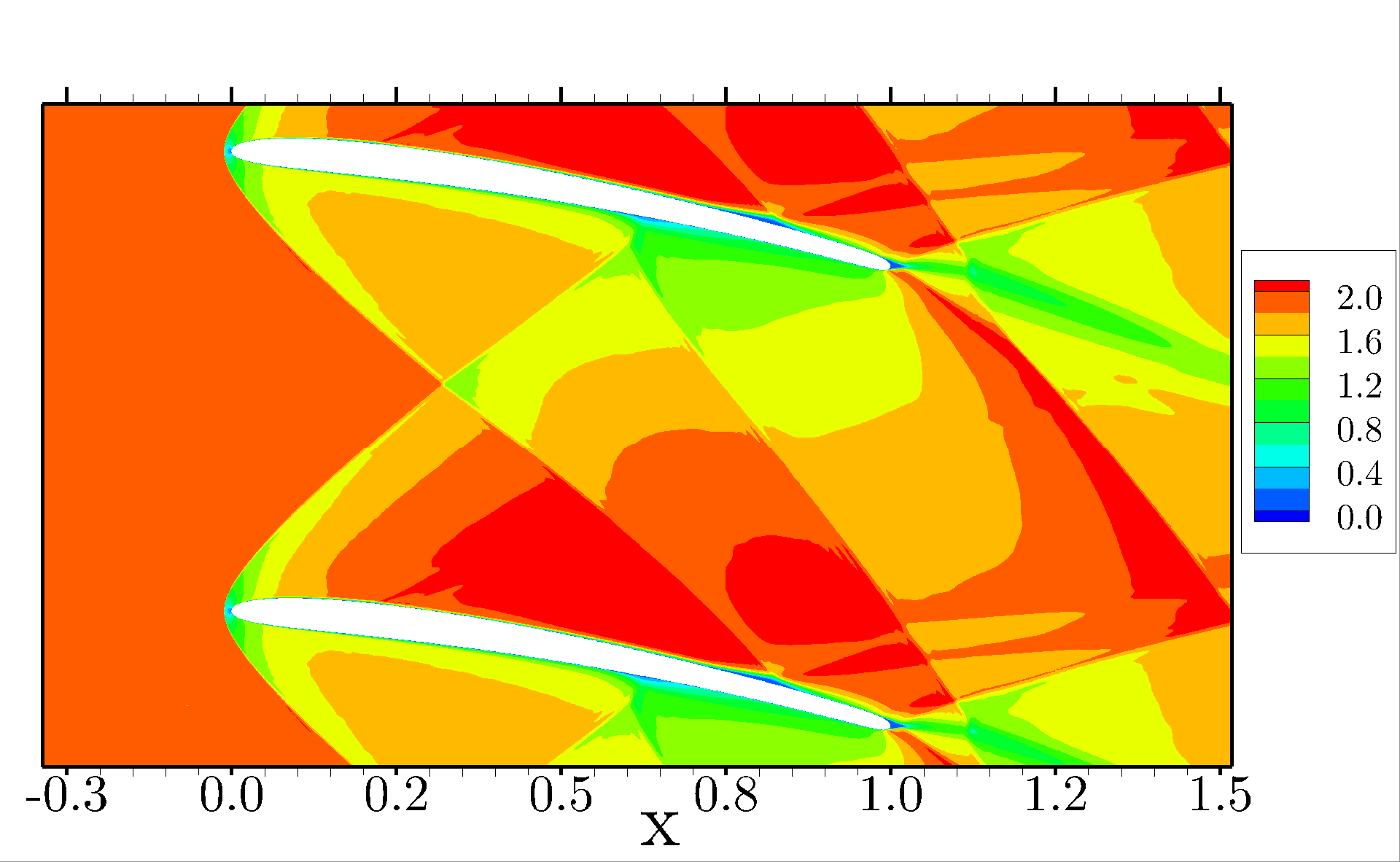}
		\put(5,47){(a)}
	\end{overpic}
	\begin{overpic}[trim = 1mm 5mm 2mm 15mm, clip,width=.49\linewidth]{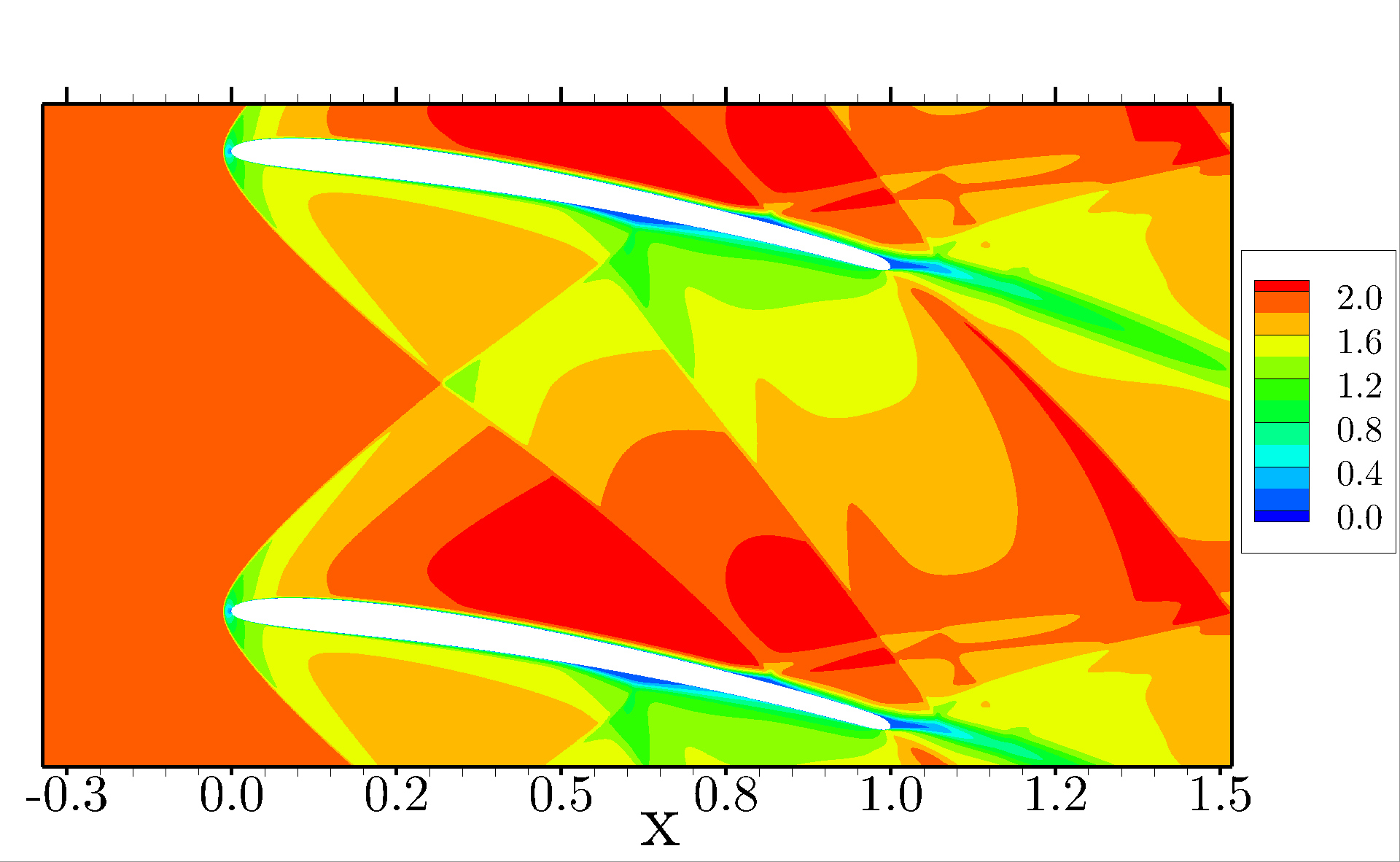}
		\put(5,47){(b)}
	\end{overpic}
	\begin{overpic}[trim = 1mm 5mm 2mm 15mm, clip,width=.49\linewidth]{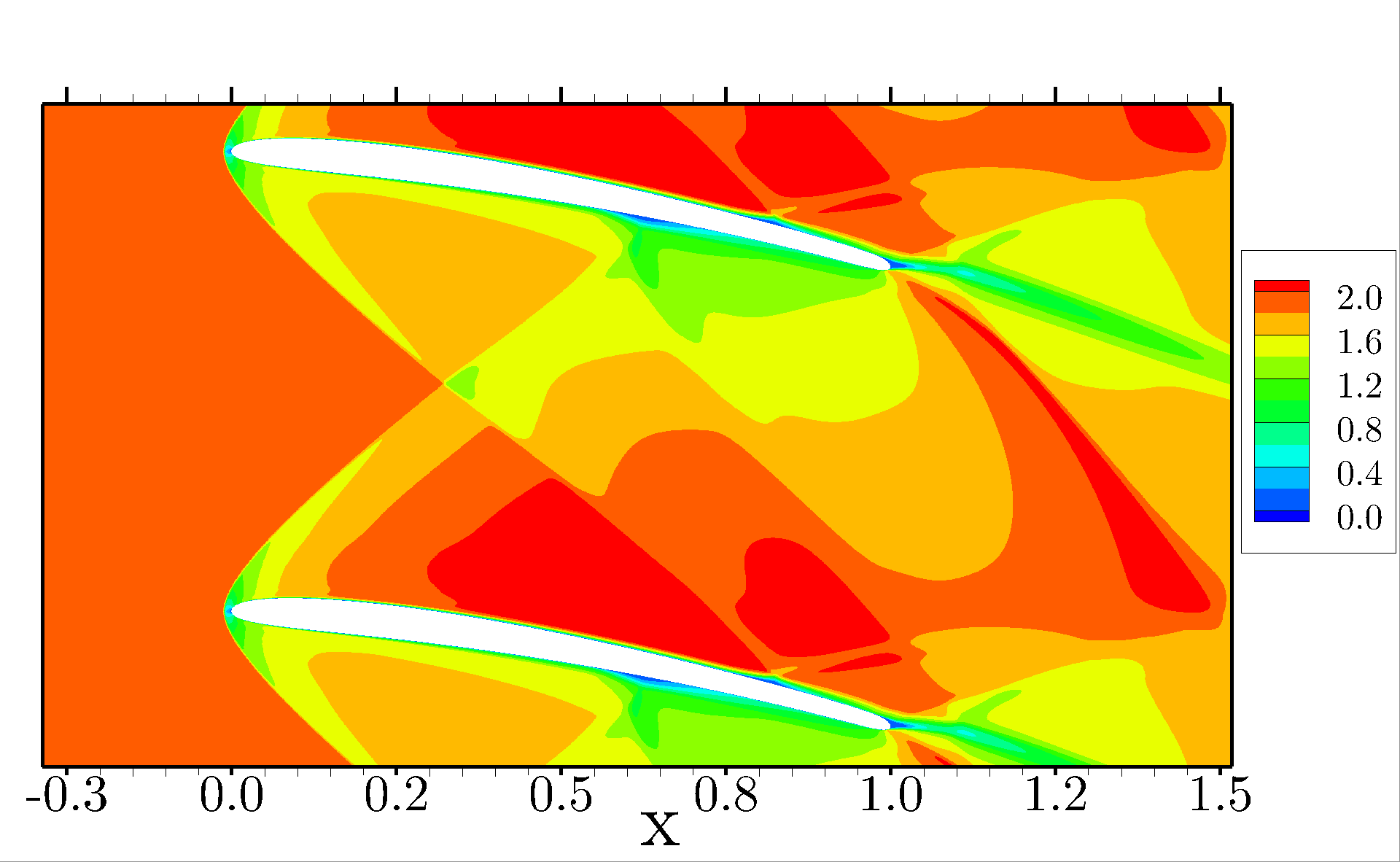}
		\put(5,47){(c)}
	\end{overpic}
	\begin{overpic}[trim = 1mm 5mm 2mm 15mm, clip,width=.49\linewidth]{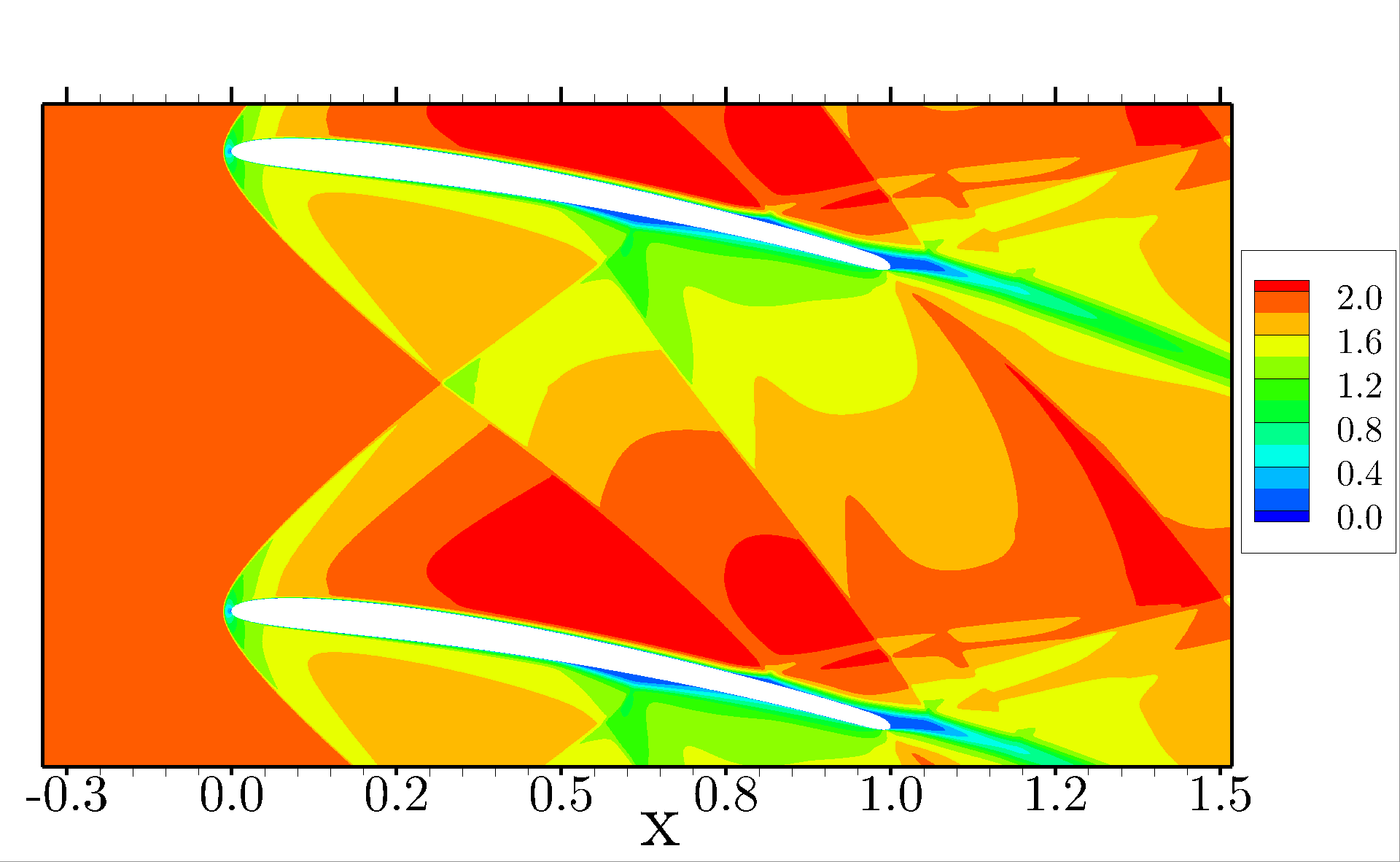}
		\put(5,47){(d)}
	\end{overpic}
	\caption{Comparison of mean Mach number contours between LES and RANS models with (a) LES, (b) SST, (c) cubic $k$-$\epsilon$, and (d) Goldberg.}
	\label{fig:rans_mach}
\end{figure}

\begin{figure}
	\begin{overpic}[trim = 1mm 1mm 1mm 1mm, clip,width=.49\linewidth]{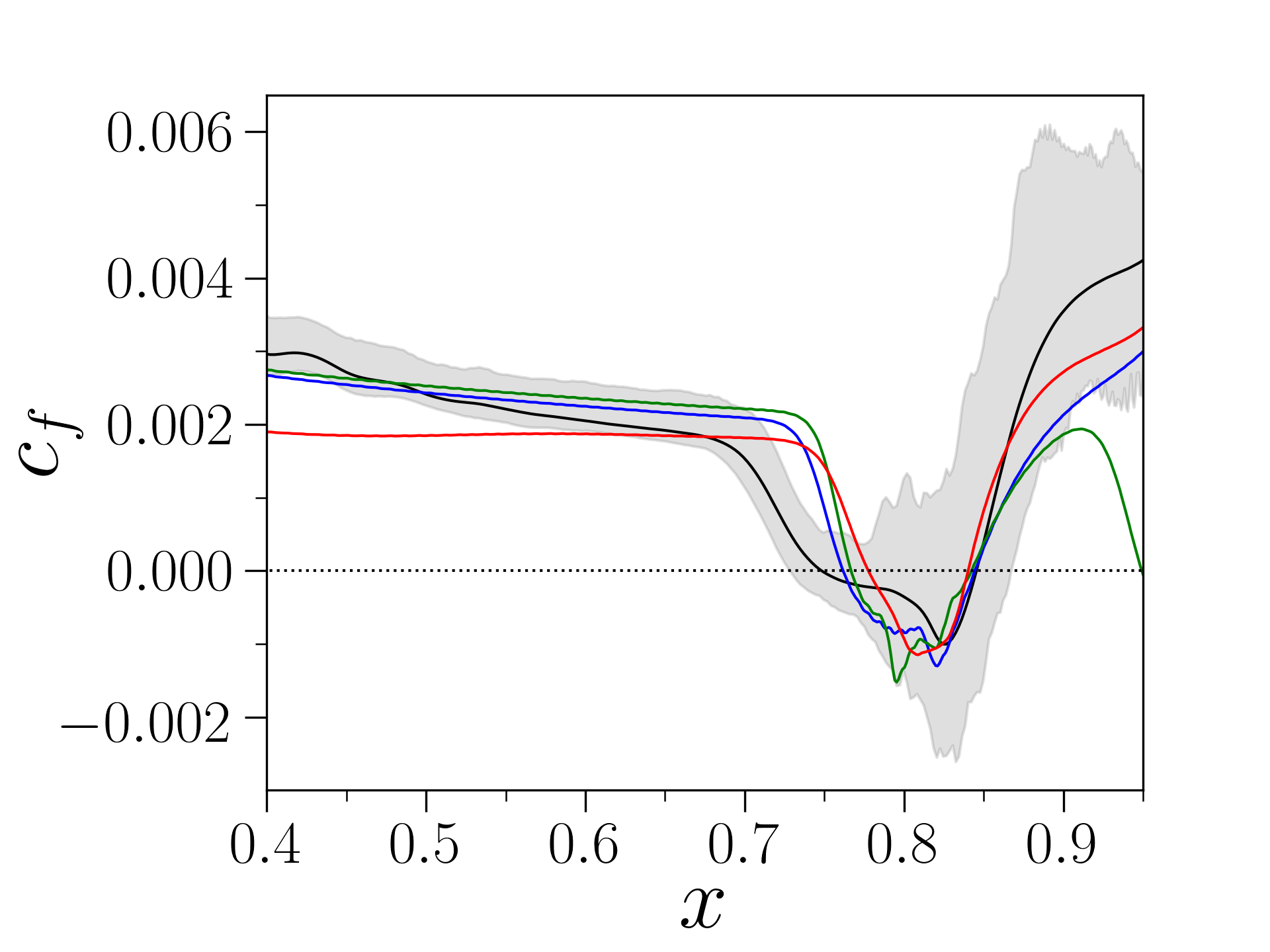}
		\put(-1,65){(a)}
	\end{overpic}
	\begin{overpic}[trim = 1mm 1mm 1mm 1mm, clip,width=.49\linewidth]{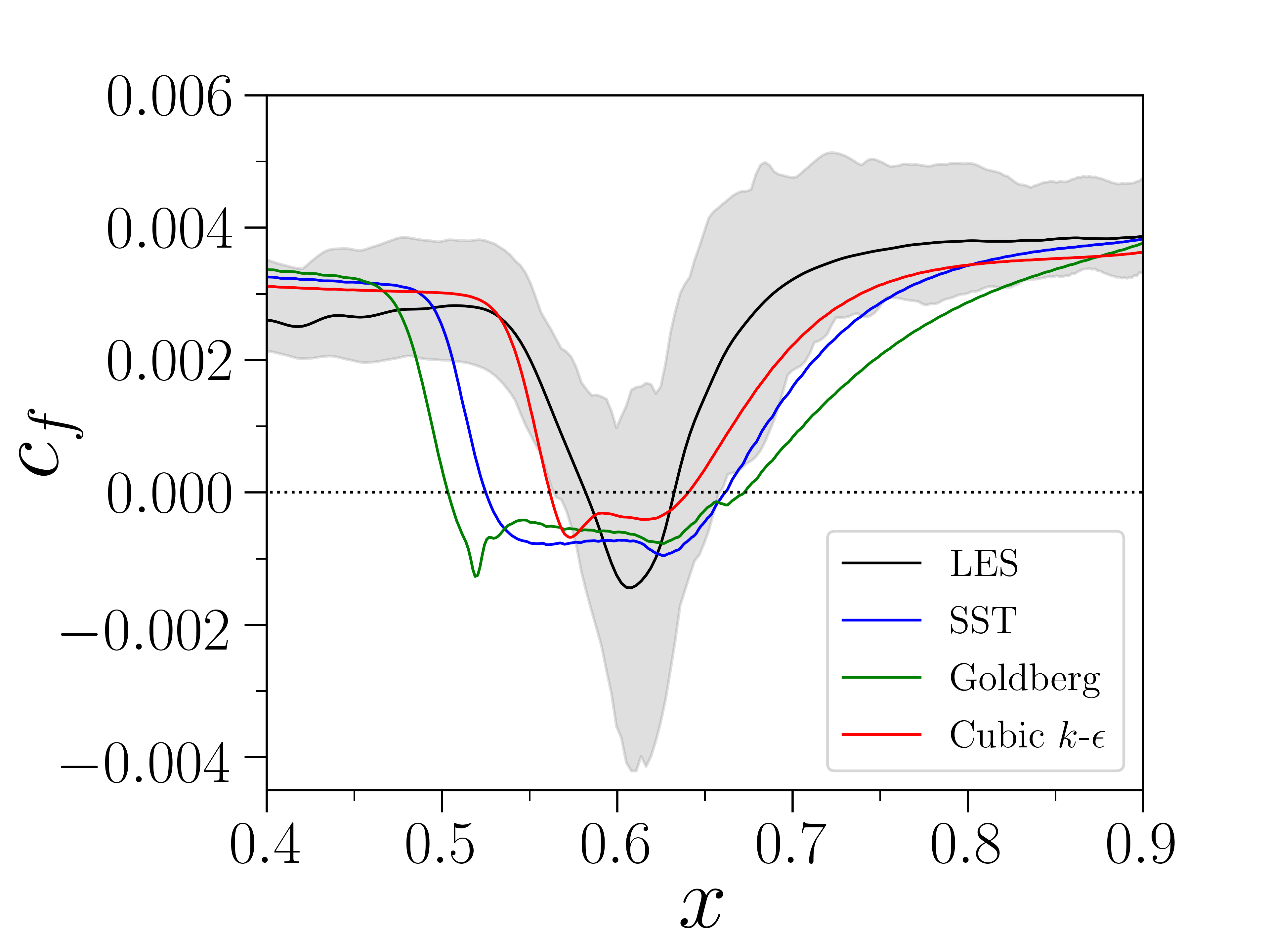}
		\put(-1,65){(b)}
	\end{overpic}
	\begin{overpic}[trim = 1mm 1mm 1mm 1mm, clip,width=.49\linewidth]{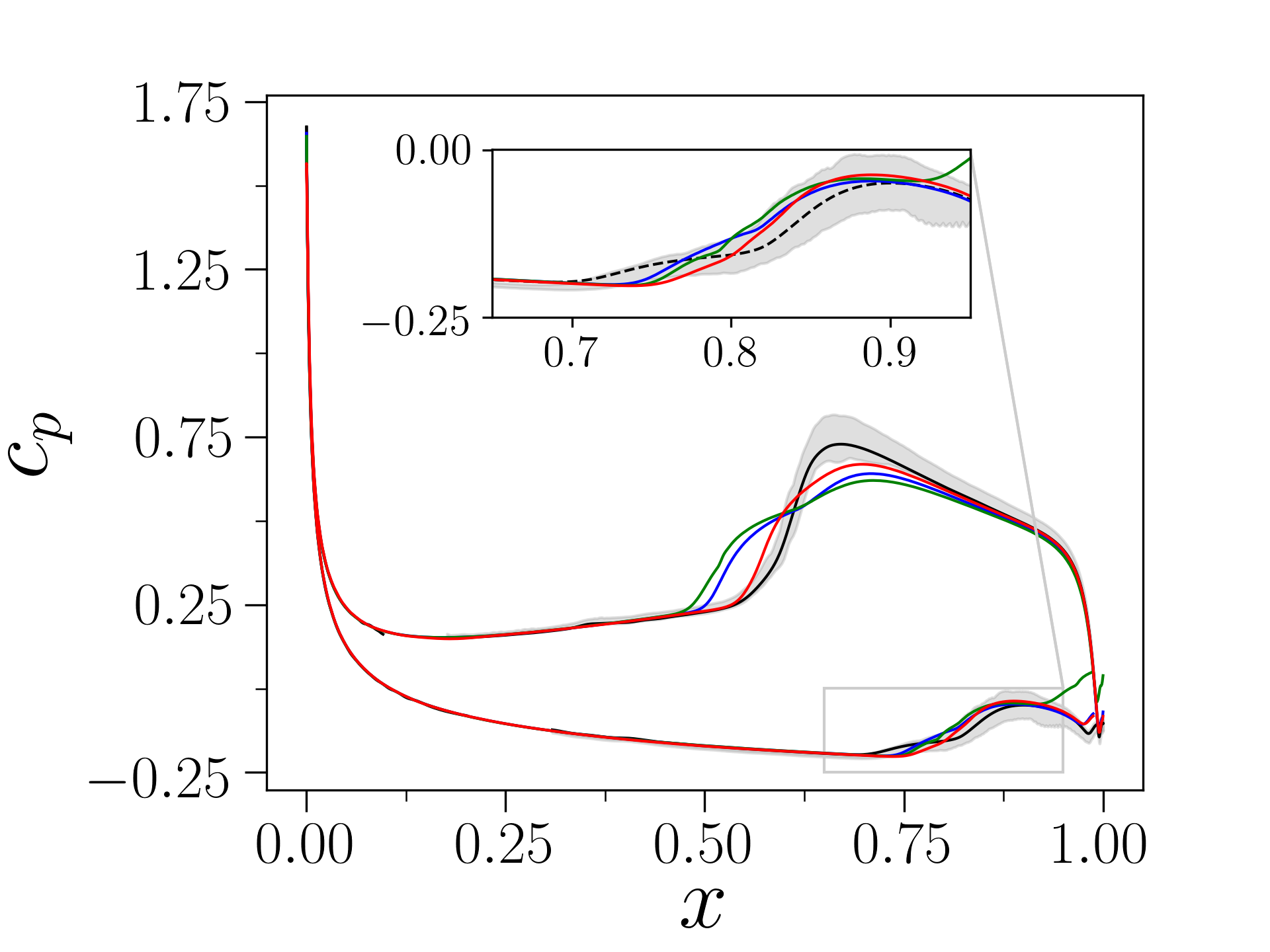}
		\put(0,65){(c)}
	\end{overpic}
	\caption{Comparison between LES and RANS models for (a) skin-friction coefficient on the suction side, (b) skin-friction coefficient on the pressure side, and (c) pressure coefficient.}
	\label{fig:rans_coefficients}
\end{figure}

Figure \ref{fig:rans_mach} shows a comparison between the LES and RANS solutions in terms of the mean Mach number contours. One can observe that the flow contours are very alike, especially on the blade suction side. Small discrepancies can be noticed on the pressure side, where a more pronounced separation region is noticed for the RANS solutions, except for the cubic $k$-$\epsilon$ model. Despite these differences, one can see that the shock wave structures are similar for all calculations. Figures \ref{fig:rans_coefficients}(a)-\ref{fig:rans_coefficients}(c) display plots of mean skin friction and pressure coefficient distributions along the chord. The gray shaded contours represent minimum and maximum values obtained by the LES. Along most of the chord, the suction side distributions of $c_f$ obtained by the RANS calculations fall in between the LES min and max solutions. However, the cubic $k$-$\epsilon$ model has a plateau upstream the recirculation bubble and the Goldberg model shows a sudden drop in $c_f$ toward the trailing edge. The RANS separation points appear downstream compared to that of the LES but the reattachment locations are similar. In general, the separation bubbles have similar lengths. The bubble sizes on the pressure side show larger discrepancies, with the cubic $k$-$\epsilon$ model achieving the best comparison with the LES. Comparisons in terms of $c_p$ show good agreement on the suction and pressure sides. However, for the latter, the SST and Goldberg models do not capture the steep pressure rise along the shock due to the larger bubble sizes predicted by these solutions.

\bibliography{bibfile}

\end{document}